\documentclass[%
reprint,
amsmath,amssymb,
pra,superscriptaddress
]{revtex4-1}
\usepackage{amsmath,amssymb,graphicx}
\usepackage{hyperref}
\usepackage{verbatim}
\usepackage{amsfonts}
\usepackage{braket}

\usepackage[usenames,dvipsnames]{xcolor}
\definecolor{darkerblue}{rgb}{0.2,0.2,0.5}

\usepackage{afterpage}

\newcommand{\bear}{\begin{array}}
\newcommand{\ear}{\end{array}}

\newcommand{\beq}{\begin{eqnarray}}
\newcommand{\eeq}{\end{eqnarray}}
\newcommand{\beqa}{\begin{eqnarray}}
\newcommand{\eeqa}{\end{eqnarray}}

\newcommand{\nn}{\nonumber}

\newcommand{\bk}[1]{\left(#1\right)}
\newcommand{\Bk}[1]{\left[#1\right]}

\newcommand{\trace}{\operatorname{tr}}

\def\be{\begin{equation}}
\def\ee{\end{equation}}
\def\bea{\begin{eqnarray}}
\def\eea{\end{eqnarray}}

\def\la{\langle}
\def\ra{\rangle}

\def\sJ{\mathcal{J}}
\def\sK{\mathcal{K}}
\def\sL{\mathcal{L}}

\def\tildepsi{\tilde{\psi}}
\def\tildephi{\tilde{\phi}}
\def\tildechi{\tilde{\chi}}

\def\OMIT#1{{}}
\newcommand{\lsim}{\mathrel{\rlap{\lower4pt\hbox{\hskip1pt$\sim$}}
    \raise1pt\hbox{$<$}}}         
\newcommand{\gsim}{\mathrel{\rlap{\lower4pt\hbox{\hskip1pt$\sim$}}
    \raise1pt\hbox{$>$}}}         

\begin{document}
\title{Quantum limits of parameter estimation in long-baseline imaging}

\author{Aqil Sajjad}
\email{aqilsajjad@arizona.edu}
\affiliation{James C. Wyant College of Optical Sciences, University of Arizona, Tucson, AZ 85721, USA}
\author{Michael R Grace}
\affiliation{James C. Wyant College of Optical Sciences, University of Arizona, Tucson, AZ 85721, USA}
\author{Saikat Guha}
\affiliation{James C. Wyant College of Optical Sciences, University of Arizona, Tucson, AZ 85721, USA}
\affiliation{Department of Electrical and Computer Engineering, University of Arizona, Tucson, AZ 85721, USA}

\begin{abstract}
Telescope systems with distributed apertures are a well-established approach for boosting resolution in astronomical imaging. However, theoretical limits on quantitative imaging precision, and the fundamentally best possible mode-sorting, beam-combining and detection schemes to use with such arrays, remain largely unexplored. Using mathematical tools of the quantum and classical Cr\'amer-Rao bounds, we perform analyses showing the fundamental origins of the enhancement provided by distributed imaging systems, over and above a single monolithic telescope, and consider the precision with which one can estimate any desired parameter embedded in a scene's incoherent radiation with a multi-aperture imaging system. We show how quantum-optimal measurements can be realized via beam-combination strategies of two classes: (1) {\em multi-axial}: where light from different apertures is directed to a common focal plane, e.g., of a segmented-aperture telescope; and (2) {\em co-axial}: where light collected at each aperture, e.g., telescope sites of a long-baseline array, is routed to an optical interferometer. As an example, we show an explicit calculation of the quantum Fisher information (QFI) for estimating the angular separation between two point emitters using two identical apertures separated by a baseline distance. We show that this QFI splits instructively into additive contributions from the single apertures and from the baseline. We quantify the relative benefits of intra-telescope (e.g., spatial-mode) optical processing and inter-telescope beam combination. We show how both receiver designs can be used to capture both sources of information and discuss how similar methods could be extended to more general imaging tasks. We discuss translating QFI-attaining measurements to explicit receiver designs, and the use of pre-shared entanglement to achieve the QFI when it is impractical to co-locate and combine light collected by the apertures.
\end{abstract}
\maketitle

\section{Introduction}

In recent years, studies based on quantum information theory have started to shed new light on Sub-Rayleigh imaging and parameter estimation with interesting insights. At the heart of these developments are the so-called classical and quantum Cramer-Rao bounds. According to the former, the inverse of a quantity known as the Classical Fisher Information (CFI) gives the lower bound on the variance of an unbiased estimator for estimating a parameter for a given measurement on multiple copies of a physical system. It therefore serves as a measure for precision~\cite{VanTrees2013}.
The quantum Cramer-Rao bound, in turn, introduces a quantity called the Quantum Fisher Information (QFI), as the maximum possible CFI in a physical system for estimating the same parameter~\cite{Helstrom1976}, Tsang {\em et al.}. In other words, the QFI quantifies the amount of physically accessible information about a parameter in a system that can be accessed through {\em any optical receiver measurement}. If we can find a receiver measurement whose CFI is equal to the QFI, then no other method can outperform it for estimating that parameter.

While the classical and quantum Cramer-Rao bounds have been known for several decades, a major breakthrough for sub-Rayleigh parameter estimation was made in~\cite{Tsang2016b} which considers the problem of estimating the separation between two equally bright incoherent light sources. If we employ ideal direct imaging, where we measure the incoming photons for their position on the imaging screen through a pixel array, we run into ``Rayleigh's curse". This is reflected in the Classical Fisher Information approaching zero as the separation goes to zero. However, Tsang {\em et al.} 
showed that this is only an artifact of direct detection in the traditional image plane, and that the Quantum Fisher Information for the separation between two uniformly bright incoherently-radiating quasi-monochromatic point sources is a {\em constant} independent of the value of the separation, and does not fall to zero even in the sub-Rayleigh limit. They went on to show that for a Gaussian point spread function (PSF), a spatial-mode demultiplexing (SPADE) measurement where we sort the incoming photons in terms of Hermite-Gaussian (HG) modes pointed at the center of the two-point constellation and obtain their photon counts, attains this constant QFI, i.e. its CFI is equal to the QFI. These results were generalized to arbitrary (non-Gaussian) PSFs in ~\cite{KGA17} and~\cite{Rehacek2017a}, 
where, among other things, it was shown that the sinc-Bessel mode basis provides an optimal measurement for 2-point separation estimation for a 1-dimensional hard aperture.

All this has sparked considerable interest in how the classical and quantum Cram\'er-Rao bounds can be employed to evaluate the performance of various receiver designs and come up with new approaches that can either surpass the performance of traditional techniques or are easier to implement. This perspective has led to the discovery of substantial quantitative performance improvements in estimating the size of extended objects~\cite{Zachary2019, Prasad2020B}, estimating distances between sources in two and three dimensions~\cite{Ang2016, Prasad2018}, simultaneously estimating multiple parameters encoded in sub-diffraction scenes~\cite{Prasad2019,Rehacek2017b, rehacek2018,Prasad_2020}, estimating spatial moments of arbitrary objects~\cite{Tsang2019,Tsang2019a,Zhou2019}, discriminating between multiple candidate objects~\cite{Lu2018,Huang2021,Grace2021c}, and adaptively estimating the locations and brightnesses of several point emitters in a sub-Rayleigh field of view~\cite{Bao2021,Matlin2022,Lee2022}.

In light of all these exciting developments, it is worth investigating how these quantum information theory methods can offer us new insights on fundamental limits, and optimal receiver techniques for long baseline interferometry, where we combine the signals from distant receivers in a way that creates the effect of a giant telescope whose size is equal to the distances between the individual telescopes.
An interesting first step in this direction was taken in Ref.~\cite{Cosmo2020}, which considers a number of small light collectors in a plane that can be thought of as pinholes. The light received by these is then fed into a linear interferometer, whose outputs are then measured with photodetectors. This set up is used to estimate the 3-dimensional positions of $n$ emitters in some arbitrary arrangement. 
They work out the theory for such a system to attain the quantum limit, and also consider the 2 equally bright point separation problem as a special example.
Other works have compared the performance of phase-sensitive interferometry between distant, pinhole-like apertures against the quantum Cr\'amer-Rao bound for estimating the separation between two point sources in the context of bright thermal emitters ~\cite{Wang2021} and in a contextualized comparison against other standard astronomical interferometric techniques~\cite{Bojer2022}, albeit under a quasi-monochromatic approximation.
Additional works such as~\cite{Gottesman2012, Khabiboulline2018} take a hard core quantum mechanical approach, showing how shared entanglement between two distant point-like receivers can be used to estimate the mutual coherence function of a sub-diffraction scene.

In this paper, we analyze the quantum-estimation-theoretic limits of long baseline interferometry for quantitative imaging and evaluate various receiver designs that can be employed for different regimes of baseline distances. Unlike previous analyses that consider only point-like apertures, our work relaxes this assumption and accounts for the additional information residing in the multi-spatial-mode optical field within each aperture in a long-baseline interferometric context. We quantify the relative value of the aperture-local high-order modal information content vis-a-vis the information content extractable from optimized interferences among light from the multiple apertures, in various regimes, in order to approach the {\em quantum limit} of a multi-aperture imaging system---be it a segmented mirror where light from the apertures can be diffractively co-located in a common image plane, or of a long-baseline system where light from the individual telescope must be brought together via optical fibers or `light pipes'~\cite{Shaklan1987, Guyon2022, Jovanovic}.

We give the general framework for an arbitrary one-dimensional parameter estimation problem with a multiple-telescope system, as well as illustrate it in detail for Tsang {\em et al.}'s problem of separation estimation between two equally bright, quasi-monochromatic incoherent sources. A full theoretical treatment accounting for broadband radiation will be presented in a future work.

In the process of laying down the ground work for analyzing the various receivers, we give a fully self-contained introduction to all the relevant mathematical tools and concepts, so that this work is completely accessible to a reader without any prior background in quantum estimation theory or quantum imaging and sensing, such as researchers in the astronomical imaging community who may be interested in engaging with these methods.

In the remainder of this work, we organize our proposed long-baseline interferometric schemes into three different categories based on the distances between the telescopes:
\begin{figure}
	\centering
	\includegraphics[width=0.8\columnwidth]{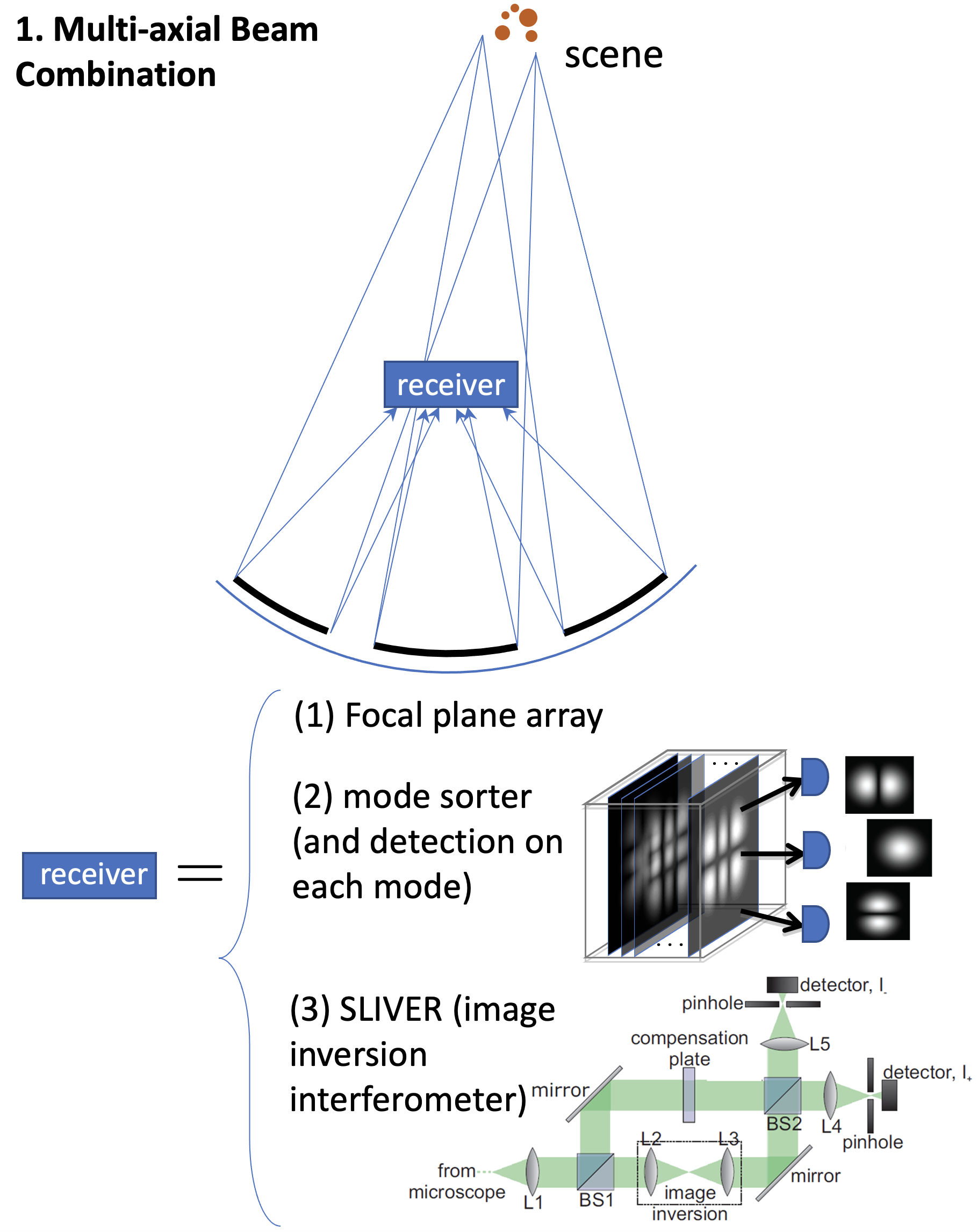}
	\caption{Schematic diagram of receivers based on multi-axial beam combination, e.g., between light reflected off of sub-apertures of a segmented-aperture telescope.}
	\label{fig:MultiAxial_Schematic}
\end{figure}
\begin{enumerate}
\item {\textbf{Multi-axial beam combination}}---When the different apertures are sufficiently close to each other, we can focus all the light collected at different apertures on to a common imaging screen (see fig.~\ref{fig:MultiAxial_Schematic}). This is often referred to as multi-axial beam combination in the literature (i.e., bringing all the spatial frequencies from multiple telescopes together at a common screen at different angles or axes)~\cite{Glindemann2011}.
One simple form of this can be to have at each aperture, a carved out piece of what would be a hypothetical giant parabolic mirror covering the entire baseline region.
We can then employ any of the measurement schemes that can be used for a single aperture system, such as ideal direct imaging or a SPADE.
The advantage of building such a system relates to the fact that it would save the cost and effort associated with constructing a full parabolic mirror spanning the entire baseline.
\begin{figure}
	\centering
	\includegraphics[width=\columnwidth]{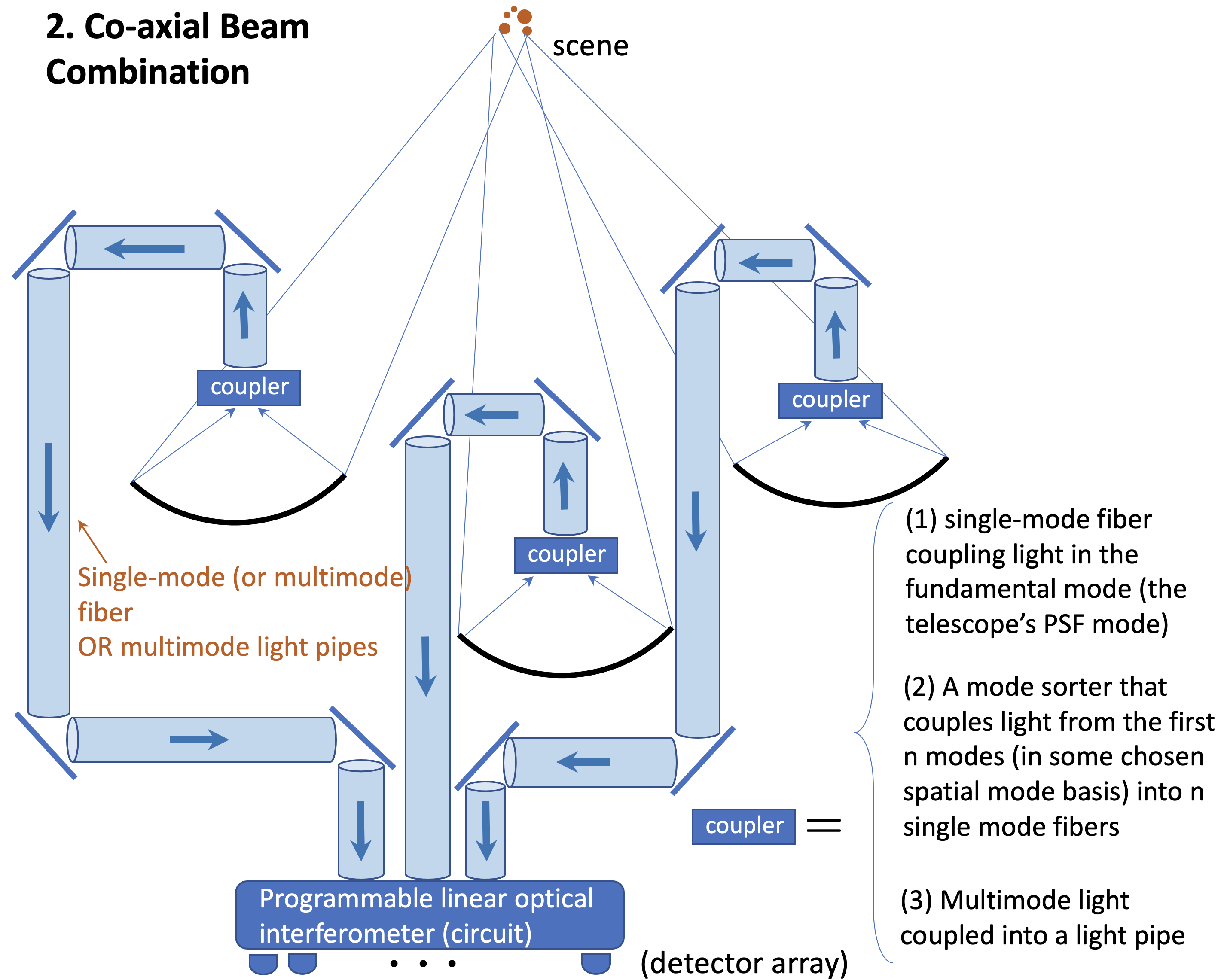}
	\caption{Schematic diagram of receivers based on co-axial beam combination, e.g., between light collected at the telescope sites of a long baseline telescope system.}
	\label{fig:CoAxial_Schematic}
\end{figure}
\item {\textbf{Co-axial beam combination}}---When the individual telescopes are too far from each other to allow focusing the light on to a common imaging screen, we can bring the light collected at the individual apertures through single mode fibers (SMFs) or multimode `light pipes' to a central location where we combine it interferometrically through a linear-optical circuit made up of beamsplitters and phase shifters. This is often referred to as co-axial beam combination in the astronomy literature (i.e., bringing the beams from individual telescopes together and combining them with an overlapping axis with a beam splitter)~\cite{Glindemann2011}. We can then obtain the photon counts at the output of the interferometer. See Fig.~\ref{fig:CoAxial_Schematic} for a schematic. The key is to use a beam splitter arrangement that gives the optimal performance for the parameter of interest. By `light pipes,' we mean any method that can bring the entire (multi-spatial-mode) beam from one location to another one with low loss; this is usually done by most current optical interferometry systems by using evacuated vacuum tubes.

\begin{figure}
	\centering
	\includegraphics[width=\columnwidth]{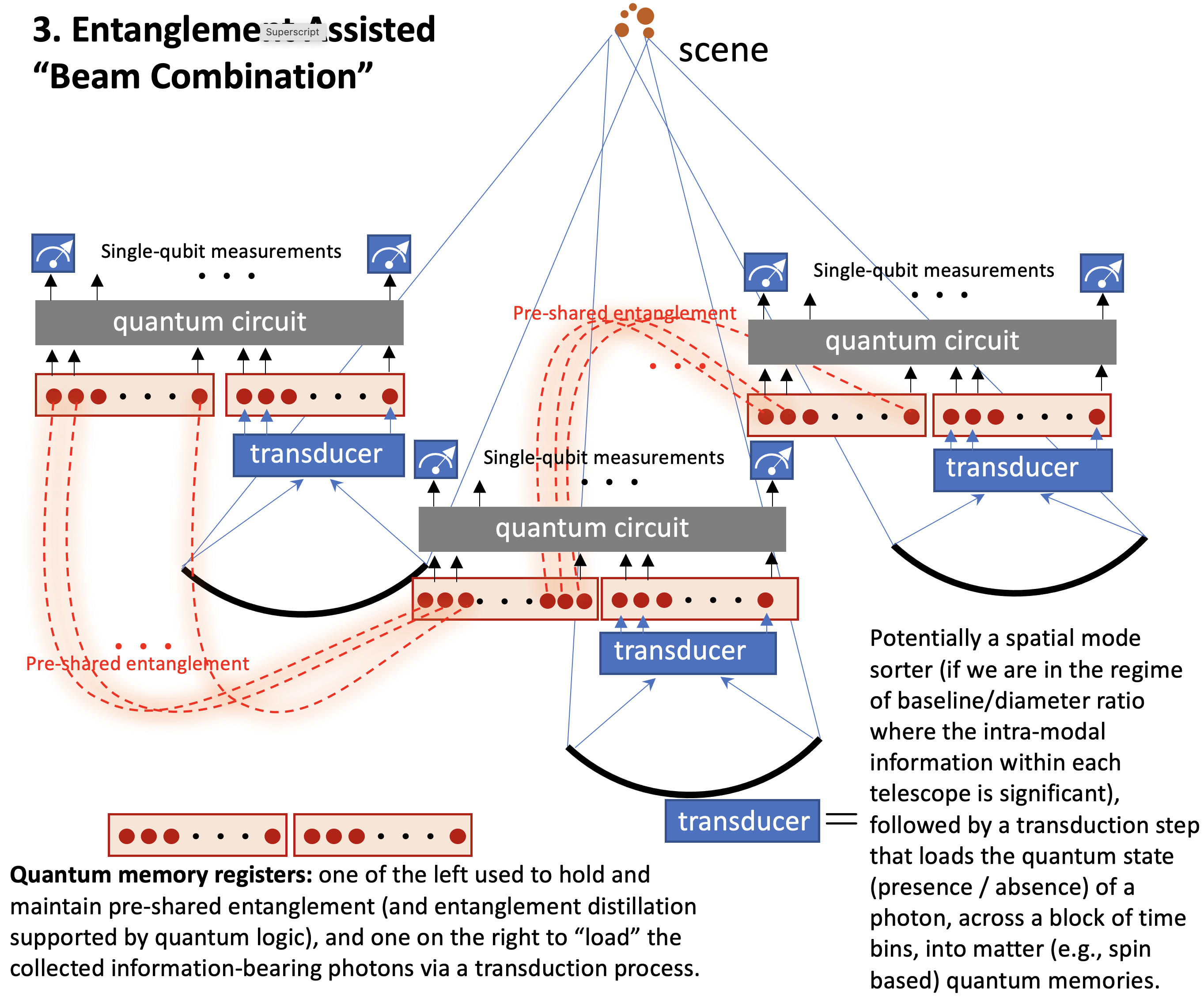}
	\caption{Schematic diagram of receivers based on pre-shared entanglement, where one can still in principle attain the full `quantum limit' of precision of any quantitative imaging task, without ever interfering light collected at different telescopes. But, such a receiver system will require high-fidelity high-rate long distance entanglement distribution, e.g., using quantum repeaters or a satellite, quantum-logic-capable memories, and a transducer (e.g., similar to the scheme in Ref.~\cite{Zheng2022}) to load the photons collected at each telescope site, `compressed' efficiently into a quantum memory register.}
	\label{fig:Entanglement_Schematic}
\end{figure}
\item {\textbf{Pre-distributed-entanglement based receiver (no direct optical interference)}}---when the receivers are so far that even bringing all the light to a central location through single mode fibers or light pipes is impractical, we can employ an entanglement based approach~\cite{Gottesman2012, Khabiboulline2018}. For example, if we wish to combine the light collected at two separate locations, then one possibility is to use a quantum protocol to transfer the part of the quantum state from one of the telescopes to the other one (via teleportation, which consumes pre-shared entanglement), where the two can then be combined locally in a quantum circuit that mimics what an actual interferometer would have measured. Since any multiport interferometer can be broken up into 50-50 two-port beamsplitters and single mode phases~\cite{Clements2016}, and since we know how to mimic the action of a nonlocal 50-50 beamsplitter (meaning one acting on a pair of single modes at distant telescope sites) using pre-shared Bell states, without actually bringing light together at one location~\cite{Khabiboulline2018}, in principle any co-axial or multi-axial receiver design can be realized using sufficient amount of pre-shared entanglement between each pair of telescopes, without bringing light from the individual telescopes into an interferometer. See Fig.~\ref{fig:Entanglement_Schematic} for an illustration.
\end{enumerate}

Our paper is organized as follows. In section 2, we describe the physical set up and Tsang {\em et al.}'s model for describing the quantum state for the incoming photons, along with an introduction to the classical and quantum Cramer-Rao bounds for a general parameter estimation problem. We then outline the basics of the separation estimation problem for 2 uniformly bright point sources as a special example. In section 3, we develop our mathematical foundation for working with multiple apertures and obtain some results which we use later for the study of the various receiver designs.
In section 4, we describe the various types of receivers for the three distance regimes outlined above and how their performance can be calculated for the general estimation problem. In this process, we show, in terms of the CFI, where the performance enhancement from combining multiple telescopes arises from. In section 5, we discuss the calculation of the quantum Fisher Information for the same general estimation problem, again showing the enhancement from combining distant telescopes. In section 6, we illustrate the special example of the equally bright two-point separation estimation problem with our various receivers.
In section 7, we give our conclusions.

\section{Physical model and preliminaries for a single-aperture system}
\label{physical-model-preliminaries-section}

\subsection{The weak source model}
 \label{weak-source-model}

We use the weak source model introduced in~\cite{Tsang2016b}, where we assume that our scene emits nearly monochromatic light and that the average number of photons $\epsilon$ arriving in each coherence time interval is much less than one, requiring a large number of photons to be measured over many such time intervals to extract any useful information~\cite{Goo85Statistical, MW95, Labeyrie2006, Gottesman2012, Tsang2011}. We can write the quantum density operator in each time interval for the photon field on the image plane as
\be
\rho = (1-\epsilon) \rho_0
+\epsilon\rho_1
+O(\epsilon^2),
\ee
where $\rho_0 = |{\rm vac}\ra\la {\rm vac}|$ is the zero-photon or ``vacuum" state, $\rho_1$ is a one-photon state, and $O(\epsilon^2)$ are higher order terms which we can ignore since $\epsilon\ll1$.

The one-photon mixed state $\rho_1$ is the sum of the contributions from each of the emitters, which me model as point sources. If we have a total of $n_e$ emitters, it is: 
\be
\rho_1 =
\sum_{s=1}^{n_e}
b_s \,|\psi_s, 1ap\ra\la\psi_s, 1ap|,
\label{rho1_definition}
\ee
Here $b_s$ are relative brightnesses associated with each of the sources satisfying the normalization property $\sum_{s=1}^{n_e} b_s =1$,
and $|\psi_s, 1ap\ra$ are the corresponding kets: 
\be
|\psi_s, 1ap\ra = \int_{-\infty}^\infty dx \psi_{\rm 1ap}(x -x_s) |x\ra,
\label{psi_s-ket-definition}
\ee
where $\psi_{\rm 1ap}(x)$ is the point spread function (PSF) of our aperture, $x_s$ is the position of the $s$-th light source, and $|x\ra$ is the position $x$ eigenket---the state of one photon in a delta-function spatial mode located at position $x$ on the image plane.
We have explicitly put $1ap$ in the ket $|\psi_s, 1ap\ra$
and the subscript of the PSF here because later we will be going to multiple aperture configurations, and therefore we will need to distinguish between single and multiple aperture functions. We should also point out that all our equations in this section will hold for any PSF unless we explicitly state otherwise. Therefore, we will for the most part not specify any specific PSF function here. In later sections though, when we consider multiple aperture configurations, we will specifically focus on hard rect shaped apertures and will specify $\psi_{\rm 1ap(x)}$ accordingly.

If we measure the incoming photons for some observable $\mathcal Y$, then we obtain random samples from the probability distribution $P(\mathcal Y) = \la\mathcal Y|\rho|\mathcal Y\ra$ for each copy of the received state $\rho$. If the measurement observable corresponds to a passive mode transformation (followed by photon counting on the transformed modal basis), $|\mathcal Y\ra$ is the single photon eigenket corresponding to the mode $\mathcal Y$, and detection of a photon---which happens with probability $\epsilon$---heralds that $\rho_1$ was received, and the measurement outcome, the appearance of a single photon detection in mode $|\mathcal Y\ra$, is a random sample from $\la\mathcal Y|\rho|\mathcal Y\ra$~\cite{Ram2006, pawley2006, Labeyrie2006,Zmuidzinas2003CramrRaoSL}.
If our incoming light has optical bandwidth $W$ and the integration time is $T$, then it has roughly $M \approx WT$ mutually-orthogonal modes in that time-bandwidth window.
Since we are assuming nearly monochromatic light, our $W$ is very small, and therefore, all these $M$ orthogonal modes correspond to different temporal modes at nearly the same frequency. The quantum description of the entire collected light during the integration time will be taken to be $\rho^{\otimes M}$. The average number of photons received over the integration time, $N = M\epsilon$. If we constrain ourselves to SPADE based receivers, where detection (of a single photon in a coherence time interval) heralds $\rho_1$, we can take the quantum description of the collected light to be $\rho_1^{\otimes N}$, with $N = \lfloor{M\epsilon}\rfloor$.

\subsection{The classical and Quantum Cramer-Rao bounds}
\label{classical-quantum-Cramer-Rao-bound-section}

Let us say that we have $\rho_1^{\otimes N}$, i.e., $N$ copies of a quantum state $\rho_1$ that depends on a set of parameters $\left\{\theta_\mu\right\}$, which we wish to estimate by measuring some observable $\hat{\mathcal Y}$ on each copy of $\rho_1$.
The error covariance matrix with a set of estimators $\tilde\theta_\mu(\mathcal Y)$ is:
\be
\Sigma_{\mu\nu} \equiv \int d\mathcal Y P(\mathcal Y)
\Bk{\tilde\theta_\mu(\mathcal Y)-\theta_\mu}
\Bk{\tilde\theta_\nu(\mathcal Y)-\theta_\nu}.
\ee
If $\tilde\theta_\mu(\mathcal Y)$ are unbiased estimators, then this covariance matrix obeys the Cramer-Rao bound 
\be
\Sigma \geq \mathcal J^{-1},
\label{Cramer-Rao-bound}
\ee
where $\mathcal J_{\mu\nu}$
is the classical Fisher Information matrix~\cite{VanTrees2013} (for the measurement $\hat{\mathcal Y}$) and is given, for measuring the same observable on $N$ copies of $\rho_1$, by:
\be
\mathcal J_{\mu\nu}
\equiv N \, \int d\mathcal Y
\frac{1}{P(\mathcal Y)}
\frac{\partial P(\mathcal Y)}{\partial\theta_\mu}
\frac{\partial P(\mathcal Y)}{\partial\theta_\nu}.
\label{FI-formula}
\ee
The Cramer-Rao bound is saturated by the maximum likelihood estimator for large $N$. If $\mathcal Y$ is continuous, then $P(\mathcal Y)$ becomes a probability density, with the sum over $\mathcal Y$ in (\ref{FI-formula}) being replaced by an integral.

The quantum Cramer-Rao bound gives us the maximum possible CFI that can be achieved by any measurement scheme for a given physical system and parameter:
\be
\Sigma \geq \mathcal J^{-1}
\geq \mathcal K^{-1},
\ee
where $\mathcal K$ is the QFI~\cite{Helstrom1976}.
For $N$ measurements on a system with a quantum density operator $\rho_1$, the QFI is given by
\be
\mathcal K_{\mu\nu}(\rho_1^{\otimes N})
\equiv N \frac{1}{2} \trace\left(\rho_1 \{\sL_\mu(\rho_1),\, \sL_\nu(\rho_1)\} \right),
\label{QFI-formula}
\ee
where $\mathcal L_\mu(\rho_1)$ is a symmetric logarithmic derivative (SLD) of $\rho_1$ with respect to $\theta_\mu$.
It is a Hermitian operator defined by the relation
\be
\frac{\partial\rho_1}{\partial\theta_\mu}
=\frac{1}{2}\left(\rho_1 \sL_\mu(\rho_1) + \sL_\mu(\rho_1) \rho_1\right)
\label{SLD-relation}
\ee
If $\rho_1 = \sum_j D_j |e_j\ra\la e_j|$
is the decomposition of $\rho_1$ in terms of its eigenvalues $D_j$ and eigenvectors $|e_j\ra$,
then the SLD is given by
\be
\sL_\mu(\rho_1) = \sum_{j,k; D_j + D_k \neq 0}
\frac{2}{D_j+D_k}
\la e_j| \frac{\partial\rho_1}{\partial\theta_\mu}| e_k\ra
| e_j\ra\la e_k|.
\label{sld}
\ee
Note that the QFI is independent of the measurement choice and quantifies the amount of information available in the quantum state of light (the physical system) about a parameter. If the CFI for a given measurement is equal to the QFI, then we know that measurement is optimal for estimating the given parameter, since by the Quantum Cramer-Rao bound, no other measuring scheme can perform better. Our goal is therefore to come up with a measurement that comes as close to the QFI as possible with the least (physical-implementation) cost.

It is worth noting that based on the above construction of the QFI, a projective measurement in the eigenbasis of the SLD will achieve the QFI~\cite{PhysRevLett.72.3439, Barndorff_Nielsen_2000, Paris2008}, though it is not necessarily the only optimal measurement. The problem, however, is that in general, the SLD depends on the unknown parameter which we are trying to estimate itself. One solution to this is a two stage approach proposed by~\cite{Barndorff_Nielsen_2000}
\begin{enumerate}
\item First, use a small part of the total integration time with a mean photon count of $N^\alpha$ with any $0 < \alpha < 1$, to obtain an initial maximum likelihood estimate $\hat \theta_1$ for the unknown parameter $\theta$ by employing any measurement that has a non-zero CFI for estimating $\theta$. 
\item Based on the estimate $\hat \theta_1$ from stage one, measure each of the remaining photons in the eigenbasis of the SLD of $\theta$, evaluated at $\hat \theta_1$. From the results of this measurement, we then obtain a maximum likelihood estimate for $\theta$.
\end{enumerate}
We should mention that this procedure has also been described in~\cite{2005atqs.book162H, PhysRevA.61.042312}
and section 6.4 of ~\cite{Hayashi2006}, with the specific choice of $\alpha = 1/2$.

This two-stage method asymptotically reaches the QFI as $N$ tends to infinity. While a detailed proof for this can be found in the above references, here is a short version of the argument which we also gave in~\cite{Sajjad2021}.
If $N$ is large, $N^\alpha$ (or $\sqrt{N}$ for $\alpha=1/2$) will also be large.
Therefore, the variance of the estimate $\hat \theta_1$ scales as $1/(\mathcal{J}_1 N^\alpha)$, where $\mathcal{J}_1 > 0$ is the CFI of the stage-one measurement.
With $N^\alpha$, photons, this variance will therefore approach zero for large $N$.
The Fisher information in stage-two will be $(N-N^\alpha) \mathcal K$, if we measure in the eigenbasis of the SLD based on the exactly correct value of $\theta$.
But in reality, since we will carry out this measurement at the estimated value $\hat \theta_1 =\theta +\theta_{1,\rm err}$,
we need to replace $\mathcal K$ by the CFI for the SLD eigenbasis measurement evaluated at this value rather than the true $\theta$. 
Taylor expanding this CFI around the true value $\theta$, we obtain
\be
\mathcal{J}_{\rm SLD}(\theta +\theta_{1,\rm err}) 
= \mathcal{J}_{\rm SLD}(\theta) \,+\, \theta_{1,\rm err}^2 \frac{\partial^2 \mathcal{J}_{\rm SLD}(\theta_1)}{\partial\theta_1^2} \,+\,\ldots
\ee
Here we do not have a first derivative term because the CFI $\mathcal{J}_{\rm SLD}$ is equal to the QFI at $\theta_1$, which is a maximum. Therefore, the first derivative must be zero.
Secondly, since this is a maximum, the second derivative will be a negative constant with respect to $\theta_{1,\rm err}$.
All in all, we can rewrite the stage-two CFI as
\be
\begin{split}
	\mathcal{J}_{\rm SLD}(\theta +\theta_{1,\rm err}) = & \mathcal K\left(1 -O\left(\theta_{\rm err}^2\right)\right) \\
= &\mathcal K \left(1-O\left(\frac{1}{N^\alpha \mathcal{J}_1}\right)\right)
\end{split}
\ee
where in the last step, we have used the fact that the mean squared error of the initial centroid estimate is approximately equal to the inverse of the CFI for stage-one.
The total CFI accumulated over stage-two is therefore $(N-N^\alpha) \mathcal K\left(1 -O\left(\frac{1}{N^\alpha J_1}\right)\right)$.
And when $N$ is large
\be
\frac{1}{(N-N^\alpha) \mathcal K\left(1-O\left(\frac{1}{N^\alpha \mathcal{J}_1}\right)\right)}
\approx \frac{1}{N \mathcal{K}},
\ee
the variance approaches that of the optimal measurement.
In reality since $N$ will be a finite number, we could optimize the choice of $\alpha$ such that the overall CFI attained at the end of stage-two is maximized, for a given fixed $N$ based on the choice of the stage-one measurement and its CFI.
A multi-stage adaptive quantum estimation algorithm is given in~\cite{Fujiwara_2006} where the result of each stage is used as input for the next one, which could lead to further improved performance in this non-asymptotic setting. Lastly, In case the SLD depends on more than one unknown parameter, then we would need to get an initial estimate for all such unknown parameters in stage one.

\subsection{How to calculate the CFI and QFI}
\label{general-parameter-estimation-problem-section}

Let us say that we wish to estimate some parameter $\theta_g$ on which the brightnesses $b_s$ and positions $x_s$ in the scene depend~\footnote{In this paper, we will not consider the multi-parameter estimation problem, e.g., {\em localization} of an unknown number of point emitters: ${\boldsymbol \theta} = (\left\{x_s, b_s\right\}, n_e)$, for which dynamically-adaptive modal measurement techniques is known to outperform direct imaging~\cite{Lee2022}}. To calculate the CFI for a single scalar parameter $\theta_g$ for a given receiver, we need to write down the probability distribution of measurement outcome in terms of $\theta_g$ and then apply (\ref{FI-formula})).
For ideal direct imaging, we measure the incoming photons for their position on the image plane through a grid of shot-noise limited photon counting detectors with infinitesimally small pixellation, unity fill factor, and infinite spatial extent; no direct imaging instrument could ever outperform this idealized model.
For this receiver measurement, the probability of finding a received photon to be at position $x$,
\be
P(x) = \la x|\rho_1| x\ra
= \sum_{s=1}^{n_e} b_s |\psi_{\rm 1ap}(x -x_s)|^2.
\label{P_x-general}
\ee
In the same spirit, consider a measurement in the image plane in a spatial basis with normalized mode functions $\phi_j(x)$ with $j= 0, 1,2, \ldots$, i.e., a spatial mode demultiplexer (SPADE) followed by photon detection on each of the $\phi_j(x)$ modes. The eigenket corresponding to an incoming photon being in the $j$-th mode is: 
\be
|j\ra = \int_{-\infty}^\infty dx \phi_j(x) |x\ra
\ee
The probability amplitude for a photon originating from a point source at position $x_s$ being measured and found to be in mode $\phi_j(x)$ is given by the correlation function,
\be
\Gamma_j(x_s) \equiv
\la j|\psi_s, 1ap\ra = \int_{-\infty}^\infty dx \phi_j ^*(x) \,\psi_{\rm 1ap}(x-x_s).
\label{correlation-function-mode-j-1ap}
\ee
Naturally, the probability is the magnitude squared of this amplitude, and the total probability for a photon from our scene being found in the $j$th mode is
\be
P_j = \la j|\rho_1|j\ra
= \sum_{s=1}^{n_e} b_s |\Gamma_j(x_s)|^2.
\label{P_j-general}
\ee 
The CFI for estimating our parameter $\theta_g$ from ideal direct imaging or this SPADE basis is then given by (\ref{FI-formula})):
\begin{align}
\sJ_{\rm DI}(\theta_g) &= N \int_{-\infty}^\infty dx \frac{1}{P(x)}\left(\frac{\partial P(x)}{\partial\theta_g}\right)^2 \nn, \\
\sJ_{\rm SPADE}(\theta_g) &= N \sum_{j=0}^\infty \frac{1}{P_j} \left(\frac{\partial P_j}{\partial\theta_g}\right)^2.
\end{align}
The alert reader might have noticed that we are calculating the CFI for $N$ copies of the single-photon density matrix $\rho_1$ instead of the full density matrix $\rho$, for which we have $M$ copies over our integration time. This is reasonable because the latter effectively translates into the former. For example, for direct imaging, the probability of obtaining a photon at position $x$ for a given copy of the density matrix $\rho$ (which corresponds to a given temporal mode) is $P_{\rm full}(x) = \epsilon P(x)$. The total CFI for $M$ copies of $\rho$ is given by
\begin{equation}
	\begin{aligned}
		M \int_{-\infty}^\infty dx\frac{1}{P_{\rm full}(x)}& \left(\frac{\partial P_{\rm full}(x)}{\partial\theta_g}\right)^2\\
		&= M \epsilon \int_{-\infty}^\infty dx \frac{1}{P(x)} \left(\frac{\partial P(x)}{\partial\theta_g}\right)^2 \\
		&= N\int_{-\infty}^\infty dx \frac{1}{P(x)} \left(\frac{\partial P(x)}{\partial\theta_g}\right)^2.
	\end{aligned}
\end{equation}
A somewhat similar argument holds for the QFI. The partial derivative of the density matrix with respect to $\theta_g$ does not contain any contributions from the vacuum term $(1-\epsilon) \rho_0$, and therefore,
\be
\frac{\partial\rho}{\partial\theta_g} = \epsilon \frac{\partial\rho_1}{\partial\theta_g}.
\ee
Consequently, the symmetric logarithmic derivative $\sL$ only contains contributions from the single-photon part of the density matrix. Since the QFI is given by $\trace(\rho \sL^2)$, and the vacuum states $|{\rm vac}\ra$ are orthogonal to single photon states, the QFI only gets non-zero contributions from the single photon state $\rho_1$ and is proportional to $N =M\epsilon$ times $\trace(\rho_1 \sL_1 ^2)$, where
$\sL_1$ is the SLD calculated from (\ref{sld}).

Since $\rho_1$ is a brightness-weighted sum of the contributions from the individual point sources $|\psi_s, 1ap\ra\la\psi_s, 1ap|$,
it acts on the $n_e$ dimensional vector space spanned by the states $|\psi_s, 1ap\ra$. The eigenstates of $\rho_1$ provide a natural basis for this purpose, especially since the SLD is also computed in terms of these. The SLD also involves the derivatives of the density matrix with respect to $\theta_g$, and therefore acts on the space spanned by the $|\psi_s, 1ap\ra$ as well as their derivatives with respect to $\theta_g$. Therefore, the overall space we need is $2n_e$ dimensional.
If we have only two point sources, such as in Tsang {\em et al.}'s equally bright 2-point separation estimation problem, then we can explicitly write down the 4-dimensional orthogonal basis kets~\cite{Tsang2016b}. However, this becomes increasingly difficult and messy as the number of point sources in the scene increases.
Therefore, one alternative is to take a numerical approach and expand the states $|\psi_s, 1ap\ra$ in terms of the spatial mode kets $|j\ra$, and truncate at some maximum value that is large enough to give us the desired level of precision~\cite{Zachary2019}.
The single-photon density matrix $\rho_1$ in the spatial mode basis is given by
\be
\rho_1 = \sum_{s=1}^{n_e} b_s \,\sum_{j,l=0}^\infty \Gamma_j(x_s) \,\Gamma_l ^*(x_s) \,|j\ra\la l|.
\label{rho1-mode-expansion}
\ee
The partial derivative of this with respect to $\theta_g$ is
\begin{equation}
	\begin{aligned}
		\frac{\partial\rho_1}{\partial\theta_g} =& \Bigg( \sum_{s=1}^{n_e} \bigg(\frac{\partial b_s}{\partial\theta_g}\bigg) \,\sum_{j,l=0}^\infty \Gamma_j(x_s) \,\Gamma_l ^*(x_s)\\
		&+ b_s \left(\frac{\partial x_s}{\partial\theta_g}\right)
		\, \sum_{j,l=0}^\infty \Gamma_j ^\prime(x_s) \,\Gamma_l ^*(x_s)
		\\
		&+ \Gamma_j(x_s) \,\Gamma_l ^{*\prime}(x_s)\Bigg)
		\,|j\ra\la l|.
	\end{aligned}
	\label{rho1-derivative-mode-expansion}
\end{equation}
After that, calculating the QFI is a matter of truncating the modes at some maximum number, and applying equations (\ref{QFI-formula}) and (\ref{sld}).

\subsection{The equally-bright two-point separation estimation problem}
\label{two-point-separation-one-aperture-section}

For two equally bright point sources, $b_1 = b_2 = \frac{1}{2}$. We will follow the conventions of~\cite{KGA17} to assume that the locations of the two point sources are
\begin{eqnarray}
x_1 &=& \theta_c +\theta, {\text{and}} \nn\\
x_2 &=& \theta_c -\theta,
\end{eqnarray}
where $\theta_c$ is the centroid and we are taking the separation to be $2\theta$. We will be interested in estimating $\theta$ instead of the separation $2\theta$ because it makes some of the formulas simpler. This is different from the original paper by Tsang {\em et al.}, where the point sources were assumed to be at $\theta_c \pm\theta/2$ and they calculate the CFI and QFI for estimating the separation $\theta$~\cite{Tsang2016b}.
In light of this, our QFI for estimating $\theta$ will be 4 times that obtained by Tsang {\em et al.} for the separation. We outline the calculation in appendix \ref{2pt-QFI-calculation-apendix},
and the result is
\be
\sK_{\theta} = 4N \, \int_{-\infty}^\infty dx \left|\frac{\partial\psi(x)}{\partial x}\right|^2.
\label{QFI-Tsang1}
\ee
Here we did not include the subscript $1ap$ with $\psi(x)$ because this will also hold for any multiple aperture configuration, and we will revert to it later, so we wish to specifically state this as a general result. 

It is however convenient to express this in terms of the autocorrelation function of the PSF introduced by~\cite{KGA17},
\be
\Gamma_{\rm PSF}(a)
= \int_{-\infty}^\infty
\, \psi^* (x) \,\psi(x -a) \, dx.
\label{Gamma-def}
\ee
Using integration by parts, the QFI result in (\eqref{QFI-Tsang1}) can be written in terms of the second derivative of the autocorrelation function as
\be
\sK_{\theta} = -4N \Gamma_{\rm PSF}^{\prime\prime}(0).
\label{QFI-autocorrelation}
\ee
Here the double prime indicates the second derivative.
As a special example, for the specific case of a rect aperture with PSF $\psi_{\rm 1ap}(x) = \frac{\sqrt{\sigma} \sin(\pi x/\sigma)}{\pi x}$, we obtain the QFI~\cite{KGA17, Rehacek2017a}:
\be
\sK_{\text{h,1ap}, \theta} = \frac{4\pi^2 N}{3\sigma^2},
\label{single-hard-aperture-QFI-2-points},
\ee
where h in the subscript denotes a hard aperture, and ``1ap" stands for single aperture. We will have a lot more to say on the above PSF and where it comes from later in the paper.
However, for now, we should mention that this PSF as well as its autocorrelation function and its derivatives give a 0/0 division when evaluated at $x=0$, so they are evaluated as the limit when $x\to 0$. 

Coming to the CFI, if we are using ideal direct detection of the incoming photons on a pixel array on the image screen, then the probability of finding an incoming photon to be incident on the imaging screen at position $x$ is
\be
P(x) = \la x|\rho_1| x\ra
= \frac{1}{2} \left[ |\psi_{\rm 1ap}(x -x_1)|^2 + |\psi_{\rm 1ap}(x-x_2)|^2\right],
\ee
where the factor of $1/2$ represents the equal probability for an incident photon to have originated from one of the two point sources due to their equal brightness (i.e., namely $b_1 = b_2 = \frac{1}{2}$), and $|\psi(x-x_s)|^2$ is the probability density for a photon coming from source $s$ (with $s=1, 2$) to hit the image screen at position $x$.
Applying the formula (\ref{FI-formula}) for the CFI with $P(x)$ as the probability density then gives the CFI. The integral over $x$ needs to be computed numerically at least for a Gaussian or rect aperture.

Now, consider a SPADE measurement with mode functions $\phi_j(x)$ with $j=0, 1, 2,\ldots$ as discussed in the previous section. Assume that the SPADE is pointed perfectly at the centroid such that the two positions are $\pm\theta$. This requires prior knowledge of the centroid, for which we can use direct imaging~\cite{Tsang2016b, Sajjad2021, Grace2020c}.
The probability for an incoming photon to be found in mode $j$ is then 
\be
P_j = \frac{1}{2}\left(|\Gamma_j(\theta)|^2 +|\Gamma_j(-\theta)|^2\right).
\ee
If the aperture is symmetric and all the mode functions are real-valued having definite parity, then the correlation functions are also purely real valued with definite parity, so this becomes:
\be
P_j = \Gamma_j ^2(\theta).
\ee
It is straightforward to see that the CFI is then
\begin{align}
\sJ_{\theta} &= N\sum_{j=0}^\infty \frac{1}{P_j(\theta)}\left(\frac{\partial P_j(\theta)}{\partial\theta}\right)^2 \nn \\
&= 4N \sum_{j=0}^\infty \left(\Gamma_j ^\prime(\theta)\right)^2.
\label{CFI-2-point-all-modes}
\end{align}
Tsang {\em et al.} show that for a Gaussian PSF (with the SPADE pointed at the centroid), the HG basis attains the QFI.
This was generalized to arbitrary PSFs in~\cite{KGA17} and~\cite{Rehacek2017a}, with the latter work proving that for a symmetric aperture, any basis of purely real-valued orthonormal functions with definite parity will attain the QFI.
That is, for any symmetric aperture and mode basis with definite parity, we obtain:
\be
\sJ_{\theta}
= 4N \sum_j \left(\Gamma_j ^\prime(\theta)\right)^2
= \sK_{\rm 1ap, \theta}
= -4N \Gamma_{\rm PSF}^{\prime\prime}(0).
\label{single-aperture-QFI-2-points}
\ee

\subsection{CFI invariance under rotations of SPADE basis}
\label{invariance-section}

Note that the formula (\ref{CFI-2-point-all-modes}), when summed over $j$, has the form of an $L_2$-norm dot product on a real-valued vector space. Therefore, it must be invariant under a rotation of basis with purely real-valued transformation coefficients that are independent of the parameter $\theta$. That is, let us say we have a SPADE basis with purely real-valued functions $\phi_j(x)$ with $j=0, 1, 2,\ldots j_{\rm max}$, where $j_{\rm max}$ may be finite or $\infty$. Now, consider a change of basis of the form
\be
\phi_{{\text{mod}}, j} (x) = \sum_{l=0}^{j_{\rm max}} U_{jl} \phi_l(x),
\ee
where the coefficients $U_{jl}$ are real-valued and independent of the parameter $\theta$. The correlation functions $\Gamma_j(x)$ and their derivatives $\Gamma_j ^\prime(x)$ inherit the same change of basis coefficients: 
\be
\Gamma_{\text{mod}, j}^\prime(\theta) = \sum_{l=0}^{j_{\rm max}} U_{jl} \Gamma_l ^\prime(\theta).
\ee
Therefore, the CFI is invariant under this rotation of basis
\be
\sJ_{\theta} = 4N\sum_{j=0}^{j_{\rm max}} \left[\Gamma_j ^\prime(\theta)\right]^2 
= 4N\sum_{j=0}^n \left[\Gamma_{\text{mod}, j}^\prime(\theta)\right]^2.
\ee 
 This invariance however does not hold if the change of basis coefficients $U_{jl}$ are complex (i.e., having both real and imaginary parts) or depend on $\theta$.
In the former case, the CFI for mode $j$ no longer has the nice form 
$\left[\Gamma_j ^\prime(\theta)\right]^2$, and in the latter scenario, the derivatives of the modified correlation functions in the new basis also contain derivatives of $U_{jl}$. 

It is worth noting that this invariance result is not only valid when $j_{\rm max}$ is infinite, but also holds if we have a SPADE where we sort the incoming photons between modes $\phi_j(x)$ with $j=0,1\ldots j_{\rm max}$ and their orthogonal complement. Such a SPADE can be practically implemented using a physical device that directs the specified modes up to $j= j_{\rm max}$ into their separate photon detectors, while collecting and counting the remaining modes in a bucket detector~\cite{Boucher2020}.
In that case, the total CFI from the $j = 0\ldots n$ modes is invariant as argued above, and the orthogonal complement remains unchanged under rotations within the $j=0\ldots n$ modes, so its contribution to the CFI is trivially constant under such transformations. We will consider such $n+1$ SPADEs later in the paper.

Lastly, we would like to end this section by stating this invariance result in its most general form that is not specific to the 2-point problem or even imaging. Consider a mixed quantum state $\rho$ that is an equal-probability sum of $n_e$ pure states $|\psi_s\rangle$ depending on some parameter $\theta$, with $s=1\ldots n_e$.
Assume that these pure states have a symmetry such that measuring them in some basis $|j\ra$ gives the same purely real-valued probability coefficients $\Gamma_j(\theta)$ up to a possible minus sign. The CFI for estimating $\theta$ from such a measurement is then given by:
\be
\sJ_\theta
= \sum_j 4\left(\Gamma_j ^\prime(\theta)\right)^2,
\ee
and is invariant under any change of the $|j\ra$ basis with purely real valued transformation coefficients. Here we have not explicitly given the limits of the sum over $j$ because the $|j\rangle$ basis could be of any dimensionality including $\infty$. The above statement, of course,  also equally well holds if the probability amplitudes $\Gamma_j(\theta)$ are purely imaginary because we can absorb the $i$ into the definition of the states $|j\ra$ to make the probability coefficients real-valued.

Note that our two-point separation problem satisfies the conditions of this generally stated invariance result  if the aperture is symmetric so that the PSF is a symmetric real-valued function. Any point at the same distance from the origin then has the same probability for even or odd spatial mode functions. 

\section{Basics for working with multiple telescopes}
\label{multiple-aperture-framework}

We are now ready to introduce our framework for parameter estimation with a multiple telescope system where we generalize some of the tools introduced in the previous section. We will do this in several steps.

\subsection{The combined aperture function and PSF}
\label{multiple-aperture-PSF}

If we have several separate telescopes which we wish to combine light from, then we can take the sum of the individual aperture functions as our total aperture function, and take its Fourier transform in order to obtain the combined PSF, just as we get the PSF for a single telescope by Fourier transforming its aperture function.
Specifically, consider a hard aperture of width $d$.
It is convenient when working with angular object- and image-plane coordinates to nondimensionalize the aperture size by scaling with the wavelength $\lambda$, such that we obtain $\delta = d/\lambda$ as the size of our aperture in momentum space (i.e., transverse axes of the aperture plane of the imaging system)~\cite{Goodman2005}. Our aperture function is then
  \be
\tildepsi_{\rm 1ap}(k) = \frac{1}{\sqrt{\delta}} \textrm{rect}(k/\delta),
\label{psi_k-def}
\ee
Here, we are defining the rect function such that $\textrm{rect}(x)$ is equal to unity in the interval $-1/2 < x < 1/2$ and therefore gives unity if integrated over all $x$.
Taking the inverse Fourier transform of this aperture function gives the PSF
\be
\psi_{\rm 1ap}(x)
= \frac{1}{\sqrt{2\pi}} \int dk \exp(i k x) \tildepsi_{\rm 1ap}(k)
= \sqrt{\sigma}\,
\frac{\sin(\pi x/\sigma)}{\pi x},
\label{PSF-sinc-def}
\ee
where
\be
\sigma = 2\pi/\delta
\label{delta-sigma-relation}
\ee
is the Rayleigh separation, and our convention is to have ${1}/{\sqrt{2\pi}}$ in the Fourier transform and its inverse in order to maintain normalization.

If we place the same aperture at a shifted location, say at position $\alpha$ in the momentum plane instead of the origin, then we get the same PSF but with a phase factor $\exp(i \alpha x)$. This does not have any physical relevance as long as we only have a single telescope. However, if we have several apertures at separate locations, then they all acquire different phases, and therefore, if we combine the light collected by them interferometrically in some form, then we get interference between these phases. It is this interference that results in the gain from combining the light collected at distant locations, creating the effect of a giant telescope whose size is nearly equal to the distance between the different apertures.

To study this phenomenon, we will take our aperture function as the sum of all the individual aperture functions, and define their combined PSF as the sum of their Fourier transforms.
Regardless of what measurement is used,
this combined PSF gives the profile of the incoming light falling on an imaginary imaging plane and can be used for calculating the QFI of our physical set up for parameter estimation as well as the CFI for different measurement schemes.
Consider $n$ apertures of equal size centered at locations $\alpha_\mu$ in momentum space (i.e. aperture plane coordinates) with $\mu = 1\ldots n$.
We assume that the average position of these apertures is zero, that is,
\be
\sum_\mu \alpha_\mu = 0
\label{mean-aperture-position-zero}
\ee
If we express $\alpha_\mu$ in terms of the single aperture size $\delta$ and the ratio $R_\mu = {\alpha_\mu}/{\delta}$, then we have
\be
\alpha_\mu = R_\mu \delta = 2\pi R_\mu/\sigma,
\label{alpha_mu-sigma-relation}
\ee
where in the second step, we have used (\ref{delta-sigma-relation}). The full aperture function for the entire set up is
\be
\tildepsi_{\rm comp}(k)
= \frac{1}{\sqrt{n}} \sum_\mu \tildepsi_{\rm 1ap}(k-\alpha_\mu).
\label{aperture-function-n-apertures}
\ee
The compound PSF, which is obtained by taking the inverse Fourier transform of this, is then
\be
\psi_{\rm comp}(x)
= \frac{1}{\sqrt{n}} \psi_{\rm 1ap}(x)
\times \sum_\mu \exp(i \alpha_\mu x).
\label{compound-PSF-n-apertures}
\ee
Or if we like, we can express this in terms of the single aperture Rayleigh scale $\sigma$ and the ratios $R_\mu$ of the positions of the individual apertures to the single telescope size, writing the exponentials as $\exp(i \alpha_\mu x) = \exp(2\pi i R_\mu x/\sigma)$.

As a specific example let us say we have two hard apertures $(n=2)$ with their centers at positions $\pm b/2$. In our aperture plane coordinates, we obtain the positions $\alpha_1 = -\beta/2$.
 and $\alpha_2 = \beta/2$, with $\beta = b/\lambda$. 
We thus have the aperture function
\be
\tildepsi_{\rm 2ap}(k) = 
\frac{1}{\sqrt{2\delta}} 
\left[\textrm{rect}\left(\frac{k -\beta/2}{\delta}\right)
\,+\, \textrm{rect}\left(\frac{k +\beta/2}{\delta}\right)\right].
\label{psi-compound-2-aperture-k-def}
\ee
The inverse Fourier transform gives the combined PSF
\be
\psi_{\rm 2ap}(x)
= \sqrt{2\sigma}
\cos(\pi x r/\sigma) \, \frac{\sin(\pi x/\sigma)}{\pi x},
\label{compound-psf-2-apertures}
\ee
where
\be
r \equiv \beta/\delta,
\label{r-def}
\ee
and hence,
\be
\beta = 2\pi r/\sigma.
\label{beta-in-terms-of-sigma}
\ee

\subsection{Local modes for an aperture}
\label{local-mode-section}

The other ingredient we need for our analysis is a convenient set of basis modes for a multiple aperture system. For this, it is natural to think in terms of a basis of local mode functions defined on each aperture
\be
\tildephi_j(k)
= \frac{f_j(ik)}{\sqrt{\delta}} \textrm{rect}(k/\delta),
\label{tilde-psi-j-general-definition}
\ee
spanning the space of all functions $f_j(i k)$ defined on the aperture, i.e., in the region $-\delta/2 <k <\delta/2$. (The reason for writing $f_j(i k)$ as functions of the imaginary $i k$ instead of just $k$ will become clearer in section \ref{sinc-bessel-modes-introduction-section}.)
For each hard aperture at $\alpha_\mu$, we will get its set of mutually orthonormal `local' modes $\tildephi_j(k-\alpha_\mu)$.
These will be (trivially) orthogonal to all the local modes of other apertures at different locations due to their rect functions being zero in non-overlapping regions:
\be
\int dk \tildephi_j(k-\alpha_\mu) \,\tildephi_l(k-\alpha_\nu)
= \delta_{\mu\nu} \delta_{jl}.
\label{orthogonality-local-modes-momentum-space}
\ee
These local modes thus form a complete orthonormal basis on the union of all the apertures.
This mutual orthogonality will naturally also hold if we shift to the (joint) image plane, where the modes will be given by the inverse Fourier transforms of the aperture plane functions:
\begin{equation}
	\begin{aligned}
\phi_{j\mu}(x) \equiv &\frac{1}{\sqrt{2\pi}} \int_{-\infty}^\infty dk \,\exp(i k x) \tildephi_j(k-\alpha_\mu)\\
		= &\exp(i \alpha_\mu x) \phi_j(x),
	\end{aligned}
\label{local-modes-in-position-space}
\end{equation}
where $\phi_j(x)$ is the $j$th mode function for an aperture located at the origin of the aperture plane, and is given by the inverse Fourier transform of the momentum space function 
\be
\phi_j(x) \equiv \frac{1}{\sqrt{2\pi}} \int_{-\infty}^\infty dk \,\exp(i k x) \tildephi_j(k).
\label{phi_j_def}
\ee
The functions $\phi_{j\mu}(x)$ will be defined on the entire imaging screen, and will form a complete orthonormal basis for the image plane:
\be
\int_{-\infty}^\infty dx \, \phi_{j\mu}^*(x) \, \psi_{\nu l}(x)
= \delta_{\mu\nu} \delta_{jl}.
\label{orthogonality-local-modes-position-space}
\ee
Now, let us say we wish to calculate the correlation function of the combined PSF $\psi_{\rm comp}(x)$ given in (\ref{compound-PSF-n-apertures})
with some arbitrary function which can be expressed in terms of the local mode functions as
\be
\psi_{\rm arb}(x)
= \sum_{\mu=1}^n \sum_{j=0}^\infty a_{j\mu} \phi_{j\mu}(x)
= \sum_{\mu=1}^n \sum_{j=0}^\infty a_{j\mu} \exp(i \alpha_\mu x)\, \phi_j(x)
\label{psi-arb-definition}
\ee
where $a_{j\mu}$ are some constant coefficients.
Since each $\exp(i\alpha_\mu x) \phi_j(x)$ term corresponds to a rect function centered at $\alpha_\mu$ in momentum space,
only products of terms from the same aperture will give non-zero contributions to the correlation function.
These contributions will involve correlation functions of the single aperture local modes $\phi_{j\mu}(x)$ with the single aperture PSF phased shifted according to its location. Therefore,
\begin{align}
\Gamma_{j\mu}(a)
\equiv &\int_{-\infty}^\infty dx \,\phi_{j\mu}^*(x) \, \exp(i\alpha_\mu (x-a)) \,\psi_{\rm 1ap}(x-a) \nn \\
= &\int_{-\infty}^\infty dx\, \left[\exp(-i\alpha_\mu x) \phi_j ^*(x)\right] \\
&\times\left[\exp(i\alpha_\mu (x-a)) \,\psi_{\rm 1ap}(x-a)\right] \nn \\
= &\exp(-i\alpha_\mu a) \Gamma_j(a),
\label{correlation-function-shifted}
\end{align}
where 
\be
\Gamma_j(a) = \int_{-\infty}^\infty dx \phi_j ^*(x) \,\psi(x-a)
\ee
is the correlation function between the $j$th mode and the PSF for a single aperture located at the origin of the aperture plane, as defined in (\ref{correlation-function-mode-j-1ap}).

The correlation function of our arbitrary function $\psi_{\rm arb}(x)$ with the compound aperture PSF (\ref{compound-PSF-n-apertures}) is then given in terms of the unshifted correlation functions $\Gamma_j(a)$ as
\begin{align}
\Gamma_{\rm arb} &= \int_{-\infty}^\infty dx \psi_{\rm arb}^*(x) \, \psi_{\rm comp}(x-a) \nn \\
&= \frac{1}{\sqrt{n}} \sum_{\mu=1}^n \sum_{j=0}^\infty a_{j\mu}^* \,\Gamma_{j\mu}(a) \nn \\
&= \frac{1}{\sqrt{n}} \sum_{\mu=1}^n \sum_{j=0}^\infty a_{j\mu}^* \,\exp(-i\alpha_\mu a) \, \Gamma_{\alpha_\mu, j}(a).
\label{arbitrary-function-correlation-function-multiple-apertures}
\end{align}
As a special application of the same idea, case, the autocorrelation function of the compound aperture PSF will be
\begin{align}
\Gamma_{\rm PSF, comp}(a)
&= \int_{-\infty}^\infty dx \,\psi_{\rm comp}^*(x) \,\psi_{\rm comp}(x-a) \nn \\
&= \frac{1}{n} \Gamma_{\rm PSF, 1ap}(a) 
\, \sum_\mu \exp(-i \alpha_\mu a),
\label{autocorrelation-function-compound-PSF}
\end{align}
where $\Gamma_{\rm PSF,1ap}(a)$ is the autocorrelation function for our single hard aperture PSF (\ref{PSF-sinc-def})
\begin{equation}
\Gamma_{\rm PSF,1ap}(a) = \int_{-\infty}^\infty dx \, \psi_{\rm 1ap}^*(x) \,\psi_{\rm 1ap}(x-a)= \frac{\sigma \sin(\pi x/\sigma)}{\pi x}.
\label{correlation-function-sinc}
\end{equation}
Before closing this section, we would like to point out that the individual aperture correlation functions $\Gamma_j(a)$ for an unshifted hard aperture, i.e., one centered at $k=0$ are proportional to the mode functions $\phi_j ^*(a)$. This follows from writing out the correlation function integrals in terms of the momentum space forms
\begin{align}
\Gamma_j(a) &= \int_{-\infty}^\infty dx \, \phi_j ^*(x) \,\psi_{\rm 1ap}(x-a) \nn \\
&= \int_{-\infty}^\infty dx \,dk \,dp\,  
\exp(i k (x-a) -i p x) \tildepsi_{\rm 1ap}(k) \, \tildephi_j ^*(p) \nn \\
&= \int dk\,
\exp(-i k a) \tildepsi_{\rm 1ap}(k) \, \tildephi_j ^*(k) \nn \\
&= \int dk \exp(-i k a) \frac{f_j ^*(i k)}{\delta} \textrm{rect}^2(k/\delta) \nn \\
&= \sqrt{\sigma} \phi_j ^*(a).
\label{mode-functions-as-correlation-functions-on-hard-aperture}
\end{align}
The third equation above was obtained by integrating over $p$ and $x$ by using the Fourier representation of the delta function, and the fourth one by writing $\tildepsi_{\rm 1ap}(k)$ and $\tildephi_j(k)$
in terms of the rect functions according to (\ref{psi_k-def}) and (\ref{tilde-psi-j-general-definition}).
In the last step, we note that the square of the rect function gives the rect function itself and compare with (\ref{tilde-psi-j-general-definition}). Here we also use (\ref{delta-sigma-relation}) to write $\delta$ in terms of $\sigma$. Lastly, note that for real valued mode functions, such as in the sinc-bessel basis,  $\phi_j ^*(a) = \phi_j(a)$.

Note that this property that the correlation functions for a mode basis defined on a hard aperture is proportional to the corresponding mode function (or its complex conjugate if the function is complex) will also hold for a hard aperture in 2 dimensions.
In that case, the exact same steps that we wrote above will apply except that the rect function will be replaced by our two dimensional step function that is unity on the aperture and zero everywhere else, and the proportionality constant will arise from the square root of the area of the aperture instead of length. For example, for a circular aperture, we will have a circ function instead of the rect function.

\subsection{The single-photon state of the PSF mode and local modes in the bra-ket notation}
\label{local-modes-bra-ket-notation-section}

We now summarize our results from the previous two sections on a multiple aperture system in a convenient bra-ket notation. Recall from Section \ref{physical-model-preliminaries-section} that the single-photon density matrix is given by $\rho_1 =\sum_{s=1}^{n_e} b_s |\psi_s\ra\la\psi_s|$, where $|\psi_s\ra = \int_{-\infty}^\infty dx \psi(x-x_s) |x\ra$.
This still holds for the multiple aperture case, except that the PSF is now the compound aperture function (\ref{compound-PSF-n-apertures}) associated with our multiple aperture system. Since this compound PSF is a sum of the individual contributions from all the apertures, the quantum state $|\psi_s\ra$ is also a sum:
\begin{align}
|\psi_s\ra & = \int_{-\infty}^\infty dx \psi(x-x_s) |x\ra \nn \\
&= \frac{1}{\sqrt{n}} \sum_{\mu=1}^n \int_{-\infty}^\infty dx \exp(i \alpha_\mu (x-x_s)) \, \psi_{\rm 1ap}(x-x_s) |x\ra \nn \\
&= \frac{1}{\sqrt{n}} \sum_{\mu=1}^n |\psi_{\mu, s}\rangle,
\label{compound-aperture-s-point-state}
\end{align}
where in the second line, we have used our formula for the compound aperture function (\ref{compound-PSF-n-apertures}) and in the third line, we have defined the contribution from each aperture to be the state
\be
|\psi_{\mu, s}\ra = \int_{-\infty}^\infty dx \,\exp(i \alpha_\mu (x-x_s)) \, \psi_{\rm 1ap}(x-x_s) \,|x\ra.
\label{PSF-state-s-point-contribution}
\ee
Physically, the state $|\psi_{\mu,s}\ra$ is the quantum state of a photon originating from a point source at position $x_s$ and arriving at the image plane through the $\mu ^{\rm th}$ aperture of our multiple-telescope system.

As for our local modes, we have the states
\be
|j, \mu\ra = \int_{-\infty}^\infty dx \,\exp(i\alpha_\mu x) \, \phi_j(x) \,|x\ra.
\label{local-mode-ket-definition}
\ee
That is, this is the quantum state for an incoming photon to be found in the local mode number $j$ of the $\mu$th aperture of our system. As discussed in the previous section, these local modes are mutually orthonormal, that is
\be
\la j, \mu|l, \nu\ra = \delta_{jl} \,\delta_{\mu\nu}.
\ee
For the contributions of the single aperture pieces to correlation functions, recalling (\ref{correlation-function-shifted}) and the discussion around it, we have
 \be
\la j,\mu |\psi_{\nu, s}\ra
= \delta_{\mu\nu} \exp(-i\alpha_\mu a) \Gamma_j(a).
\label{correlation-function-individual-aperture-pieces}
\ee

\subsection{The sinc-bessel modes}
\label{sinc-bessel-modes-introduction-section}

So far, we have refered to local modes of apertures in general terms without specifying a particular basis. It is worth pointing out, however, that the sinc-bessel modes provide a natural basis for a rect aperture. In this basis, the zeroth mode is simply the PSF (associated with a rect function), and subsequent modes can be obtained by taking derivatives of the PSF and carrying out a Gram-Schmidt orthogonalization in order to subtract their overlap with the previous modes~\cite{KGA17, Rehacek2017a}.
For example, mode one is simply proportional to the derivative of the PSF, since the latter is an even function and the former is odd, and hence their inner product is zero. The second derivative, however, does have a non-zero overlap with the PSF, and therefore we need to subtract it to obtain mode two. Likewise, the third derivative overlaps with the first derivative (but not the PSF and the second derivative), so we need to subtract that piece to get mode three.
If we instead carry out this procedure in the aperture plane, then the rect aperture function (\ref{psi_k-def}) is the fundamental mode, and the derivatives are replaced by moments on which we carry out the Gram-Schmidt orthogonalization. We then obtain functions of the imaginary $i k$ and this is why we have been writing the functions $f_j(i k)$ in (\ref{tilde-psi-j-general-definition}) with $i$. It turns out that these have the form of $j$th degree Legendre polynomials having only even (odd) terms for even (odd) $j$ \cite{Rehacek2017a}.

\subsection{Apertures other than hard ones}

Most of the above discussion also holds for arbitrary apertures, especially the phase shift in the PSF and the correlation functions arising from the position of an aperture.
The tricky part may however arise when  two apertures are close to each other and the aperture functions overlap, thus not being orthogonal. This is of course not exactly a realistic situation and normally we have hard apertures. Smooth functions like Gaussians are usually used as approximations or proxies for hard apertures because that sometimes makes calculations easier. Therefore, a realistic approximation should also capture the fact that two (hard) apertures at different positions should not overlap with each other even when the distance between them is small (i.e., the center to center distance is larger than, but not much larger than the sum of their radii). We can therefore make an ad-hoc fix of this overlap problem and assume that two apertures and their local modes will be mutually orthogonal.
With that done, we can use the rules of Section \ref{local-modes-bra-ket-notation-section} with the amplitude functions $\Gamma_j(x_s)$  being that of the new aperture type and the basis functions being considered.
This makes sense when we are doing mode sorting on our multiple aperture system. It is easy to see this for measurement schemes where light collected at different receivers is brought to a common location through single mode fibers or multimode `light pipes.' However, it also holds when the light from different apertures is being focused on to a common imaging screen, e.g., individual mirrors of a large segmented telescope, and we apply a spatial mode  measurement on that imaging screen. The key in the latter case is to express the spatial modes on the common imaging screen in terms of the local modes of the individual apertures, and drop the overlap contributions involving products of terms arising from different apertures.
The situation will however be more complicated for direct imaging on light focused on to a common imaging screen.
In that case, approximating the individual apertures with Gaussians or some other smooth function type does not make sense to begin with, unless there is a really sound reason to justify such an approach.

\section{Multi-aperture receiver designs}

We are now in a position to discuss our different multi-aperture receiver categories and the general parameter estimation problem from Section \ref{general-parameter-estimation-problem-section}.

\subsection{Multi-axial beam combination}

When the different apertures are sufficiently close to each other, then we can use carved out mirror segments of a giant parabolic mirror that we would use if we had a large telescope. This situation is encountered in binocular telescopes, such as the Large Binocular Telescope (LBT) in Arizona, and many planned next generation ground-based telescopes, such as the Giant Magellan Telescope (GMT). These mirror segments can focus the light on to a common imaging screen where an appropriate measurement can be made using any of the techniques employed for a single-aperture telescope. We will consider ideal direct imaging, a SPADE and the so-called SLIVER (Super Localization by Image Inversion Interferometry), which sorts the incoming light into even and odd parts, is experimentally easier to implement than a SPADE, and attains the QFI for the 2-point problem in the small separation limit~\cite{Nair2016}.
 
\subsubsection{Ideal direct imaging}

Direct imaging is a natural imaging strategy when using multiple apertures in a multi-axial combination configuration with a common Fourier-conjugate image plane, provided that the telescope array can be precisely phased~\cite{Glindemann2011}.
As in section \ref{general-parameter-estimation-problem-section}, we will consider the idealized case with infinitesimally small pixellation, unity fill factor, and infinite spatial extent of the detection screen. The probability density for an incoming photon to be found at position $x$ on the imaging screen will be calculated just as in section \ref{general-parameter-estimation-problem-section}, except that we now use the combined PSF (\ref{compound-PSF-n-apertures}).
\begin{align}
P(x) = &\frac{1}{n} \sum_{s=1}^{n_e}
 b_s \bigg|\psi_{\rm comp}(x-x_s)\bigg|^2 \nn \\
= &\frac{1}{n} \sum_{s=1}^{n_e}
 b_s \bigg|\sum_{\mu=1}^n \exp(i \alpha_\mu (x-x_s))\bigg|^2 \, \psi_{\rm 1ap}^2(x-x_s) \nn \\
= &\frac{1}{n} \sum_{s=1}^{n_e}
 b_s \left(n +\sum_{\mu> \nu, \mu, \nu=1}^n \cos\big((\alpha_\mu-\alpha_\nu)(x-x_s)\big)\right)\nn \\
 &\times \psi_{\rm 1ap}^2(x-x_s).
\label{P_x-general-compound-aperture}
\end{align}
Here we are taking the square of the single-aperture PSF since it is a purely real-valued function due to the individual apertures being symmetric.
The partial derivative with respect to the unknown parameter $\theta_g$ then has three parts
\begin{widetext}
\begin{align}
\frac{\partial P(x)}{\partial \theta_g}
&= \frac{1}{n} \sum_{s=1}^{n_e}
\frac{\partial b_s}{\partial\theta_g} \left(n +\sum_{\mu> \nu, \mu, \nu=1}^n \cos((\alpha_\mu-\alpha_\nu)(x-x_s))\right) \, \psi_{\rm 1ap}^2(x-x_s) \nn \\
&+ \frac{1}{n}b_s \frac{\partial x_s}{\partial\theta_g} 
\Bigg(\sum_{\mu> \nu, \mu, \nu=1}^n 
(\alpha_\mu-\alpha_\nu) \,\sin((\alpha_\mu-\alpha_\nu)(x-x_s)) \,\psi_{\rm 1ap}^2(x-x_s)
\nn \\
&-2\sum_{\mu> \nu, \mu, \nu=1}^n\left(n +\cos((\alpha_\mu-\alpha_\nu)(x-x_s))\right) \,\psi_{\rm 1ap}^\prime(x-x_s)\Bigg)\, \psi_{\rm 1ap}(x-x_s).
\label{P_x-general-compound-aperture-derivative}
\end{align}
\end{widetext}
Here $\psi_{\rm 1ap}^\prime(x-x_s)$ is the partial derivative of $\psi_{\rm 1ap}(x-x_s)$ with respect to $x$.

Recall from (\ref{FI-formula}) and the discussion around it that the CFI is given by 
\be
\sJ_{\rm DI}(\theta_g) = N \int_{-\infty}^\infty dx \frac{1}{P(x)} \left(\frac{\partial P(x)}{\partial\theta_g}\right)^2.
\ee
Note that $\left(\partial P(x)/\partial \theta_g\right)^2$ will have a term proportional to the square of the distance between two apertures $(\alpha_\mu-\alpha_\nu)^2$.
It is this term that gives us the gain from baseline interferometry and creates the effect of a giant telescope whose size is approximately equal to the distance between the individual telescopes.
If we only have a single telescope, then we only obtain $\left(\psi_{\rm 1ap}^\prime(x-x_s)\right)^2$, which is  of the order of the square of the individual aperture size i.e. $1/\sigma^2 = \delta^2/(4\pi^2)$. Note in (\ref{P_x-general-compound-aperture-derivative}), however, that we only obtain this $(\alpha_\mu-\alpha_\nu)^2$ contribution to the CFI from the partial derivative of the positions $x_s$ with respect to the parameter $\theta_g$. It is not there in the term involving the derivatives of the brightnesses. This may mean that we do not get such an enhancement if only the brightnesses depend on the parameter $\theta_g$. But it is not immediately clear if that is so, as that dependence may still show up through an implicit dependence. The framework presented in this paper allows for studying quantum limits of generalized parameter estimation for a general-configuration multi-aperture system, but such an investigation is out of scope of this manuscript.

\subsubsection{SPADE}
\label{SPADE-common-imaging-screen-section}

We can also carry out a SPADE measurement on our common image plane, in lieu of the direct-detection imaging screen. The details of this depend on the exact mode basis and therefore cannot be written in general terms, except reproducing the formulas for a SPADE from Section \ref{general-parameter-estimation-problem-section}, where now the correlation functions will be those associated with the mode basis of our SPADE.
Recall that correlation functions of an arbitrary function in a multiple aperture set up with the PSF having the form (\ref{arbitrary-function-correlation-function-multiple-apertures}).
Since the probability of finding an incoming photon to be in a given mode function in a SPADE measurement will involve the square of such correlation functions, we will get terms with factors of the form $|\exp(i (\alpha_\mu-\alpha_\nu)x_x)|^2$. Therefore, if $x_s$ depends on the parameter $\theta_g$,
the derivative of the probability with respect to $\theta_g$ will contain factors of the form $(\alpha_\mu-\alpha_\nu)^2$, giving us the performance enhancement from combining light from distant apertures just as we found for direct imaging.

\subsubsection{SLIVER}
\label{SLIVER-introduction-section}

The SLIVER, which was introduced in~\cite{Nair2016}, is a simple setup that separates the even and odd parts of the collected light. The input light is first split into two parts using a 50-50 beam splitter. One of the beamsplitter outputs is then spatially inverted about the central axis of the apparatus. The two beams are then recombined using a second 50-50 beam splitter. The outputs then contain the sum and difference of these beams, with the former being the even part of the incoming signal, and the latter being the odd piece. These are then measured using photon counters.
The details of the performance of a SLIVER for a general parameter estimation problem are not very insightful, and therefore we will skip them. The SLIVER, however, attains the QFI in the small separation limit of the uniformly bright 2-point problem, and therefore we will calculate its performance during our discussion of that problem in Section \ref{2-point-separation-section}.

\subsection{Co-axial beam combination}

When the distances between the individual apertures get larger, it is no longer feasible to focus the light on to a common imaging screen. However, we can bring the light collected at distant telescopes to a central location through single mode fibers that couple in light from the `local modes', or via multimode `light pipes'.

\subsubsection{Receivers employing mode sorting}

\noindent {\textbf{(a) Universal co-axial receiver}}

Recall from section \ref{multiple-aperture-framework} 
that the local modes $\tildephi_j(k-\alpha_\nu)$ associated with the different telescopes form a complete orthonormal basis for all the regions in the aperture plane covered by the apertures.
Now, if we have an incoming photon in some arbitrary state
\be
\tildepsi_{\rm arb}(k)
= \sum_{\mu=1}^n \sum_{j=0}^\infty a_{j\mu} \, \tildephi_j(k-\alpha_\mu),
\label{tilde-psi-arb-definition}
\ee
then the coefficients $a_{j\mu}$
are the probability amplitudes associated with measuring an incoming  photon in this basis and finding it to have arrived in aperture $\mu$ and been sorted into the local mode $j$.
If we have a mode sorter at each aperture coupling the light in $\tildephi_j(k-\alpha_\mu)$ into the Gaussian mode $T_{00}$ of a single mode fiber, then we get
the probability amplitude $a_{j\mu} T_{00}$ in that fiber.
If we bring a collection of such fibers to a central location and feed them into a linear optical interferometer with some combination of coefficients $d_{\xi \mu j}^*$, where $\xi$ is an index for the different outputs of the interferometer, then we obtain the amplitudes 
$\sum_{\mu=1}^n \sum_{j=0}^{j_{\rm max}} d_{\xi \mu j}^* a_{j\mu} T_{00}$
in the output ports of that interferometer, where $j_{\rm max}$ is the maximum-$j$ local modes collected. The total photon probability associated with output $\xi$ is then:
\be
P_{\xi} = \left|\sum_{\mu=1}^n \sum_{j=0}^{j_{\rm max}} d_{\xi \mu j}^* a_{j\mu}\right|^2.
\ee
This is exactly the probability for finding a photon coming into our multiple aperture system to be in the aperture-plane mode
\be
\tildechi_\xi(k)
= \sum_{\mu=1}^n \sum_{j=0}^{j_{\rm max}} d_{\xi \mu j} \tildephi_j(k-\alpha_\mu).
\ee
Thus, measuring the outputs of the interferometer allows us to obtain the (Poisson shot noise limited) photon counts in the $\tildechi_\xi(k)$ basis. We have therefore shown that coupling the different local modes at the distant telescopes into single mode fibers (via local SPADEs), bringing them into a linear interferometer at a central location, and measuring the outputs with photon counters is equivalent to carrying out a SPADE measurement on the multiple aperture system in the $\tildechi_\xi(k)$ basis. While we have explained all this in terms of the aperture plane functions, this remains equally valid in the image plane where $\tildechi_\xi(k)$ will be replaced by its Fourier transform
\be
\chi_\xi(x) = \sum_{\mu=1}^n \sum_{j=0}^{j_{\rm max}} d_{\xi \mu j} \exp( i\alpha_\mu x) \phi_j(x).
\ee
\\ 

\noindent {\textbf{(b) Groupwise receiver}}
\label{groupwise-receiver-section}

While we can in principle combine the various local modes from the different receivers in any arbitrary combinations, one simpler approach is to bring the light in each local mode together and interfere it in the same linear combination. That may for example render local SPADEs unnecessary. Specifically, for each local mode number $j$, we have $n$ signals from the different apertures labeled by the index $\mu$. We can feed these into a beam splitter circuit with linear transformation coefficients $c_{\gamma\mu}^*$,
where $\gamma = 0, 1,\ldots n$ is an index denoting the $n$ outputs of each such interferometer. These coefficients $c_{\gamma\mu}^*$ thus form a unitary matrix describing the linear transformation between the $n$ input ports of the beam splitter enumerated by the index $\mu$ and the $n$ outputs denoted by the index $\gamma$.
The light in the output ports of these beam splitters can then be measured with photon counters.
This approach amounts to measuring the incoming photons in the basis functions
\be
\phi_{j\gamma}(x) = \left(\sum_\mu c_{\gamma\mu} \,\exp(i \alpha_\mu x)\right) \,
\phi_j(x).
\label{groupwise-basis-modes}
\ee
To calculate the performance for estimating our general parameter $\theta_g$ by measuring in this basis, we need the correlation functions of these modes with the $n$-aperture PSF as discussed earlier. Recall from Section \ref{local-mode-section} that the inner products of the contributions from two different apertures are zero, and for the same aperture, we get the single aperture correlation functions up to a phase. Therefore, we obtain
\begin{equation}
	\begin{aligned}
		\Gamma_{j\gamma}(x_s) \equiv &\int_{-\infty}^\infty dx \phi_{j\gamma}^*(x) \psi_{\rm comp}(x-x_s)\\
		= &\frac{1}{\sqrt{n}}
		\left(\sum_{\mu=1}^n c_{\gamma\mu}^* \,\exp(-i\alpha_\mu x_s)\right)\, \Gamma_j(x_s).
	\end{aligned}
\end{equation}
The probability for an incoming photon to be measured in a detector associated with basis function $\phi_{j\gamma}$ is therefore
\be
P_{j\gamma} = \frac{1}{n} \sum_{s=1}^{n_e} b_s\,
B_\gamma(x_s) \, \Gamma_j ^2(x_s),
\label{P-gamma-j-def}
\ee
where we have taken the regular square for $\Gamma_j (x_s)$ assuming that it is purely real-valued, hence the magnitude squared is equal to the regular square, and
\begin{align}
B_\gamma(x_s) &\equiv \bigg|\sum_{\mu=1}^n c_{\gamma\mu}^* \,\exp(-i\alpha_\mu x_s)\bigg|^2 \nn \\
&= \sum_{\mu,\nu=1}^n c_{\gamma\mu}^* c_{\gamma_\nu}\,\exp(-i(\alpha_\mu-\alpha_\nu) x_s).
\label{B-def}
\end{align}
The CFI is then given by the usual formula
\be
\sJ_{gw}(\theta_g)
= N \sum_{\gamma =1}^n \sum_{j=0}^\infty \frac{1}{P_{j\gamma}} \left(\frac{\partial P_{j\gamma}}{\partial\theta_g}\right)^2.
\label{CFI-groupwise-receiver-general-problem}
\ee

In a realistic setting, it will only be possible to collect up to a certain number of local modes, so $j$ will go from $0$ to some maximum number $j_{\rm max}$.
We can however deploy local SPADEs at each telescope, which collect and fiber-couple modes indexed $0$ through $j_{\rm max}$ (for transmission to the central location), and detect any photons in the remainder of higher-order modes with a bucket detector. We will then only add over $j$ up to $j_{\rm max}$ in equation (\ref{CFI-groupwise-receiver-general-problem}), and just add the CFI contribution from the orthogonal complement of this modal plan for which the photon count is obtained from the local bucket detectors. This will be given by:
\be
P_r = \sum_{s=1}^{n_e} b_s \left(1-\sum_{j=0}^{j_{\rm max}} \Gamma_j ^2(x_s)\right).
\ee 

The $j_{\rm max}=0$ case corresponds to the special situation where we simply have a binary mode sorter at each aperture location, coupling the local PSF (zero) mode into a single mode fiber, while allowing the rest of the light to go into a bucket detector. As the zero-mode signals from the $n$ apertures are combined in a beam splitter circuit, we obtain $n$ outputs corresponding to different linear combinations of the zero mode from each aperture. The total photon count from the bucket detectors at the local sites give the number of photons in the orthogonal complement of the function space, and thus we have an effective $n+1$ SPADE on the compound aperture.

One special example of a groupwise receiver would be a pairwise measurement scheme involving two apertures. For each mode $j$, we would have two signals from the two locations, and we would combine them in a 50-50 beam splitter. So the outputs would be the sum and difference of the input signals.
If the apertures are at locations $\alpha_1=-\beta/2$ and $\alpha_2=\beta/2$ in the aperture plane, as we discussed in Section \ref{multiple-aperture-PSF} starting around equation (\ref{psi-compound-2-aperture-k-def}), then measuring the output with the symmetric linear combination of the $j$th mode signals corresponds to measuring in the mode
\begin{align}
 \phi_{j+}(x) &= \frac{1}{\sqrt{2}} \left(\exp(-i\beta x/2) +\exp(i\beta x/2)\right) \psi_{\textrm{h},j}(x) \nn \\
&= \sqrt{2}\cos(\pi r x/\sigma) \psi_{\textrm{h},j}(x),
\end{align}
where we have used (\ref{r-def}) and (\ref{delta-sigma-relation}) to express the argument in terms of the single aperture Rayleigh parameter $\sigma$. The $j=0$ case of this corresponds to measuring in the compound aperture PSF.
In the same way, measuring the anti-symmetric outputs of the 50-50 beam splitter for mode $j$ corresponds to measuring in the spatial mode function
\be
 \phi_{j-}(x) = \sqrt{2}\sin(\pi r x/\sigma) \psi_{\textrm{h},j}(x).
\ee
The functions $B_\gamma(x_s)$ for the symmetric and anti-symmetric combinations are then 
\begin{eqnarray}
B_+(x_s) &=& 2\cos^2(\pi r x_s/\sigma), \,{\text{and}} \nn \\
B_-(x_s) &=& 2\sin^2(\pi r x_s/\sigma).
\label{B-symmetric-and-antisymmetric}
\end{eqnarray} 
We will use this simplification in our treatment of the 2-point-source problem in \ref{2-point-separation-section}.

We can also extend this pairwise scheme to more than two apertures that are in a symmetric arrangement about the center of our multiple aperture system. Say we have apertures at $\pm\beta_1/2$, $\pm\beta_2/2$, $\pm\beta_3/2$, and so on. Then for each equidistant oppositely located pair, we can mix the signals for each local mode using a 50-50 beam-splitter to obtain their symmetric and anti-symmetric combinations. We can of course also have an odd number of apertures such that in addition to a bunch of equidistant pairs, we also have one in the middle.
In that case, we would still combine the signals from all the equidistant pairs as mentioned above, and the $j$th mode of the aperture at the center will just be one other mode, which we will not mix with any light collected at other locations. 

\subsubsection{Receivers without mode sorting: using light pipes}
\label{light-pipe-receiver-section}

In this receiver type, we do not use any mode sorters. Instead, we carry all the (full, multimode) light collected at the individual apertures through `light pipes' to a central location in a way that maintains the spatial structure of the collected optical fields, combine the signals interferometrically through a beam splitter circuit, and count the total photons in each output port through a photon detector.

The probability for an incoming photon to be found in the $\gamma$th output of our beam splitter circuit can be obtained by summing (\ref{P-gamma-j-def}) over all the local modes. Since $\sum_j \Gamma_j ^2(x_s) =1$ from the normalization property, the sum over $j$ will remove the local mode dependence from the probabilities, and we obtain
\begin{align}
P_\gamma &= \sum_j P_{j\gamma} \nn \\
&= \frac{1}{n} \sum_{s=1}^{n_e} b_s\,
B_\gamma(x_s),
\label{P_gamma-def}
\end{align}
where $B_\gamma(x_s)$ was defined in (\ref{B-def}).
The CFI is calculated from this probability as
\be
\sJ_{lp}(\theta_g) = \sum_{\gamma=1}^n \frac{1}{P_\gamma}\left(\frac{\partial P_\gamma}{\partial\theta_g}\right)^2.
\ee
We should mention that this is also the CFI for a groupwise receiver in the limit where the individual aperture size shrinks to zero. In that case, $\Gamma_0(x_s) \to 1$ and all the other $\Gamma_j(x_s)$ functions approach zero. This limit is the subject of~\cite{Cosmo2020}.

Like the groupwise mode sorter based design described above, the simplest example of the light pipe receiver is to combine the light pairwise when the apertures are in a symmetric configuration about the center of the baseline.
That is, for each pair of oppositely located telescopes, we just bring the light collected at both locations and carried through light pipes into a 50-50 beam splitter and place photon counters at the output ports. 
So the output of one port is the symmetric sum of all the input from the two telescopes, and the other output is the anti-symmetric combination. The probabilities are given in terms of the functions $B_+(x_s)$ and $B_-(x_s)$ given in (\ref{B-symmetric-and-antisymmetric}).

\subsubsection{Obtaining a SLIVER: adding a reflection in a pairwise light pipe receiver}
\label{SLIVER-lightpipe-reflection-section}

It turns out that we can obtain the SLIVER with inserting a small tweak in the above-mentioned pairwise lightpipe receiver.
That is, consider having an even number of equidistant oppositely located telescopes about the center of the baseline, and are combining the signals pairwise without any spatial mode sorting. If we add a reflection to one of the input ports of the beam splitter for each such pair, then we can obtain the SLIVER introduced in section \ref{SLIVER-introduction-section}. This sifts the incoming light into the even and odd parts of the electric field for each telescope pair.

To see this explicitly, say we have two apertures of size $\delta$ centered at locations $\alpha_1=-\beta/2$ and $\alpha_2=\beta/2$, and the photon-unit field in the aperture plane is $E(k)$. In the two apertures, this field can be parameterized in terms of a variable $\kappa$:
\be
E_\pm(\kappa) = E\left(\frac{-\delta \pm\beta}{2} +\kappa\right), \,\,\,\, 0<\kappa<\delta.
\ee 
The outputs of the 50-50 beam splitter give $1/2$ times the sum and difference of these two.
However, if we put a reflection on one of the interferometer inputs around its center, say the one located at $-\beta/2$, then its signal becomes
\be
E\left(\frac{\delta-\beta}{2} -\kappa\right), \,\,\,\, 0<\kappa<\delta.
\ee
Then the beam splitter outputs contain:
\be
\frac{1}{2}\left(
E\left(\frac{\beta-\delta}{2}+\kappa\right)
\pm E\left(-\frac{\beta-\delta}{2}-\kappa\right)\right), \,\,\,\, 0<\kappa<\delta.
\ee
That is, the even and odd parts of $E(k)$ in the regions covered by the two apertures.
Hence adding a reflection on one of the input ports of the 50-50 beam splitter turns the light pipe receiver into the equivalent of the so-called SLIVER which sorts the incoming light into its odd and even parts.
The generalization of this to any arbitrary number of equi-distant aperture pairs from the origin is trivial. In that case, we would be combining the signals for each equidistant oppositely located pair in a fifty-fifty beam splitter with a reflection in one of the input ports. The total photon cont of sum of the symmetric combination outputs of all the fifty-fifty beam splitters would then give the even part of the incoming light for the whole system, and the sum of the anti-symmetric outputs will be the odd part. But since we will also have access to the photon counts for each individual output port instead of just the total sums of the symmetric and anti-symmetric counts, this receiver would have some additional information as well and would therefore subsume the SLIVER. In case we also have an aperture at the center, then we would not combine the light received by it to any other aperture's signal, and implement just a local SLIVER there in order to get a complete SLIVER for the compound-aperture of our multiple aperture system. Naturally, this connection with the SLIVER will not hold if the aperture configuration is not symmetric about the center of the origin in the $k$-space.

\subsection{Entanglement based receivers}

When the apertures are so far away that we cannot bring the light collected by them to a central location through single mode fibers or multimode light pipes, we can still attain the QFI, a.k.a. the ultimate limit of the resolving power of the multi-aperture baseline, but we would need to use pre-shared entanglement among the telescope sites, a method to couple the quantum state of the incoming light into a bank of quantum memories, and local quantum operations at each telescope site, along the lines of Refs.~\cite{Gottesman2012} and~\cite{Khabiboulline2018}, but: (1) with the addition of local mode sorters at each telescope site, and (2) `compiling' the unitary corresponding to the quantum-optimal interferometer---described in our receiver designs above (where it was possible to bring the local modes to a central location)---into an appropriate entanglement based scheme that uses teleported logic on the so-called unary-encoded (single-photon in many modes) qudit basis. Fully fleshing out this idea for concrete imaging problems will be left as the subject of future research.

\section{The QFI for the general parameter estimation problem}
\label{QFI-general-parameter-section}

In this Section, we show how to calculate the QFI for the general parameter estimation problem for a multi-aperture baseline. We provide a scaling argument for how the QFI of a multi-aperture system improves compared to that of one constituent aperture.

Recall from Section \ref{general-parameter-estimation-problem-section} that the density matrix for $n_e$ emitters has dimensions of $n_e \times n_e$, and when we include the derivative of the density matrix, then we have a $2n_e$ dimensional Hilbert space. As the number of points increases beyond 2 or 3, calculating the density matrix and its derivative becomes increasingly complicated. As this happens, analytical calculations become impossible and one way to calculate the QFI is to expand the density matrix in a modal basis as in (\ref{rho1-mode-expansion}), truncate at some large enough mode number, and carry out the calculations numerically.
We can follow the same approach for calculating the QFI for a multiple aperture system, but must now work in terms of an appropriate basis of our multiple-aperture system.
An obvious basis for this can be the local modes of the individual apertures defined in section \ref{local-mode-section}.
Recall that the state associated with a photon in the local mode $j$ of the $\mu$th aperture is given by $|j, \mu\ra$
which we defined in (\ref{local-mode-ket-definition}).
Also recall the result (\ref{correlation-function-shifted})
that the correlation functions of the PSF of a shifted aperture at location $\alpha_\mu$ with its local modes is given by the corresponding correlation function at position 0 with a phase factor.
The density matrix in the basis of the local modes of all the individual apertures is then given by
\begin{equation}
	\begin{aligned}
		\rho_1 = &\sum_{s=1}^{n_e} b_s \,\sum_{j,l=0}^\infty \sum_{\mu, \nu=1}^n
\exp(-i(\alpha_\mu -\alpha_\nu) x_s) \\
&\times \Gamma_j(x_s) \,\Gamma_l ^*(x_s)
		\,|j, \mu\ra\la l,\nu|.
	\end{aligned}
\label{density-matrix-multiple-apertures-general-problem}
\end{equation}
The QFI can then be calculated from this by using the formula (\ref{QFI-formula}) and the SLD equation (\ref{sld}). To do this numerically, we of course truncate the modes $j, l$ at some maximum value $j_{\rm max}$.
Recalling from (\ref{sld})
that the SLD involves the derivative of the density matrix, we can see that the contributions arising from the derivative acting on the phases will again give contributions proportional to $(\alpha_\mu -\alpha_\nu)$ in the SLD. Since the QFI involves the square of the SLD, we will therefore obtain the by now familiar $(\alpha_\mu -\alpha_\nu)$ enhancement from baseline interferometry.

We should mention that the above density matrix simplifies significantly when all of the sources are within a deeply sub-Rayleigh region (i.e. $x_s \ll \sigma \, \forall\,s\in[1,n_e]$). In this limit,
$\Gamma_0(x_s) \to 1$ and
$\Gamma_j(x_s) \to 0$ for $j\neq 0$.
Therefore,  the density matrix only has contributions from the $0$th spatial mode of each local aperture and becomes
\begin{equation}
\begin{aligned}
\rho_1 ^{({\text{dsr}})} = &\sum_{s=1}^{n_e} b_s \,
\sum_{\mu, \nu=1}^n
\exp(-i(\alpha_\mu -\alpha_\nu) x_s) \\
& \times |0, \mu\ra\la 0, \nu|.
\end{aligned}
\label{density-matrix-multiple-apertures-general-problem-dsr}
\end{equation}
Here the superscript (dsr) denotes the deep sub-Rayleigh regime.
This is the limit considered in~\cite{Cosmo2020} and is the regime of interest for diffraction-limited imaging, and it also makes the calculation of the QFI relatively simpler. In this case, we can even carry out some calculations analytically for a small number of apertures (e.g., two apertures) for more complex imaging problems, if we like.

Recall from section \ref{classical-quantum-Cramer-Rao-bound-section} that measuring in the eigenbasis of the SLD of the density operator with respect to the unknown parameter attains the QFI. But since the SLD generally depends on the unknown parameter $\theta$ as well, we cannot straightaway measure in the SLD basis. We described a two-stage method where first we measure a small fraction of the incoming photons in some other basis to obtain an initial estimate $\hat\theta_1$ for the parameter, and then measure in the SLD basis obtained by setting $\theta$ equal to $\hat\theta_1$. The performance of this two-stage method asymptotically approaches the QFI for large $N$. We can also apply this method for our multiple aperture arrangement. This would require building an adaptive version of one of the receiver designs described in the previous section, which would measure in some pre-determined basis in stage one, and then switch to the SLD basis evaluated at the initial estimate of the parameter in stage two. To measure in the SLD basis in stage two, we would need to have an adaptive SPADE for a multi-axial receiver. For the co-axial receiver, we would need to engineer the beam splitter circuit so that the linear combinations can be varied and set to match the SLD basis measurement as closely as possible.

\section{The equally bright two-point-source separation problem}
\label{2-point-separation-section}

We now illustrate some of the ideas developed in the previous sections in terms of the equally bright two-point separation problem. Here we remind the reader of Section~\ref{two-point-separation-one-aperture-section} where we stated that we are assuming that our SPADEs are pointed perfectly at the centroid and therefore we can take the angular positions of the two point sources to be $\pm\theta$.
 
\subsection{The QFI}

For simplicity, we will limit ourselves to considering aperture arrangements that are symmetric about the origin. Recalling (\ref{QFI-autocorrelation}), the QFI is $-4N$ times the second derivative of the autocorrelation function of the PSF evaluated at $\theta=0$.
From section \ref{local-mode-section} and specifically equation (\ref{autocorrelation-function-compound-PSF}), the autocorrelation function for the zero mode of our compound aperture will be
\be
\Gamma_{\rm PSF,comp}(\theta)
= \frac{1}{n} \Gamma_{\rm PSF, 1ap}(\theta) 
\, \sum_\mu \exp(-i \alpha_\mu \theta).
\ee
Calculating the second derivative of this combined aperture correlation function and evaluating it at $\theta=0$, we obtain the QFI: 
\begin{align}
\sK_{\textrm{full}, \theta}
=& -4N\Gamma_{\rm PSF,comp}^{\prime\prime}(0) \nn \\
=& -4N \Gamma_{\rm PSF,1ap}^{\prime\prime}(0)
-\frac{4i N}{n} \Gamma_{\rm PSF,1ap}^\prime(0) \sum_{\mu=1}^n \alpha_\mu\nn\\
&
+\frac{4N}{n}\Gamma_{\rm PSF,1ap}(0) \sum_{\mu=1}^n \alpha_\mu ^2 \nn \\
= &\sK_{\textrm{1ap}, \theta}
+\sK_{\textrm{lb}, \theta}.
\label{QFI-2-contributions}
\end{align}
Here the first term is the single-aperture QFI we obtained in Section \ref{two-point-separation-one-aperture-section} (recall (\ref{single-aperture-QFI-2-points}))
\be 
\sK_{\textrm{1ap}, \theta}
= -4N \Gamma_{\rm PSF,1ap}^{\prime\prime}(0)
\label{single-aperture-QFI-2-points2}.
\ee
The term proportional to $\sum_{\mu=1}^n \alpha_\mu$ is zero because we took the mean position of the apertures to be zero (recall (\ref{mean-aperture-position-zero})).
The piece involving the sum of the squares of $\alpha_\mu$ is the contribution arising from baseline interferometry
\begin{align}
\sK_{\textrm{lb}, \theta}
&= \frac{4N}{n} \Gamma_{\rm PSF,1ap}(0) \sum_\mu \alpha_\mu ^2 \nn \\
&= \frac{4N}{n} \sum_\mu \alpha_\mu ^2,
\label{QFI-lb}
\end{align}
where we have used the normalization property of the single aperture PSF to see that $\Gamma_{\rm PSF,1ap}(0) = 1$.

Note that if the aperture function is a Dirac delta function, i.e., $\tilde{\psi}(k)=\delta(k)$, then the single-aperture term in the QFI will be zero. This can be shown by rewriting the Dirac delta aperture model as an infinitesimally small hard aperture $\tilde{\psi}(k)=\delta(k)\equiv\textrm{lim}_{\delta\to0}(1/\sqrt{\delta})\textrm{rect}(k/\delta)$ (\ref{psi_k-def}) and evaluating the QFI in (\ref{single-hard-aperture-QFI-2-points}) in the limit $\delta\to0$, which is equivalent to $\sigma\to\infty$ via the definition in (\ref{delta-sigma-relation}). This confirms that if the imaging system is modeled with point-like apertures, as in Refs.~\cite{Cosmo2020,Wang2021,Bojer2022}, the QFI will only depend on the positions of the various apertures. However, this neglects the contribution to the QFI that will depend on the sizes and features of the individual apertures, which we include in our analysis.

\begin{figure}[t]
	\centering
	\includegraphics[width=\columnwidth]{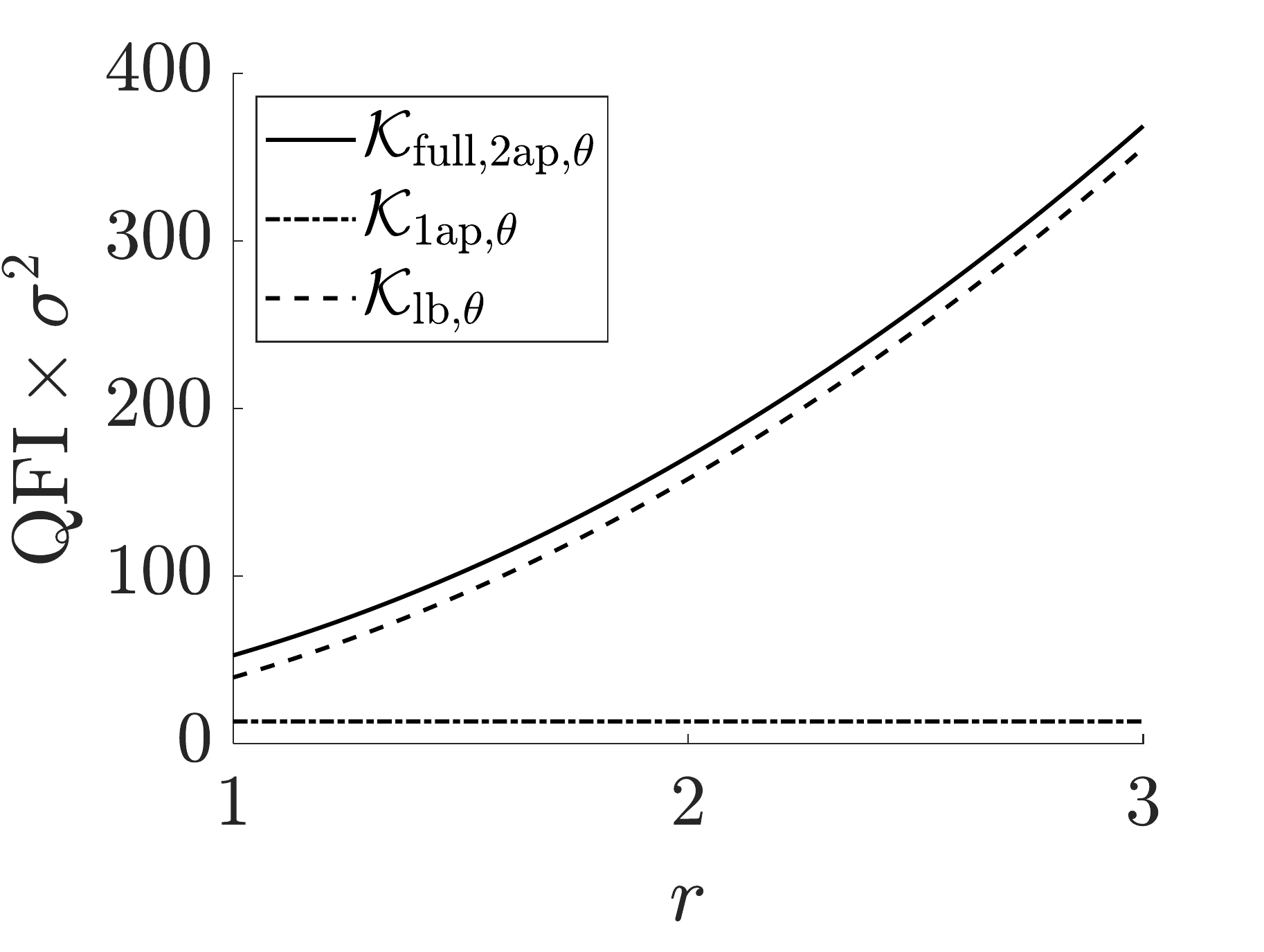}
	\caption{QFI and constituent contributions from single-aperture term and long-baseline term for estimating the separation $\theta$ between two point sources with two telescopes exhibiting baseline-to-diameter ratio $r$.}
	\label{fig:QFI}
\end{figure}

For our special example of just two hard apertures at locations $\alpha_1=-\beta/2$ and $\alpha_2=\beta/2$, we obtain the QFI
\be
\sK_{\textrm{full,2ap,} \theta}
= \frac{4\pi^2 N}{3\sigma^2}\left(3r^2 +1\right).
\label{QFI-2-apertures}
\ee 
In this, the single aperture QFI contribution is: 
\be
\sK_{\textrm{1ap,} \theta} = \frac{4\pi^2 N}{3\sigma^2},
\ee 
(recall (\ref{single-hard-aperture-QFI-2-points})), and the contribution from long baseline interferometry is: 
\be
\sK_{\textrm{lb,2ap,} \theta} = \frac{4\pi^2 N}{3\sigma^2}\left(3r^2\right),
\ee 
which scales quadratically with the baseline distance for a constant aperture diameter (recall~\eqref{r-def}). We plot the QFI as well as the contribution from the two constituent terms as a function of $r$ in Fig.~\ref{fig:QFI}. As the ratio between the baseline and the aperture diameter increases, the relative information contained in the structure of the individual apertures decreases compared with that obtained by the long-baseline aperture configuration.

If we have more than two apertures in a symmetric arrangement, then we have many such pairs with their contributions adding nicely in the same way.
If $n$ is odd where we also have an aperture at $k=0$ in addition to a bunch of equidistant symmetric pairs, the middle aperture contributes zero to the baseline term. In this way, the two-point-source problem is a special case where the various contributions to the QFI separate very nicely, additively, into a piece totally associated with a single aperture and pairwise contributions from oppositely located pairs of equidistant apertures.

\subsection{Multi-axial receivers}

In this category, we will consider three measurements: 
\begin{enumerate}
\item Ideal direct imaging in the compound aperture's image plane.
\item A SPADE measurement. We will consider the Gram-Schmidt recipe for obtaining a basis independently proposed by Kerviche {\em et al.} and Rehacek {\em et al.}~\cite{KGA17}. This, according to the latter work, attains the QFI for the uniformly bright 2 point separation problem. We will also discuss two binary SPADEs associated with the Gram-Schmidt basis, which attain the QFI in the small separation limit.
\item The SLIVER, which separates the incoming light into its even and odd pieces, and was shown to overcome Rayleigh's curse for our 2-point problem~\cite{Nair2016}.
\end{enumerate}

\subsubsection{Direct imaging}

For the performance of ideal direct imaging, we need the probability function for an incoming photon to be detected at position $x$ on the imaging screen. This is given by
\be
P(x) = \frac{1}{2}
\left[|\psi_{\rm comp}(x-\theta)|^2
+|\psi(x+\theta)|^2\right],
\ee
where $\psi_{\rm comp}(x)$ is the compound aperture PSF associated with our multiple apertures. The CFI is then
\be
\sJ_{\rm DI} = \int_{-\infty}^\infty dx \frac{1}{P(x)} \left(\frac{\partial P(x)}{\partial\theta}\right)^2.
\ee
For the special case of 2 apertures at positions $\alpha_1=-\beta/2$ and $\alpha_2=\beta/2$ in the aperture plane, we use the compound aperture PSF (\ref{compound-psf-2-apertures}),
and plot the CFI for the special case of 2 apertures for three different values of $r$.
We also show the performance we would obtain if we only used direct imaging measurements at the two apertures locally without any interferometry.

\subsubsection{The Gram-Schmidt basis and associated binary SPADEs}
\label{Gram-Schmidt-section}

We now work with the basis proposed independently by Kerviche {\em et al.} and 
Rehacek {\em et al.} ~\cite{KGA17, Rehacek2017a}.

Focusing on aperture configurations that are symmetric about the center of the baseline region, we will have a purely real-valued and symmetric PSF. Therefore, we will derive our formulae for a real-valued and symmetric PSF $\psi(x)$.
We take the zeroth mode function of the Gram-Schmidt basis to be the PSF
\be
A_0(x) = \psi(x).
\ee
Moreover, since we are only considering aperture arrangements that are symmetric around the origin of the aperture plane, mode 1 will be proportional to the derivative. Applying the normalization condition, it is straightforward to see that mode 1 will be 
\be
a_1(x)
= \frac{1}{\sqrt{-\Gamma_{\rm PSF}^{\prime\prime}(0)}} \frac{\partial\psi(x)}{\partial x},
\label{a1-def}
\ee
where $\Gamma_{\rm PSF}^{\prime\prime}(0)$ is the second derivative of the autocorrelation function of the PSF defined in (\ref{Gamma-def}). For a derivation of (\ref{a1-def}), check Appendix \ref{normalization-first-Gram-Schmidt-mode}.
For higher order modes, we will get non-zero overlap functions with previous modes of the same parity, and we will therefore have to subtract the non-orthogonal parts by following the Gram-Schmidt orthogonalization procedure.

Rehacek {\em et al.} proved that such a Gram-Schmidt basis or for that matter, any other complete basis of functions with definite parity, attains the QFI for our 2-point separation problem for any symmetric aperture function. Therefore, the Gram-Schmidt basis will also be optimal for any symmetric compound aperture set ups.
However, since such a full SPADE measurement is difficult to implement, Kerviche {\em et al.} have proposed that in the small separation limit, we can also use a binary SPADE that only sifts the incoming photons between the zero or the first mode of the Gram-Schmidt basis and their orthogonal complements. They call these binary SPADEs the 0-BinSPADE and the 1-BinSPADE, respectively, and show that their CFIs reach a non-zero constant in the zero separation limit and hence bypasses Rayleigh's curse. It turns out that this constant equals the QFI, as we show in Appendix \ref{binary-SPADE-proof}.

Here, we obtain expressions for the CFI of these two binary SPADEs. 
This was done by Kerviche {\em et al.}, but we obtain slightly simpler expressions as we are only considering aperture arrangements that are symmetric about the origin.
The probability for finding an incoming photon to be in mode $0$ will be 
\be
P_0 = \Gamma_{\rm PSF}^2(\theta).
\label{P_i-general-perfectly-centered-repeat}
\ee
The probability for the orthogonal complement will be
\be
P_{0r}
= 1-P_0.
\ee
The Fisher information for the 0-BinSPADE (when $\psi(x)$ is real) is given (after a bit of simplification) by
\begin{align}
\sJ_{\rm 0-Bin}(\theta)
&= N \frac{1}{P_0}\left(\frac{\partial P_0}{\partial\theta}\right)^2 + \frac{1}{P_{0r}}\left(\frac{\partial P_{0r}}{\partial\theta}\right)^2 \nn \\
&= \frac{N}{P_0(1-P_0)}\left(\frac{\partial P_0}{\partial\theta}\right)^2 \nn \\
&= 4N \frac{
\left[\Gamma_{\rm PSF}^\prime(\theta)\right]^2}
{1-\Gamma_{\rm PSF}^2(\theta)},
\label{CFI-0-BinSPADE}
\end{align}
where as previously defined, $\Gamma_{\rm PSF}^\prime(\theta)$ is the derivative of $\Gamma_{\rm PSF}(\theta)$,
and $\Gamma_{\rm PSF}^2(\theta) = \left[\Gamma_{\rm PSF}(\theta)\right]^2$.

For the 1-BinSPADE, the probability for finding an incoming photon to be in mode $A_1$ will be
\be
P_1 = \left[ \int_{-\infty}^\infty dx A_0(x-\theta) \, A_1(x)\right]^2
= -\frac{\left[\Gamma_{\rm PSF}^\prime(\theta)\right]^2}
{\Gamma_{\rm PSF}^{\prime\prime}(0)},
\ee
and the probability for an incoming photon to be in the orthogonal complement will be
\be
P_{1r}
= 1-P_1.
\ee
The Fisher information for the 1-BinSPADE is then given by (again, this is for real $\psi(x)$)
\begin{align}
\sJ_{\rm 1-Bin}(\theta)
&= N \frac{1}{P_1}\left(\frac{\partial P_1}{\partial\theta}\right)^2 + \frac{1}{P_{1r}}\left(\frac{\partial P_{1r}}{\partial\theta}\right)^2 \nn \\
&= \frac{N}{P_1(1-P_1)}\left(\frac{\partial P_1}{\partial\theta}\right)^2 \nn \\
&= -\frac{4N \left[\Gamma_{\rm PSF}^{\prime\prime}(\theta)\right]^2}
{\Gamma_{\rm PSF}^{\prime\prime}(0) +\left[\Gamma_{\rm PSF}^\prime(\theta)\right]^2}.
\label{CFI-1-BinSPADE}
\end{align}
The proof that (\ref{CFI-0-BinSPADE}) and (\ref{CFI-1-BinSPADE}) approach the QFI as $\theta\to 0$ is given in appendix \ref{binary-SPADE-proof}.

These are the general expressions for any symmetric aperture. For our multiple aperture set up, we would use the autocorrelation functions $\Gamma_{\rm PSF, comp}(\theta)$ of the combined aperture from (\ref{autocorrelation-function-compound-PSF}) in place of $\Gamma_{\rm PSF}(\theta)$. For two hard apertures at positions $\alpha_1=-\beta/2$ and $\alpha_2=\beta/2$, we can calculate the autocorrelation function either by brute force from the compound aperture PSF (\ref{compound-psf-2-apertures}) or from (\ref{autocorrelation-function-compound-PSF}) and (\ref{correlation-function-sinc}). We obtain
\begin{align}
\Gamma_{\rm PSF, comp} (\theta)
&= \int_{-\infty}^\infty dx\,\psi_{\rm 2ap}^*(x) \, \psi_{\rm 2ap}(x-\theta) \nn \\
&= \sigma \frac{\cos(\pi r \theta/\sigma) \sin(\pi \theta/\sigma)}{\pi \theta}.
\label{autocorrelation-function-2-apertures}
\end{align}
Applying (\ref{CFI-0-BinSPADE}) and (\ref{CFI-1-BinSPADE}), we obtain the CFI for the two binary SPADEs. The expressions for these are rather tedious and not very instructive, so we will not write them down explicitly. However, we plot them in Fig.~\ref{fig:MultiAxial}.

\subsubsection{The SLIVER}
\label{SLIVER-section-multi-axial-2-pt-problem}

As explained in Section \ref{SLIVER-introduction-section}, the SLIVER sorts the incoming light into its even and odd parts~\cite{Nair2016}.

It is a simple exercise to calculate the CFI for our 2-point separation problem using a SLIVER. While this was done in~\cite{Nair2016} for the 2-dimensional case, we re-derive the same expression for the 1-dimensional problem using our conventions and the language of correlation functions.

The (image plane) amplitudes associated with the two point sources are $\psi(x\pm\theta)$. If we split it into two parts, and spatially invert one of them, then it means taking $x\to -x$, which gives
\be
\psi_{\rm comp}(x\pm\theta) \to \psi_{\rm comp}(-x\pm\theta) = \psi_{\rm comp}(x\mp \theta),
\ee
where in the last step, we have assumed $\psi_{\rm comp}(x) = \psi_{\rm comp}(-x)$ since we are limiting ourselves to symmetric aperture configurations. Now, when we recombine the beams through the second 50-50 beam splitter, then the normalized outputs will be
\begin{eqnarray}
	\psi_{\rm E, comp}(x, \pm \theta)
	&=& \frac{1}{2} \left[\psi_{\rm comp}(x\pm\theta) +\psi_{\rm comp}(x\mp \theta)\right]\nn\\
	&=& \frac{1}{2} \left[\psi_{\rm comp}(x+\theta) +\psi_{\rm comp}(x- \theta)\right], \,{\text{and}}	
	\nn \\	
	\psi_{\rm O, comp}(x, \pm \theta)	
	&=& \frac{1}{2} \left[\psi_{\rm comp}(x\pm\theta) -\psi_{\rm comp}(x\mp \theta)\right]	\nn\\
	&=& \pm \frac{1}{2} \left[\psi_{\rm comp}(x+\theta) -\psi_{\rm comp}(x- \theta)\right],	
\end{eqnarray}
for the light emanating from the two point sources.

The probability for finding an incoming photon to be in these even or odd modes will be the (brightness-weighted) sum of the contributions from the two points
\begin{align}	
	P_{\rm E} &= \frac{1}{2} \int_{-\infty}^\infty dx \, \psi_{\rm E, comp}^2(x, \theta) + \psi_{\rm E, comp}^2(x, -\theta)	\nn \\
	&= \frac{1+\Gamma_{\rm PSF, comp}(2\theta)}{2} \\	
	P_{\rm O} &= \frac{1}{2} \int_{-\infty}^\infty dx \, \psi_{\rm O, comp}^2(x, \theta) + \psi_{\rm O, comp}^2(x, -\theta) \nn \\	
	&= \frac{1-\Gamma_{\rm PSF, comp}(2\theta)}{2}	
\end{align}
where we have simply expanded both $\psi_{\rm E, comp}(x, \pm\theta)$ and $\psi_{\rm O, comp}(x, \pm\theta)$ in terms of $\psi_{\rm comp}(x\pm\theta)$ and written out the integrals in terms of the autocorrelation function of $\psi_{\rm comp}(x)$ defined in (\ref{autocorrelation-function-compound-PSF}). The unity terms arise from the inner product of the same function with itself.

The CFI is then
\begin{align}	
	\sJ_{\rm Sliv}(\theta)	
	&= \frac{1}{P_{\rm E}} \left(\frac{\partial P_{\rm E}}{\partial\theta}\right)^2	
	+\frac{1}{P_{\rm O}} \left(\frac{\partial P_{\rm O}}{\partial\theta}\right)^2 
	\nn \\	
	&= \frac{4N\left[\Gamma_{\rm PSF, comp}^\prime(2\theta)\right]^2}	
	{1-\Gamma_{\rm PSF, comp}^2(2\theta)}
\label{SLIVER-two-point-two-aperture-CFI}
\end{align}
This is equivalent to the result in~\cite{Nair2016}, and it is straightforward to see from L'hopital's rule that in the $\theta\to 0$ limit, it approaches the QFI given in (\ref{QFI-autocorrelation}).
We can calculate the above CFI for the SLIVER explicitly by recalling (\ref{autocorrelation-function-2-apertures}).

\begin{figure*}[ht]
	\centering
	\includegraphics[width=2\columnwidth]{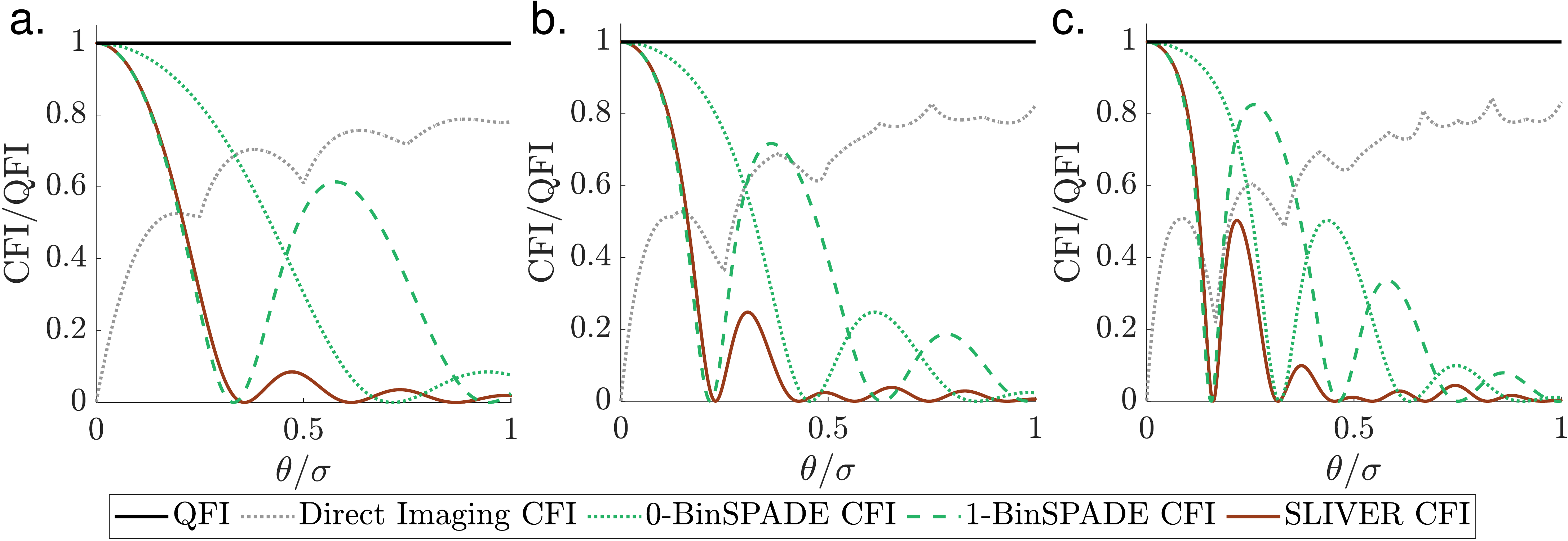}
	\caption{CFI of multi-axial receivers normalized to QFI for a. $r=1$, b. $r=2$, and c. $r=3$.}
	\label{fig:MultiAxial}
\end{figure*}

In Fig.~\ref{fig:MultiAxial} we plot the relative CFIs for our multi-axial receivers for three values of $r$, corresponding to three different baselines, for point source separation estimation. We focus on sub-Rayleigh separations $\theta<\sigma$, which corresponds to the regime where the canonical Rayleigh criterion is violated for the single-aperture case. We see that while direct imaging achieves zero information in the limit $\theta\to0$, the other three receivers we analyze all converge to the QFI in this limit, supporting their utility for sub-diffraction imaging with multiple apertures. Of the three multi-axial receivers, the 0-BinSPADE has the best performance for highly sub-Rayleigh sources, although all of the receivers exhibit oscillatory behavior in their CFIs such that no one receiver is superior for all values of $\theta<\sigma$.

\subsection{Co-axial receivers}

\subsubsection{Receiver employing mode sorting}

For the two-point source problem, as we show below, the groupwise receiver (dubbed `pairwise receiver' for this special case) achieves the QFI. Therefore, in this subsection, we will only consider the pairwise receiver.

Sticking to the case of symmetric aperture arrangements, we can attain the QFI in a number of ways, the simplest of which is to build the pairwise receiver described in Section \ref{groupwise-receiver-section}.
Let us say that we have a total of $n= 2n_{\rm pairs}$ apertures at locations $\alpha_{2i-1}=-\beta_i/2$ and $\alpha_{2i}=\beta_i/2$ with $i=1\ldots n_{\rm pairs}$.
At each aperture, we collect mode $j$, couple it into a single mode fiber to transport the signals to a central location, and combine them in a 50-50 beam splitter which gives the symmetric and anti-symmetric combinations of the $j$th modes from each equidistant aperture pair, for which we obtain the photon count.
Let us denote the photon probabilities for the $j$th mode from the aperture pair at $\pm \beta_i$ as $P_{j i\pm}$.
Just as in Section \ref{two-point-separation-one-aperture-section},
we assume that our SPADEs are pointed perfectly at the centroid so that the two positions are $x_1 = \theta$ and $x_2 =-\theta$, and the brightnesses are both $b_1 = b_2 =\frac{1}{2}$.
Recalling the discussion of the groupwise and pairwise receivers from Section \ref{groupwise-receiver-section}, we obtain the photon probabilities
\begin{align} 
P_{j i+} &= \frac{1}{n_{\rm pairs}} \cos^2(\beta_i \theta/2) \,\Gamma_j ^2(\theta) \nn \\
P_{j i-} &= \frac{1}{n_{\rm pairs}} \sin^2(\beta_i \theta/2) \,\Gamma_j ^2(\theta)
\end{align}
Here we have used equations (\ref{P-gamma-j-def}) and (\ref{B-symmetric-and-antisymmetric}) tailored for the two-point problem.
The total CFI for all the aperture combinations $i\pm$ for local mode $j$ is
\begin{align}
\sJ_{j,\theta}
&= N \sum_{i=1}^{n_{\rm pairs}} \frac{1}{P_{j i+}} \left(\frac{\partial P_{j i+}}{\partial\theta}\right)^2 
+\frac{1}{P_{j i-}} \left(\frac{\partial P_{j i-}}{\partial\theta}\right)^2 \nn \\
&= 4N \left(\Gamma_j ^\prime(\theta)\right)^2
\,+\,\frac{N}{n_{\rm pairs}} \sum_{i=1}^{n_{\rm pairs}} \beta_i ^2 \,\Gamma_j ^2(\theta)
 \end{align}

Transforming indices in the second term, this translates to
\be
\sJ_{j,\theta}
= 4N \left(\Gamma_j ^\prime(\theta)\right)^2
\,+\,\frac{4N}{n} \sum_{\mu=1}^n \alpha_\mu ^2 \,\Gamma_j ^2(\theta) 
\label{CFI-separation-j-mode}
\ee
We have derived this for an even number of apertures. It turns out that this equation is equally valid if we also add another one at the origin of the aperture plane, so that we have $n= 2n_{\rm pairs}+1$ apertures located at $0, \pm\beta_1/2, \ldots\pm\beta_{n_{\rm pairs}}/2$.
In that case, we simply measure each local mode $j$ from the central telescope without interfering it with the signals from the other apertures. The photon probabilities are then 
 \begin{align} 
P_{j i+} &= \frac{1}{2n_{\rm pairs} +1} 2\cos^2(\beta_i \theta/2) \,\Gamma_j ^2(\theta) \nn \\
P_{j i-} &= \frac{1}{2n_{\rm pairs}+1} 2\sin^2(\beta_i \theta/2) \,\Gamma_j ^2(\theta) \nn \\
P_{j 0} &= \Gamma_j ^2(\theta)
\end{align}
It is a straightforward exercise to add all the CFI contributions from these and see that the sum gives (\ref{CFI-separation-j-mode}).

The first term in (\ref{CFI-separation-j-mode}) is equal to the CFI contribution from the $j$th mode for a single aperture, and the second term is associated with baseline interferometry.
 This means that the total CFI for our groupwise receiver is equal to the QFI given in \eqref{QFI-2-contributions}, as we will argue shortly:
\begin{align}
\sJ_{\rm full, \theta} &= \sum_j \sJ_{\rm j, \theta} \nn \\
&= \sK_{\rm 1ap, \theta}
+\sK_{\textrm{lb}, \theta} \nn \\
&= \sK_{\textrm{full}, \theta}
\end{align}
where we have recalled (\ref{single-aperture-QFI-2-points}) to identify the first term as the single aperture QFI, and used the normalization property to take $\sum_{j=0}^\infty \Gamma_j ^2(\theta) = 1$ for the second term. Specifically, $\Gamma_j ^2(\theta)$ is the probability for a photon coming into a single aperture to be in the $j$th mode, and adding over $j$ gives unity.
So like the QFI, the CFI also has two parts, one equal to the single-aperture CFI, and the other associated with long-baseline interferometry which is independent of the shape and features of the individual apertures (as long as all the apertures are of the same size and shape).

It is worth mentioning that the pairwise combinations are not the only way to attain the QFI. From the invariance result in section \ref{invariance-section},
any other linear combinations of the pairwise modes involving purely real-valued coefficients will give the same CFI and hence also achieve the QFI.

\begin{figure*}[ht]
	\centering
	\includegraphics[width=2\columnwidth]{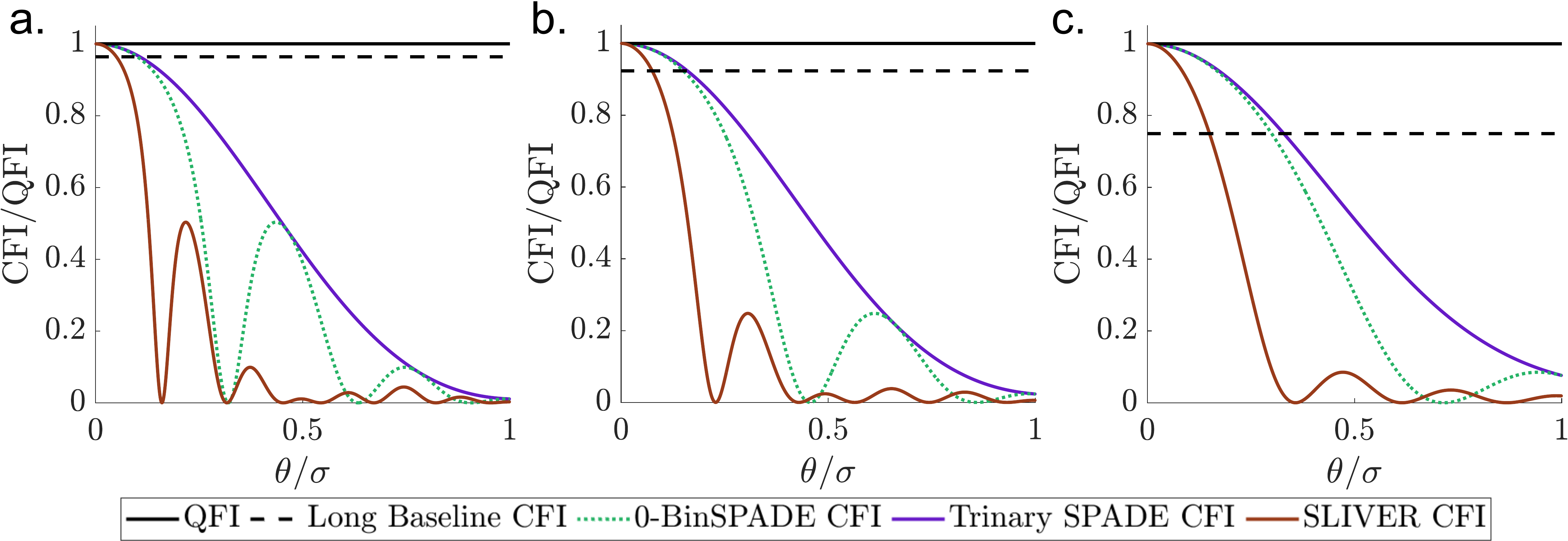}
	\caption{CFI of co-axial receivers normalized to QFI for a. $r=1$, b. $r=2$, and c. $r=3$.}
	\label{fig:CoAxial}
\end{figure*}

As mentioned earlier, in a realistic application, we will not be able to measure all the infinite modes individually. Let us say that we are only collecting up to mode number $j_{\rm max}$ from the different apertures into single mode fibers for combination in a beam splitter circuit and subsequent measurement. We can still count the total number of photons that are in the orthogonal complement (i.e. in local mode $j_{\rm max}+1$ or higher of the individual apertures) by placing a bucket detector at each location. The sum of the photon counts from these bucket detectors will give us the total photon count for the orthogonal complement.
It is easy to see that the probability for an incoming photon to be found in the global orthogonal complement (i.e., in any one of the bucket detectors at each aperture site) will be 
\begin{align}
P_{r, j_{\rm max}} &= \sum_{\mu=1}^n \frac{1}{n} \sum_{j=j_{\rm max}+1}^\infty |\exp(\pm i\alpha_\mu \theta)\,\Gamma_j(\theta)|^2 \nn \\
&= \sum_{j=j_{\rm max}+1}^\infty \Gamma_j ^2(\theta) 
\end{align}
We can calculate the CFI contribution of this orthogonal complement from the usual formula $\sJ_{r, j_{\rm max},\theta} = \frac{N}{P_{r, j_{\rm max}}} \left(\frac{\partial P_{r, j_{\rm max}}}{\partial\theta}\right)^2$.
The total CFI for such a truncated groupwise receiver will be
\begin{align}
\sJ_{j_{\rm max}, \theta} = &\sJ_{r, j_{\rm max}, \theta} + \sum_{j=0}^{j_{\rm max}} \sJ_{\rm j, \theta} \nn \\
=& N \sum_{j=0}^{j_{\rm max}} 4\left(\Gamma_j ^\prime(\theta)\right)^2\nn\\
&
+\frac{4N}{1-\sum_{j=0}^{j_{\rm max}} \Gamma_j ^2(\theta)} \left(\sum_{j=0}^{j_{\rm max}} \Gamma_j(\theta) \Gamma_j ^\prime(\theta)\right)^2\nn\\
&+\sK_{\textrm{lb}, \theta} \sum_{j=0}^{j_{\rm max}} \Gamma_j ^2(\theta)
\label{truncated-SPADE-CFI-2-points}
\end{align}
Here the first two terms are equal to the CFI for a truncated SPADE with just a single aperture, and the last term involves the contribution from baseline interferometry.
As $j_{\rm max}$ gets larger, $\sum \Gamma_j ^2(\theta)$ becomes progressively closer to 1, and the long baseline term gets closer to the QFI term associated with baseline interferometry.
In the same way, the first part (i.e. the single-aperture truncated SPADE) also gets closer to the single aperture QFI as we increase $j_{\rm max}$.

\begin{figure*}
	\centering
	\includegraphics[width=0.8\textwidth]{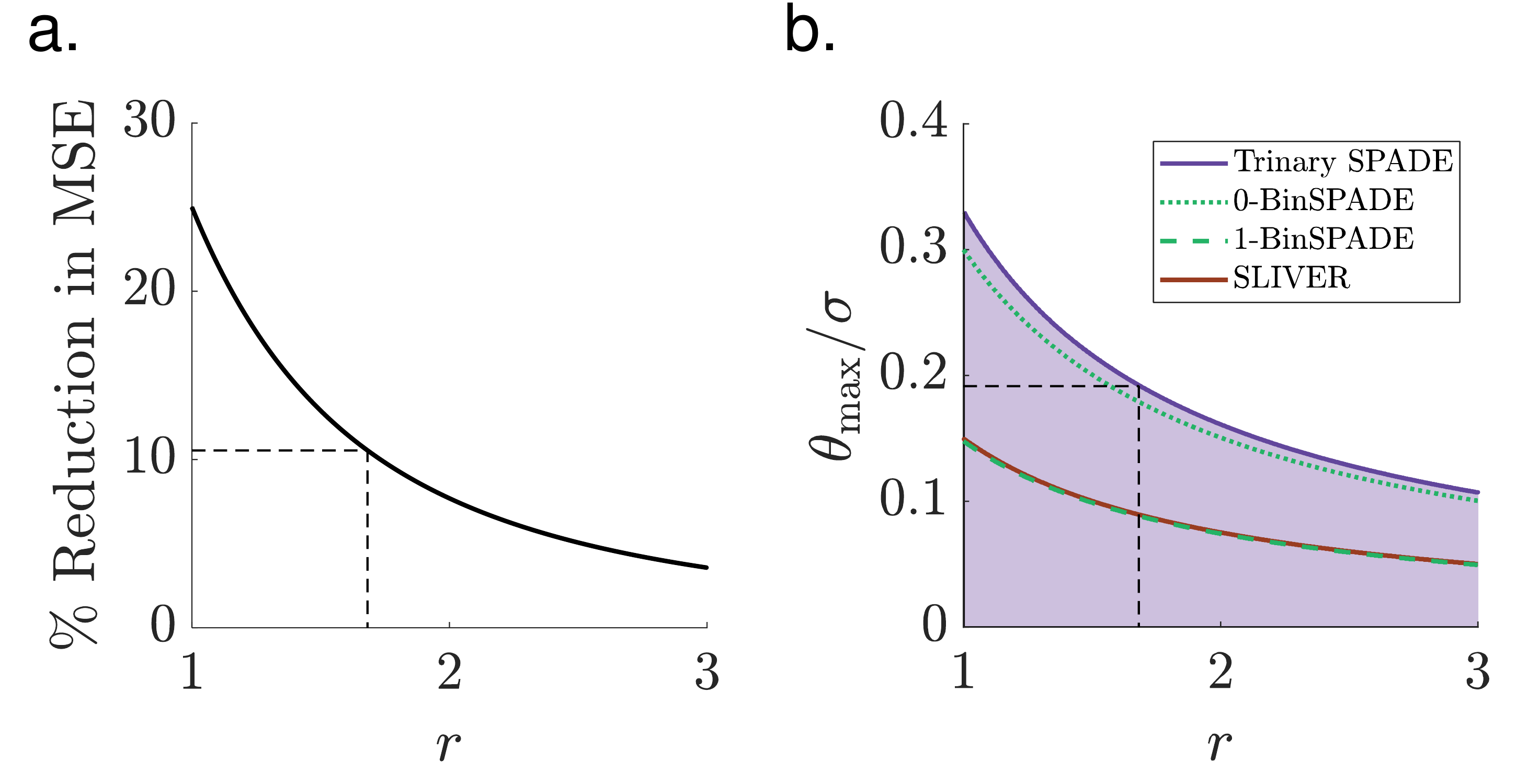}
	\caption{a. Percentage of reduction in MSE obtained by saturating the QCRB for the full QFI $\mathcal{K}_{\rm full,2ap,\theta}$ as opposed to the long-baseline  QFI $\mathcal{K}_{\rm lb,\theta}$ (computed as $100\%\times(\mathcal{K}_{\rm lb,\theta}^{-1}-\mathcal{K}_{\rm full,2ap,\theta}^{-1})/(\mathcal{K}_{\rm lb,\theta}^{-1})$). b. Maximum value of $\theta$ for which each of our receivers achieves a CFI greater than $\mathcal{K}_{\rm lb,\theta}$ for a given baseline-to-diameter ratio $r$ (e.g., Trinary SPADE outperforms the long-baseline QFI for any combination of parameters in the purple shaded region). In both a. and b. the dashed lines correspond to the LBT with $r=1.71$.}
	\label{fig:LBcomparison}
\end{figure*}

The $j_{\rm max} = 0$ case corresponds to a receiver where we simply take a binary mode sorter at each telescope cite that couples the zero mode into single mode fibers for combination at a central location, with the rest going into a bucket detector.
For our special example of two apertures at positions $\pm\beta/2$ with our pairwise scheme, and taking the local modes to be the sinc-bessel functions, this means a trinary SPADE sorting the incoming photons between the following three modes or subspaces:
\begin{enumerate}
\item The symmetric combination of the zero mode from the two apertures, corresponding to the compound aperture PSF mode $\sqrt{2} \cos(\pi r x/\sigma) \psi_0(x)$.
For hard apertures, this was given in (\ref{compound-psf-2-apertures}).
\item The anti-symmetric combination of the zero mode from the two apertures, corresponding to the state $\sqrt{2} \sin(\pi r x/\sigma) \psi_0(x)$. 
For hard apertures, this was given in (\ref{compound-psf-2-apertures}). That is, the same as the PSF mode but with $\cos(\pi r x/\sigma)$ replaced by a sine.
\item The orthogonal complement of the above two functions.
\end{enumerate}
This trinary SPADE is therefore an improvement over the 0-BinSPADE discussed in Section \ref{Gram-Schmidt-section} since it also has the anti-symmetric combination of the fundamental mode. Putting together the various pieces such as (\ref{correlation-function-sinc}) with $\Gamma_0(a) \equiv \Gamma_{\rm PSF,1ap}(a)$ and (\ref{truncated-SPADE-CFI-2-points}), we obtain the CFI for this trinary SPADE
\begin{align}
\sJ_{\rm tri, \theta}
= &4N \frac{\left(\Gamma_0 ^\prime(\theta)\right)^2}
{1-\Gamma_0 ^2(\theta)}
+\frac{4\pi^2 N r^2}{\sigma^2} \Gamma_0 ^2(\theta) \nn \\
= &4N\frac{\left(\frac{\cos(\pi \theta/\sigma)}{\theta}
-\frac{\sigma\sin(\pi \theta/\sigma)}{\pi \theta^2}\right)^2}
{1-\sigma^2 \frac{\sin^2(\pi \theta/\sigma)}{\pi^2 \theta^2}}\nn\\
&+\frac{4\pi^2 N r^2}{\sigma^2} 
\sigma^2 \frac{\sin^2(\pi \theta/\sigma)}{\pi^2 \theta^2}
\end{align}
In Fig.~\ref{fig:CoAxial}, we plot the CFI of this trinary SPADE for point source separation estimation in the sub-Rayleigh regime and observe that it approaches the QFI in the limit $\theta\to0$. Comparing the trinary SPADE against a multi-axial 0-BinSPADE, we see the advantage obtained by collecting the anti-symmetric combination between the two apertures: the oscillatory 0-BinSPADE CFI is enveloped by that of the trinary SPADE, with the latter always obtaining a CFI at least as large as the former.

\subsubsection{Receiver without mode sorting: using light pipes}

Based on Section \ref{light-pipe-receiver-section}, if instead of single mode fibers carrying individual local spatial modes, we bring all of the light from each aperture to a central location through multimode light pipes and let it interfere pairwise through one bulk 50-50 beam splitter, then we simply attain the long baseline piece of the QFI for the two-point separation problem, $\sK_{\textrm{lb}, \theta}$ given in (\ref{QFI-lb}), i.e.,
\begin{eqnarray}
\sK_{\textrm{lb,2ap,} \theta} &=& \frac{4\pi^2 N}{3\sigma^2}(3r^2)\\
&=& {4\pi^2 r^2 N}/{\sigma^2}.
\end{eqnarray}
In this scenario, the information $\sK_{\rm 1ap, \theta}$ contained in the structure of the individual apertures is lost. 

\subsection{SLIVER: Lightpipe receiver with a reflection}

Recall from section \ref{SLIVER-lightpipe-reflection-section} that if we add a reflection to one of the input ports of a fifty-fifty beam splitter in a light pipe receiver, then we obtain the so-called SLIVER. For the two-point problem, the performance of the SLIVER will be given by (\ref{SLIVER-two-point-two-aperture-CFI}), which we obtained in Section~\ref{SLIVER-section-multi-axial-2-pt-problem} while discussing multi-axial beam combination.
Recall that in the $\theta \to 0$ limit, this approaches the full QFI, rather than just the long baseline contribution achieved by the lightpipe receiver without a reflection.

In Fig.~\ref{fig:CoAxial}, we show the optimality of this co-axially implemented SLIVER receiver as $\theta\to0$, although the performance at larger angular separations is worse than the trinary SPADE.

\subsection{Discussion}
\label{sec:Discussion}
Our receiver designs in this work can be thought of as attempts to capture not only the quantum information arising from the presence of distant apertures but also the information contained in the spatial structure of the individual apertures. In Fig.~\ref{fig:LBcomparison}a. we plot for $1\leq r \leq 3$ the percent reduction in the MSE of estimating the 2-point source separation, according to the quantum Cr\'amer-Rao bound $\textrm{MSE}\geq (N\mathcal{K})^{-1}$, that could be obtained by capturing this additional information. This plot directly relates to Fig.~\ref{fig:QFI} and shows that a maximum of 25\% of the information about $\theta$ is contained in the single-aperture term of (\ref{QFI-2-apertures}) when $r=1$ (i.e., when the apertures are touching each other) and this relative contribution decreases quadratically as the baseline distance increases. For reference, the LBT has mirror diameters of $d = 8.408\textrm{ m}$ and a center-to-center baseline distance of $b = 14.4\textrm{ m}$~\cite{LBT}, giving $r\equiv b/d = 1.71$. This corresponds to a relative reduction of $10.23$\% in the quantum-limited MSE if the information contained within the individual apertures was fully captured along with the baseline interferometry information. We conclude that more carefully designing an imaging receiver to capture both of the two components of the quantum Fisher information represented in (\ref{QFI-2-contributions}) can provide a modest but notable boost to the ultimate limits for sub-diffraction imaging in a real-world, two-aperture telescope system used for extrasolar astronomical observation. In the case of the LBT, which already uses multi-axial beam combination in a Fizeau interferometry configuration as its primary mode of operation, the actual improvement over current methods (which are not optimized for the quantum limit of estimating the distance between two equal-brightness stars) can actually be much larger than 10\%. 

In Fig.~\ref{fig:LBcomparison}b. we ask the natural next question: can we design receivers to successfully access this additional single-aperture information and beat the quantum limit of standard long-baseline interferometry? We plot the maximum point source separation at which each of the receiver designs we analyzed outperforms the simple long-baseline receiver that retains no spatial information beyond the phase delay between distant point-like apertures. This maximum value of $\theta$ corresponds to the intersections with the horizontal dashed lines in Fig.~\ref{fig:CoAxial}. Equivalently, Fig.~\ref{fig:LBcomparison}b. can be interpreted as a parametrized region of improvement over long-baseline interferometry: any point down and to the left of the curve for a given receiver (e.g., the purple shaded region for the Trinary SPADE) indicates a pair of values for $\theta$ and $r$ for which that receiver's CFI is greater than the QFI of long-baseline interferometry. We see that the information in the individual aperture structure will have the largest effect over the largest range of object geometries when the baseline is moderate relative to the diameter of the individual apertures. For example, for the LBT, the standard astronomical diffraction limit at a wavelength of $\lambda=1.65$ $\mu$m is $\lambda/d=40.5$ milliarcseconds, so $\sigma = 2\pi \lambda / d=254.3$ milliarcseconds. We find that the distance between two point sources with angular separation smaller than $\theta_{\rm max}<0.195\sigma=49.6$ milliarcseconds (which includes all unresolved two-point-source objects) will be estimable to a higher precision with a Trinary SPADE measurement that leverages the spatial structure of the apertures compared against a baseline-interferometric measurement that does not take the spatial structure of the apertures into account.

Lastly, if we are operating in the large distance regime where we cannot use light pipes or single mode fibers and must use an entanglement based approach, then we would be combining the signals from distant locations modewise, and this will mean the truncated groupwise receiver discussed above. 

\section{Conclusion}

Recent and ongoing developments in quantum information theory hold interesting possibilities for the field of baseline interferometry. In this paper, we have given a description of several approaches for receiver designs for achieving quantum-optimal precisions for well-defined quantitative imaging tasks, that are appropriate for different distance regimes. We have used the example of two-point-source separation estimation for two uniformly bright incoherent sources as a toy problem to illustrate these ideas. In the process, we have also developed a mathematical framework for analyzing multiple aperture systems in terms of a combined aperture function or its Fourier transform which can be thought of as the compound PSF if we had a (possibly hypothetical) giant imaging screen spanning the entire baseline. 

This framework allows to extend the analysis presented herein to imaging more complex scenes---at their respective quantum-limited precision---made up of extended objects or constellations of point emitters. For example, for localizing multiple stars within a tight sub-Rayleigh angular field of view, one would apply the algorithm from Ref.~\cite{Lee2022} to the multi-aperture system's compound-aperture PSF, and translate the mathematical description of the adaptive pre-detection mode sorting to either a co-axial, multi-axial (or even an entanglement-assisted) receiver, by `compiling' the spatial mode demultiplexer (SPADE) to be applied to the entirety of the multi-spatial-mode light collected by all the apertures, using the tools from Ref.~\cite{Clements2016} into a collection of pairwise two-mode Mach-Zehnder Interferometers (MZIs).

Our first distance regime is when the different telescopes' apertures are close enough to each other that we can reflect the light on to a common imaging screen. This way, instead of building a large, prohibitively expensive parabolic mirror spanning the entire baseline, we only need to have carved out segments of such a mirror in the locations of our different apertures. Once the light is reflected on to a common imaging screen, we can carry out ideal direct imaging, or implement some SPADE basis measurement. For the specific problem of 2-point separation estimation, the Gram-Schmidt (GS) basis independently introduced by Kerviche {\em et al.} and Rehacek {\em et al.} attains the QFI. In the small separation limit, a binary SPADE separating the zeroth (compound PSF) or the first GS mode from the rest of the modal span also approaches the QFI. Moreover, another simple receiver that realizes the QFI in the small separation limit is the so-called SLIVER (image-inversion interferometry) introduced by Nair {\em et al.}.

Our intermediate regime is when we can bring the light collected at different locations to a central location through single mode fibers or light pipes, feed them into a linear interferometer, and measure the outputs with photon counters. For this case, a complete basis of orthogonal modes associated with all the individual apertures together can serve as a full basis for the whole set of functions involving the multiple aperture system. Any general SPADE measurement can then be implemented in terms of these local modes by choosing the appropriate combination coefficients to mix the different mode signals coming from the various apertures. 

For our equally bright 2-point separation problem, we have shown that a pairwise combination scheme, in which we only mix the same modal signals from oppositely-located equidistant pairs through a 50-50 beam splitter and obtain the photon counts for in the output ports, attains the QFI. We have also shown that the QFI for the 2-point problem for a physically symmetric aperture configuration can be expressed as a sum of the QFI for a single aperture and a second term associated with the long-baseline effect involved in combining light from distant receiver. Moreover, the latter term is proportional to the sum of the squares of the positions of all the individual apertures.

If instead of single mode fibers, we bring the light to a central location through multimode light pipes and do not use any local mode sorters, then the pairwise scheme, or for that matter any other equally optimal one, is shown to attain the second QFI term, that is, the one associated with the long-baseline effect. With a small tweak in the form of the introduction of a reflection, the light pipe based receiver can also be effectively turned into a SLIVER for a multiple aperture system.

Lastly, we have discussed the large distance regime where bringing the light to a common location through single mode fibers or light pipes is practically impossible or prohibitively expensive. In this case, we can use a protocol based on quantum entanglement to transfer the signal received at one location to another one. By doing so, the signals that need to be combined in an interferometer can be brought to a common location. This way, we can in principle use an entanglement based protocol in place of single mode fibers to carry out SPADE measurements. Our pairwise scheme seems tailor-made for such an approach.

\section{Acknowledgements}

We are truly grateful to Amit Ashok, Kwan Kit Lee, Michael Hart, Mark Neifeld and Ryan Luder for very helpful conversations and input. This research was supported by the DARPA IAMBIC Program under Contract No. HR00112090128. The views, opinions and/or findings expressed are those of the authors and should not be interpreted as representing the official views or policies of the Department of Defense or the U.S. Government.

\appendix

\section{The QFI for 2-point separation estimation}
\label{2pt-QFI-calculation-apendix}

This calculation was originally done by Tsang {\em et al.} in~\cite{Tsang2016b}. Here we summarize their method and closely follow their treatment and notation, but with the difference that we are calculating the QFI for twice the separation instead of the separation. Specifically, we are taking our two points to be at positions $x_1 = \theta$ and $x_2 =-\theta$ and are calculating the QFI for estimating $\theta$. For this reason, our QFI result will be 4 times the separation estimation QFI found by {\em et al.} as they take the points to be $x_1 =\theta_1 +\theta_2/2$ and $x_2 = \theta_1 -\theta_2/2$.

From (\ref{rho1_definition}), the single-photon part of the density matrix for the equally bright 2 point problem is
\be
\rho_1 = \frac{1}{2}
\left({|}\psi_1\ra\la\psi_1{|}
+ {|}\psi_2\ra\la\psi_2{|}\right)
\label{rho-2pt-def}
\ee
Here we have written the contributions as $|\psi_s\ra\la\psi_s|$ instead of $|\psi_s, 1ap\ra\la\psi_s, 1ap|$ because here we are calculating the general expression that does not depend on the number of apertures. The only constraint we will consider is that our aperture is symmetric, giving an real-valued and even PSF function $\psi(x)$, giving the ket $|\psi_s\ra = \int_{-\infty}^\infty dx \psi(x-x_s)|x\ra$.

For the QFI, we will need this density matrix as well as its symmetric logarithmic derivative (SLD).
For this, we first need to find an northonormal eigenbasis spanning
the whole space spanned by ${|}\psi_1\ra$ and ${|}\psi_2\ra$ as well as their partial derivatives with respect to $\theta$.
First note that ${|}\psi_1\ra$ and ${|}\psi_2\ra$
are not mutually orthogonal, and 
\begin{eqnarray}
\delta &\equiv& \la\psi_1{|}\psi_2\ra \nn \\
&=& \int_{-\infty}^\infty \psi^*(x-x_1) \psi(x-x_2) \nn \\
&=& \int_{-\infty}^\infty \psi^*(x-\theta) \psi(x+\theta) \nn \\
&=& \int_{-\infty}^\infty \psi^*(x-2\theta) \psi(x) \nn \\
&=&\la\psi_2{|}\psi_1\ra \nn \\
&\neq& 0
\label{delta-definition}
\end{eqnarray}
for a real valued $\psi(x)$.

To express $\rho_1$ in a diagonal form in an orthonormal basis, we first write down a $2\times 2$ matrix of the inner products $\la\psi_i{|}\rho_1{|}\psi_j\ra$
\be
\begin{pmatrix}
	1 & \delta \\
	\delta & 1
\end{pmatrix}
\label{dot-product-matrix}
\ee
The normalized eigenvectors of this matrix give us
an orthogonal set of functions, and the square roots of the eigenvalues give us the normalization factors.
We find that the eigenvalues are $1\pm \delta$,
with the eigenvectors $\frac{1}{\sqrt{2}}(1, \pm 1)$.
Therefore, our orthonormal basis of states spanning ${|}\psi_1\ra$ and ${|}\psi_2\ra$ is
\be
{|}e_1\ra = \frac{1}{\sqrt{2(1-\delta)}} ({|}\psi_1\ra -{|}\psi_2\ra)
\ee\be
{|}e_2\ra = \frac{1}{\sqrt{2(1+\delta)}} ({|}\psi_1\ra +{|}\psi_2\ra)
\ee
It is a straightforward exercise to see that these are also the eigenstates of our density operator $\rho_1$,
and that the eigenvalues $D_i$ of $\rho_1$
are the corresponding eigenvalues of the matrix of inner products (\ref{dot-product-matrix}) divided by 2:
\be
D_1 = \frac{1-\delta}{2}
\ee\be
D_2 = \frac{1+\delta}{2}
\ee
This division by 2 is simply the 1/2 factor in front of ${|}\psi_1\ra\la\psi_1{|} +{|}\psi_2\ra\la\psi_2{|}$.
Our density matrix is then
\be
\rho_1 = D_1{|}e_1\ra\la e_1{|} + D_2{|}e_2\ra\la e_2{|},
\ee
with the above definitions.

However, $\frac{\partial\rho_1}{\partial\theta}$
also contains the derivatives of ${|}\psi_1\ra$ and ${|}\psi_2\ra$.
Therefore, we need to extend our eigenbasis to span these states too. We therefore include the derivatives of ${|}\psi_i\ra$ and carry out a Gram-Schmidt orthogonalization. This gives us the additional states
\be
{|}e_3\ra = \frac{1}{c_3}
\Bk{\frac{\Delta k}{\sqrt{2}}\bk{{|}\psi_{11}\ra + {|}\psi_{22}\ra}
	- \frac{\gamma}{\sqrt{1-\delta}}{|} e_1\ra},
\ee\be
{|} e_4\ra = \frac{1}{c_4}
\Bk{\frac{\Delta k}{\sqrt{2}}\bk{{|} \psi_{11}\ra - {|} \psi_{22}\ra} +
	\frac{\gamma}{\sqrt{1+\delta}}{|} e_2\ra},
\ee
where
\be
\Delta k^2 = \int_{-\infty}^\infty dx \left[\frac{\partial \psi(x)}{\partial x}\right]^2
\ee
and
\be
\gamma = \int_{-\infty}^\infty dx \frac{\partial \psi(x)}{\partial x} \psi(x -\theta_2)
\ee
The other quantities defined here are
\be
{|}\psi_{11}\ra \equiv \frac{1}{\Delta k}\int_{-\infty}^\infty dx \,
\frac{\partial \psi(x-x_1)}{\partial x_1} {|} x\ra,
\ee\be
{|}\psi_{22}\ra \equiv \frac{1}{\Delta k}\int_{-\infty}^\infty dx \, \frac{\partial \psi(x-x_2)}{\partial x_2}{|} x\ra,
\ee\be
c_3 \equiv \bk{\Delta k^2 + b^2 - \frac{\gamma^2}{1-\delta}}^{1/2},
\ee\be
c_4 \equiv \bk{\Delta k^2 - b^2 - \frac{\gamma^2}{1+\delta}}^{1/2},
\ee\be
b^2 \equiv \int_{-\infty}^\infty dx\, \frac{\partial \psi(x-x_1)}{\partial x_1} \frac{\partial \psi(x-x_2)}{\partial x_2},
\ee
and $\delta$ was defined in (\ref{delta-definition}).

We now have a full four-dimensional eigenbasis for our vector space spanned by the density operator $\rho_1$ and its derivatives. It is now a simple exercise to use the formula (\ref{sld}) for the SLD and compute the QFI.

After a bit of algebra, we find that the non-zero entries of the SLD with respect to the parameter $\theta$ in the ${|}e_i\rangle$ ($i = 1\ldots 4$) basis are
\be
\sL_{2, 11}
= -\frac{2\gamma}{1-\delta}
\ee\be
\sL_{2, 13}
= -\frac{2c_3}{\sqrt{1-\delta}}
\ee\be
\sL_{2, 22}
= \frac{2\gamma}{1+\delta}
\ee\be
\sL_{2, 24}
= -\frac{2c_4}{\sqrt{1+\delta}}
\ee
From these, it is a straightforward exercise to calculate the QFI using (\ref{QFI-formula}).
We obtain 
\begin{equation}
	\begin{aligned}
		\sK_{\theta} &= 4N\Delta k^2 \nn \\
		&= 4N \, \int_{-\infty}^\infty dx \left|\frac{\partial\psi(x)}{\partial x}\right|^2
	\end{aligned}
\label{K22-2pt-result}
\end{equation}

\section{Normalization coefficient for mode 1 of the Gram-Schmidt basis}
\label{normalization-first-Gram-Schmidt-mode}

The normalization coefficient for mode 1 is the inverse square root of
\be
\int_{-\infty}^\infty
\, \left(\psi'(x)\right)^2 \, dx
\ee
If we integrate by parts and use the fact that $\psi(x)$ goes to zero at $\pm\infty$, then this is equal to
\be
=- \int_{-\infty}^\infty
\psi(x) \, \psi''(x) \, dx
\,= -\, \Gamma_{\rm PSF}^{\prime\prime}(0),
\ee
where $\Gamma_{\rm PSF}^{\prime\prime}(0)$ is the second derivative of $\Gamma_{\rm PSF}(0)$.
The normalized first mode is therefore given by
\be
a_1(x)
= \frac{1}{\sqrt{-\Gamma_{\rm PSF}^{\prime\prime}(0)}} \frac{\partial\psi(x)}{\partial x}
\label{a1-def-repeat}
\ee

\section{Proof that the 0- and 1-BinSPADEs attain the QFI}
\label{binary-SPADE-proof}

We will take the $\theta\to 0$ limit in (\ref{CFI-0-BinSPADE}).
For the numerator, note that $\Gamma_{\rm PSF}^\prime(\theta) \to 0$ since $\Gamma_{\rm PSF}(\theta)$ is maximum at $\theta =0$.
But the denominator also goes to zero since $\Gamma_{\rm PSF}(0) = 1$.
Therefore, we apply L'Hopital's rule.
For the numerator, we then get
\be
\lim_{\theta\to 0} 8N \Gamma_{\rm PSF}^\prime(\theta) \, \Gamma_{\rm PSF}^{\prime\prime}(\theta)
\ee
And for the denominator, we obtain
\begin{equation}
\lim_{\theta\to 0} -2\Gamma_{\rm PSF} (\theta) \, \Gamma_{\rm PSF}^\prime(\theta)
= \lim_{\theta\to 0} -2\Gamma_{\rm PSF}^\prime(\theta),
\end{equation}
where we have used the normalization property $\Gamma_{\text{PSF}}(0) = 1$. Dividing the numerator term by the denominator one, and simplifying, we get
\be
\lim_{\theta\to 0} I_{\text{0-BinSPADE}}(\theta)
= -4N \Gamma_{\rm PSF}^{\prime\prime}(0)
\ee,
which is equal to the QFI according to (\ref{QFI-autocorrelation}).

For the 1-BinSPADE, we have (\ref{CFI-1-BinSPADE}). In the numerator, taking the $\theta\to 0$ limit gives $\left[\Gamma_{\rm PSF}^{\prime\prime}(0)\right]^2$.
The denominator goes to $\Gamma_{\rm PSF}^{\prime\prime}(0)$
since $\Gamma_{\rm PSF}^\prime(\theta)$ approaches $0$ as $\theta\to 0$.
On simplifying, we get
\be
\lim_{\theta\to 0} I_{\text{1-BinSPADE}}(\theta)
= -4N \Gamma_{\rm PSF}^{\prime\prime}(0)
\ee
which is the QFI.

\bibliography{multiple-aperture_final_arxiv}

\begin{thebibliography}{54}%
\makeatletter
\providecommand \@ifxundefined [1]{%
 \@ifx{#1\undefined}
}%
\providecommand \@ifnum [1]{%
 \ifnum #1\expandafter \@firstoftwo
 \else \expandafter \@secondoftwo
 \fi
}%
\providecommand \@ifx [1]{%
 \ifx #1\expandafter \@firstoftwo
 \else \expandafter \@secondoftwo
 \fi
}%
\providecommand \natexlab [1]{#1}%
\providecommand \enquote  [1]{``#1''}%
\providecommand \bibnamefont  [1]{#1}%
\providecommand \bibfnamefont [1]{#1}%
\providecommand \citenamefont [1]{#1}%
\providecommand \href@noop [0]{\@secondoftwo}%
\providecommand \href [0]{\begingroup \@sanitize@url \@href}%
\providecommand \@href[1]{\@@startlink{#1}\@@href}%
\providecommand \@@href[1]{\endgroup#1\@@endlink}%
\providecommand \@sanitize@url [0]{\catcode `\\12\catcode `\$12\catcode
  `\&12\catcode `\#12\catcode `\^12\catcode `\_12\catcode `\%12\relax}%
\providecommand \@@startlink[1]{}%
\providecommand \@@endlink[0]{}%
\providecommand \url  [0]{\begingroup\@sanitize@url \@url }%
\providecommand \@url [1]{\endgroup\@href {#1}{\urlprefix }}%
\providecommand \urlprefix  [0]{URL }%
\providecommand \Eprint [0]{\href }%
\providecommand \doibase [0]{http://dx.doi.org/}%
\providecommand \selectlanguage [0]{\@gobble}%
\providecommand \bibinfo  [0]{\@secondoftwo}%
\providecommand \bibfield  [0]{\@secondoftwo}%
\providecommand \translation [1]{[#1]}%
\providecommand \BibitemOpen [0]{}%
\providecommand \bibitemStop [0]{}%
\providecommand \bibitemNoStop [0]{.\EOS\space}%
\providecommand \EOS [0]{\spacefactor3000\relax}%
\providecommand \BibitemShut  [1]{\csname bibitem#1\endcsname}%
\let\auto@bib@innerbib\@empty
\bibitem [{\citenamefont {Van~Trees}\ \emph {et~al.}(2013)\citenamefont
  {Van~Trees}, \citenamefont {Bell},\ and\ \citenamefont
  {Tian}}]{VanTrees2013}%
  \BibitemOpen
  \bibfield  {author} {\bibinfo {author} {\bibfnamefont {H.~L.}\ \bibnamefont
  {Van~Trees}}, \bibinfo {author} {\bibfnamefont {K.~L.}\ \bibnamefont {Bell}},
  \ and\ \bibinfo {author} {\bibfnamefont {Z.}~\bibnamefont {Tian}},\
  }\href@noop {} {\emph {\bibinfo {title} {Detection, Estimation, and
  Modulation Theory, Part I}}},\ \bibinfo {edition} {2nd}\ ed.\ (\bibinfo
  {publisher} {Wiley},\ \bibinfo {year} {2013})\BibitemShut {NoStop}%
\bibitem [{\citenamefont {Helstrom}(1976)}]{Helstrom1976}%
  \BibitemOpen
  \bibfield  {author} {\bibinfo {author} {\bibfnamefont {C.~W.}\ \bibnamefont
  {Helstrom}},\ }\href@noop {} {\emph {\bibinfo {title} {Quantum Detection and
  Estimation Theory}}}\ (\bibinfo  {publisher} {Academic Press, New York},\
  \bibinfo {year} {1976})\BibitemShut {NoStop}%
\bibitem [{\citenamefont {Tsang}\ \emph {et~al.}(2016)\citenamefont {Tsang},
  \citenamefont {Nair},\ and\ \citenamefont {Lu}}]{Tsang2016b}%
  \BibitemOpen
  \bibfield  {author} {\bibinfo {author} {\bibfnamefont {M.}~\bibnamefont
  {Tsang}}, \bibinfo {author} {\bibfnamefont {R.}~\bibnamefont {Nair}}, \ and\
  \bibinfo {author} {\bibfnamefont {X.-M.}\ \bibnamefont {Lu}},\ }\href
  {\doibase 10.1103/PhysRevX.6.031033} {\bibfield  {journal} {\bibinfo
  {journal} {Phys. Rev. X}\ }\textbf {\bibinfo {volume} {6}},\ \bibinfo {pages}
  {031033} (\bibinfo {year} {2016})}\BibitemShut {NoStop}%
\bibitem [{\citenamefont {Kerviche}\ \emph {et~al.}(2017)\citenamefont
  {Kerviche}, \citenamefont {Guha},\ and\ \citenamefont {Ashok}}]{KGA17}%
  \BibitemOpen
  \bibfield  {author} {\bibinfo {author} {\bibfnamefont {R.}~\bibnamefont
  {Kerviche}}, \bibinfo {author} {\bibfnamefont {S.}~\bibnamefont {Guha}}, \
  and\ \bibinfo {author} {\bibfnamefont {A.}~\bibnamefont {Ashok}},\ }in\
  \href@noop {} {\emph {\bibinfo {booktitle} {2017 IEEE International Symposium
  on Information Theory (ISIT)}}}\ (\bibinfo {organization} {IEEE},\ \bibinfo
  {year} {2017})\ pp.\ \bibinfo {pages} {441--445}\BibitemShut {NoStop}%
\bibitem [{\citenamefont {Rehacek}\ \emph {et~al.}(2017)\citenamefont
  {Rehacek}, \citenamefont {Pa{\'u}r}, \citenamefont {Stoklasa}, \citenamefont
  {Hradil},\ and\ \citenamefont {S{\'a}nchez-Soto}}]{Rehacek2017a}%
  \BibitemOpen
  \bibfield  {author} {\bibinfo {author} {\bibfnamefont {J.}~\bibnamefont
  {Rehacek}}, \bibinfo {author} {\bibfnamefont {M.}~\bibnamefont {Pa{\'u}r}},
  \bibinfo {author} {\bibfnamefont {B.}~\bibnamefont {Stoklasa}}, \bibinfo
  {author} {\bibfnamefont {Z.}~\bibnamefont {Hradil}}, \ and\ \bibinfo {author}
  {\bibfnamefont {L.~L.}\ \bibnamefont {S{\'a}nchez-Soto}},\ }\href@noop {}
  {\bibfield  {journal} {\bibinfo  {journal} {Opt. Lett.}\ }\textbf {\bibinfo
  {volume} {42}},\ \bibinfo {pages} {231} (\bibinfo {year} {2017})}\BibitemShut
  {NoStop}%
\bibitem [{\citenamefont {Dutton}\ \emph {et~al.}(2019)\citenamefont {Dutton},
  \citenamefont {Kerviche}, \citenamefont {Ashok},\ and\ \citenamefont
  {Guha}}]{Zachary2019}%
  \BibitemOpen
  \bibfield  {author} {\bibinfo {author} {\bibfnamefont {Z.}~\bibnamefont
  {Dutton}}, \bibinfo {author} {\bibfnamefont {R.}~\bibnamefont {Kerviche}},
  \bibinfo {author} {\bibfnamefont {A.}~\bibnamefont {Ashok}}, \ and\ \bibinfo
  {author} {\bibfnamefont {S.}~\bibnamefont {Guha}},\ }\href {\doibase
  10.1103/PhysRevA.99.033847} {\bibfield  {journal} {\bibinfo  {journal} {Phys.
  Rev. A}\ }\textbf {\bibinfo {volume} {99}},\ \bibinfo {pages} {033847}
  (\bibinfo {year} {2019})}\BibitemShut {NoStop}%
\bibitem [{\citenamefont {Prasad}(2020{\natexlab{a}})}]{Prasad2020B}%
  \BibitemOpen
  \bibfield  {author} {\bibinfo {author} {\bibfnamefont {S.}~\bibnamefont
  {Prasad}},\ }\href {\doibase 10.1103/PhysRevA.102.063719} {\bibfield
  {journal} {\bibinfo  {journal} {Phys. Rev. A}\ }\textbf {\bibinfo {volume}
  {102}},\ \bibinfo {pages} {063719} (\bibinfo {year}
  {2020}{\natexlab{a}})}\BibitemShut {NoStop}%
\bibitem [{\citenamefont {Ang}\ \emph {et~al.}(2017)\citenamefont {Ang},
  \citenamefont {Nair},\ and\ \citenamefont {Tsang}}]{Ang2016}%
  \BibitemOpen
  \bibfield  {author} {\bibinfo {author} {\bibfnamefont {S.~Z.}\ \bibnamefont
  {Ang}}, \bibinfo {author} {\bibfnamefont {R.}~\bibnamefont {Nair}}, \ and\
  \bibinfo {author} {\bibfnamefont {M.}~\bibnamefont {Tsang}},\ }\href
  {\doibase 10.1103/PhysRevA.95.063847} {\bibfield  {journal} {\bibinfo
  {journal} {Phys. Rev. A}\ }\textbf {\bibinfo {volume} {95}},\ \bibinfo
  {pages} {063847} (\bibinfo {year} {2017})}\BibitemShut {NoStop}%
\bibitem [{\citenamefont {Yu}\ and\ \citenamefont {Prasad}(2018)}]{Prasad2018}%
  \BibitemOpen
  \bibfield  {author} {\bibinfo {author} {\bibfnamefont {Z.}~\bibnamefont
  {Yu}}\ and\ \bibinfo {author} {\bibfnamefont {S.}~\bibnamefont {Prasad}},\
  }\href {\doibase 10.1103/PhysRevLett.121.180504} {\bibfield  {journal}
  {\bibinfo  {journal} {Phys. Rev. Lett.}\ }\textbf {\bibinfo {volume} {121}},\
  \bibinfo {pages} {180504} (\bibinfo {year} {2018})}\BibitemShut {NoStop}%
\bibitem [{\citenamefont {Prasad}\ and\ \citenamefont {Yu}(2019)}]{Prasad2019}%
  \BibitemOpen
  \bibfield  {author} {\bibinfo {author} {\bibfnamefont {S.}~\bibnamefont
  {Prasad}}\ and\ \bibinfo {author} {\bibfnamefont {Z.}~\bibnamefont {Yu}},\
  }\href {\doibase 10.1103/PhysRevA.99.022116} {\bibfield  {journal} {\bibinfo
  {journal} {Phys. Rev. A}\ }\textbf {\bibinfo {volume} {99}},\ \bibinfo
  {pages} {022116} (\bibinfo {year} {2019})}\BibitemShut {NoStop}%
\bibitem [{\citenamefont {{\v R}eha{\v c}ek}\ \emph {et~al.}(2017)\citenamefont
  {{\v R}eha{\v c}ek}, \citenamefont {Hradil}, \citenamefont {Stoklasa},
  \citenamefont {Pa{\'u}r}, \citenamefont {Grover}, \citenamefont {Krzic},\
  and\ \citenamefont {S{\'a}nchez-Soto}}]{Rehacek2017b}%
  \BibitemOpen
  \bibfield  {author} {\bibinfo {author} {\bibfnamefont {J.}~\bibnamefont {{\v
  R}eha{\v c}ek}}, \bibinfo {author} {\bibfnamefont {Z.}~\bibnamefont
  {Hradil}}, \bibinfo {author} {\bibfnamefont {B.}~\bibnamefont {Stoklasa}},
  \bibinfo {author} {\bibfnamefont {M.}~\bibnamefont {Pa{\'u}r}}, \bibinfo
  {author} {\bibfnamefont {J.}~\bibnamefont {Grover}}, \bibinfo {author}
  {\bibfnamefont {A.}~\bibnamefont {Krzic}}, \ and\ \bibinfo {author}
  {\bibfnamefont {L.~L.}\ \bibnamefont {S{\'a}nchez-Soto}},\ }\href@noop {}
  {\bibfield  {journal} {\bibinfo  {journal} {Phys. Rev. A}\ }\textbf {\bibinfo
  {volume} {96}},\ \bibinfo {pages} {062107} (\bibinfo {year}
  {2017})}\BibitemShut {NoStop}%
\bibitem [{\citenamefont {\ifmmode \check{R}\else
  \v{R}\fi{}eh\'a\ifmmode~\check{c}\else \v{c}\fi{}ek}\ \emph
  {et~al.}(2018)\citenamefont {\ifmmode \check{R}\else
  \v{R}\fi{}eh\'a\ifmmode~\check{c}\else \v{c}\fi{}ek}, \citenamefont {Hradil},
  \citenamefont {Koutn\'y}, \citenamefont {Grover}, \citenamefont {Krzic},\
  and\ \citenamefont {S\'anchez-Soto}}]{rehacek2018}%
  \BibitemOpen
  \bibfield  {author} {\bibinfo {author} {\bibfnamefont {J.}~\bibnamefont
  {\ifmmode \check{R}\else \v{R}\fi{}eh\'a\ifmmode~\check{c}\else
  \v{c}\fi{}ek}}, \bibinfo {author} {\bibfnamefont {Z.}~\bibnamefont {Hradil}},
  \bibinfo {author} {\bibfnamefont {D.}~\bibnamefont {Koutn\'y}}, \bibinfo
  {author} {\bibfnamefont {J.}~\bibnamefont {Grover}}, \bibinfo {author}
  {\bibfnamefont {A.}~\bibnamefont {Krzic}}, \ and\ \bibinfo {author}
  {\bibfnamefont {L.~L.}\ \bibnamefont {S\'anchez-Soto}},\ }\href {\doibase
  10.1103/PhysRevA.98.012103} {\bibfield  {journal} {\bibinfo  {journal} {Phys.
  Rev. A}\ }\textbf {\bibinfo {volume} {98}},\ \bibinfo {pages} {012103}
  (\bibinfo {year} {2018})}\BibitemShut {NoStop}%
\bibitem [{\citenamefont {Prasad}(2020{\natexlab{b}})}]{Prasad_2020}%
  \BibitemOpen
  \bibfield  {author} {\bibinfo {author} {\bibfnamefont {S.}~\bibnamefont
  {Prasad}},\ }\href {\doibase 10.1088/1402-4896/ab573d} {\bibfield  {journal}
  {\bibinfo  {journal} {Physica Scripta}\ }\textbf {\bibinfo {volume} {95}},\
  \bibinfo {pages} {054004} (\bibinfo {year} {2020}{\natexlab{b}})}\BibitemShut
  {NoStop}%
\bibitem [{\citenamefont {Tsang}(2019{\natexlab{a}})}]{Tsang2019}%
  \BibitemOpen
  \bibfield  {author} {\bibinfo {author} {\bibfnamefont {M.}~\bibnamefont
  {Tsang}},\ }\href {\doibase 10.1103/PhysRevA.99.012305} {\bibfield  {journal}
  {\bibinfo  {journal} {Physical Review A}\ }\textbf {\bibinfo {volume} {99}},\
  \bibinfo {pages} {012305} (\bibinfo {year} {2019}{\natexlab{a}})}\BibitemShut
  {NoStop}%
\bibitem [{\citenamefont {Tsang}(2019{\natexlab{b}})}]{Tsang2019a}%
  \BibitemOpen
  \bibfield  {author} {\bibinfo {author} {\bibfnamefont {M.}~\bibnamefont
  {Tsang}},\ }\href@noop {} {\bibfield  {journal} {\bibinfo  {journal}
  {Contemporary Physics}\ }\textbf {\bibinfo {volume} {60}},\ \bibinfo {pages}
  {279} (\bibinfo {year} {2019}{\natexlab{b}})}\BibitemShut {NoStop}%
\bibitem [{\citenamefont {Zhou}\ and\ \citenamefont {Jiang}(2019)}]{Zhou2019}%
  \BibitemOpen
  \bibfield  {author} {\bibinfo {author} {\bibfnamefont {S.}~\bibnamefont
  {Zhou}}\ and\ \bibinfo {author} {\bibfnamefont {L.}~\bibnamefont {Jiang}},\
  }\href {\doibase 10.1103/PhysRevA.99.013808} {\bibfield  {journal} {\bibinfo
  {journal} {Physical Review A}\ }\textbf {\bibinfo {volume} {99}},\ \bibinfo
  {pages} {013808} (\bibinfo {year} {2019})}\BibitemShut {NoStop}%
\bibitem [{\citenamefont {Lu}\ \emph {et~al.}(2018)\citenamefont {Lu},
  \citenamefont {Krovi}, \citenamefont {Nair}, \citenamefont {Guha},\ and\
  \citenamefont {Shapiro}}]{Lu2018}%
  \BibitemOpen
  \bibfield  {author} {\bibinfo {author} {\bibfnamefont {X.~M.}\ \bibnamefont
  {Lu}}, \bibinfo {author} {\bibfnamefont {H.}~\bibnamefont {Krovi}}, \bibinfo
  {author} {\bibfnamefont {R.}~\bibnamefont {Nair}}, \bibinfo {author}
  {\bibfnamefont {S.}~\bibnamefont {Guha}}, \ and\ \bibinfo {author}
  {\bibfnamefont {J.~H.}\ \bibnamefont {Shapiro}},\ }\href@noop {} {\bibfield
  {journal} {\bibinfo  {journal} {npj Quantum Information}\ }\textbf {\bibinfo
  {volume} {64}} (\bibinfo {year} {2018})}\BibitemShut {NoStop}%
\bibitem [{\citenamefont {Huang}\ and\ \citenamefont {Lupo}(2021)}]{Huang2021}%
  \BibitemOpen
  \bibfield  {author} {\bibinfo {author} {\bibfnamefont {Z.}~\bibnamefont
  {Huang}}\ and\ \bibinfo {author} {\bibfnamefont {C.}~\bibnamefont {Lupo}},\
  }\href {\doibase 10.1103/PhysRevLett.127.130502} {\bibfield  {journal}
  {\bibinfo  {journal} {Physical Review Letters}\ }\textbf {\bibinfo {volume}
  {127}},\ \bibinfo {pages} {130502} (\bibinfo {year} {2021})}\BibitemShut
  {NoStop}%
\bibitem [{\citenamefont {Grace}\ and\ \citenamefont
  {Guha}(2022)}]{Grace2021c}%
  \BibitemOpen
  \bibfield  {author} {\bibinfo {author} {\bibfnamefont {M.~R.}\ \bibnamefont
  {Grace}}\ and\ \bibinfo {author} {\bibfnamefont {S.}~\bibnamefont {Guha}},\
  }\href {\doibase https://doi.org/10.1103/PhysRevLett.129.180502} {\bibfield
  {journal} {\bibinfo  {journal} {Physical Review Letters}\ }\textbf {\bibinfo
  {volume} {129}},\ \bibinfo {pages} {180502} (\bibinfo {year}
  {2022})}\BibitemShut {NoStop}%
\bibitem [{\citenamefont {Bao}\ \emph {et~al.}(2021)\citenamefont {Bao},
  \citenamefont {Choi}, \citenamefont {Aggarwal},\ and\ \citenamefont
  {Jacob}}]{Bao2021}%
  \BibitemOpen
  \bibfield  {author} {\bibinfo {author} {\bibfnamefont {F.}~\bibnamefont
  {Bao}}, \bibinfo {author} {\bibfnamefont {H.}~\bibnamefont {Choi}}, \bibinfo
  {author} {\bibfnamefont {V.}~\bibnamefont {Aggarwal}}, \ and\ \bibinfo
  {author} {\bibfnamefont {Z.}~\bibnamefont {Jacob}},\ }\href@noop {}
  {\bibfield  {journal} {\bibinfo  {journal} {Opt. Lett.}\ }\textbf {\bibinfo
  {volume} {46}},\ \bibinfo {pages} {3045} (\bibinfo {year}
  {2021})}\BibitemShut {NoStop}%
\bibitem [{\citenamefont {Matlin}\ and\ \citenamefont
  {Zipp}(2022)}]{Matlin2022}%
  \BibitemOpen
  \bibfield  {author} {\bibinfo {author} {\bibfnamefont {E.~F.}\ \bibnamefont
  {Matlin}}\ and\ \bibinfo {author} {\bibfnamefont {L.~J.}\ \bibnamefont
  {Zipp}},\ }\href {\doibase 10.1038/s41598-022-06644-3} {\bibfield  {journal}
  {\bibinfo  {journal} {Scientific Reports}\ }\textbf {\bibinfo {volume} {12}}
  (\bibinfo {year} {2022}),\ 10.1038/s41598-022-06644-3}\BibitemShut {NoStop}%
\bibitem [{\citenamefont {Lee}\ \emph {et~al.}(2022)\citenamefont {Lee},
  \citenamefont {Gagatsos}, \citenamefont {Guha},\ and\ \citenamefont
  {Ashok}}]{Lee2022}%
  \BibitemOpen
  \bibfield  {author} {\bibinfo {author} {\bibfnamefont {K.~K.}\ \bibnamefont
  {Lee}}, \bibinfo {author} {\bibfnamefont {C.~N.}\ \bibnamefont {Gagatsos}},
  \bibinfo {author} {\bibfnamefont {S.}~\bibnamefont {Guha}}, \ and\ \bibinfo
  {author} {\bibfnamefont {A.}~\bibnamefont {Ashok}},\ }\href@noop {}
  {\bibfield  {journal} {\bibinfo  {journal} {IEEE J. Sel. Top. Signal
  Process.}\ ,\ \bibinfo {pages} {1}} (\bibinfo {year} {2022})}\BibitemShut
  {NoStop}%
\bibitem [{\citenamefont {Lupo}\ \emph {et~al.}(2020)\citenamefont {Lupo},
  \citenamefont {Huang},\ and\ \citenamefont {Kok}}]{Cosmo2020}%
  \BibitemOpen
  \bibfield  {author} {\bibinfo {author} {\bibfnamefont {C.}~\bibnamefont
  {Lupo}}, \bibinfo {author} {\bibfnamefont {Z.}~\bibnamefont {Huang}}, \ and\
  \bibinfo {author} {\bibfnamefont {P.}~\bibnamefont {Kok}},\ }\href {\doibase
  10.1103/PhysRevLett.124.080503} {\bibfield  {journal} {\bibinfo  {journal}
  {Phys. Rev. Lett.}\ }\textbf {\bibinfo {volume} {124}},\ \bibinfo {pages}
  {080503} (\bibinfo {year} {2020})}\BibitemShut {NoStop}%
\bibitem [{\citenamefont {Wang}\ \emph {et~al.}(2021)\citenamefont {Wang},
  \citenamefont {Zhang},\ and\ \citenamefont {Lorenz}}]{Wang2021}%
  \BibitemOpen
  \bibfield  {author} {\bibinfo {author} {\bibfnamefont {Y.}~\bibnamefont
  {Wang}}, \bibinfo {author} {\bibfnamefont {Y.}~\bibnamefont {Zhang}}, \ and\
  \bibinfo {author} {\bibfnamefont {V.~O.}\ \bibnamefont {Lorenz}},\ }\href
  {\doibase 10.1103/PhysRevA.104.022613} {\bibfield  {journal} {\bibinfo
  {journal} {Physical Review A}\ }\textbf {\bibinfo {volume} {104}},\ \bibinfo
  {pages} {1} (\bibinfo {year} {2021})}\BibitemShut {NoStop}%
\bibitem [{\citenamefont {Bojer}\ \emph {et~al.}(2022)\citenamefont {Bojer},
  \citenamefont {Huang}, \citenamefont {Karl}, \citenamefont {Richter},
  \citenamefont {Kok},\ and\ \citenamefont {von Zanthier}}]{Bojer2022}%
  \BibitemOpen
  \bibfield  {author} {\bibinfo {author} {\bibfnamefont {M.}~\bibnamefont
  {Bojer}}, \bibinfo {author} {\bibfnamefont {Z.}~\bibnamefont {Huang}},
  \bibinfo {author} {\bibfnamefont {S.}~\bibnamefont {Karl}}, \bibinfo {author}
  {\bibfnamefont {S.}~\bibnamefont {Richter}}, \bibinfo {author} {\bibfnamefont
  {P.}~\bibnamefont {Kok}}, \ and\ \bibinfo {author} {\bibfnamefont
  {J.}~\bibnamefont {von Zanthier}},\ }\href {\doibase
  10.1088/1367-2630/ac5f30} {\bibfield  {journal} {\bibinfo  {journal} {New
  Journal of Physics}\ }\textbf {\bibinfo {volume} {24}},\ \bibinfo {pages}
  {043026} (\bibinfo {year} {2022})}\BibitemShut {NoStop}%
\bibitem [{\citenamefont {Gottesman}\ \emph {et~al.}(2012)\citenamefont
  {Gottesman}, \citenamefont {Jennewein},\ and\ \citenamefont
  {Croke}}]{Gottesman2012}%
  \BibitemOpen
  \bibfield  {author} {\bibinfo {author} {\bibfnamefont {D.}~\bibnamefont
  {Gottesman}}, \bibinfo {author} {\bibfnamefont {T.}~\bibnamefont
  {Jennewein}}, \ and\ \bibinfo {author} {\bibfnamefont {S.}~\bibnamefont
  {Croke}},\ }\href {\doibase 10.1103/PhysRevLett.109.070503} {\bibfield
  {journal} {\bibinfo  {journal} {Phys. Rev. Lett.}\ }\textbf {\bibinfo
  {volume} {109}},\ \bibinfo {pages} {070503} (\bibinfo {year}
  {2012})}\BibitemShut {NoStop}%
\bibitem [{\citenamefont {Khabiboulline}\ \emph {et~al.}(2019)\citenamefont
  {Khabiboulline}, \citenamefont {Borregaard}, \citenamefont {De~Greve},\ and\
  \citenamefont {Lukin}}]{Khabiboulline2018}%
  \BibitemOpen
  \bibfield  {author} {\bibinfo {author} {\bibfnamefont {E.~T.}\ \bibnamefont
  {Khabiboulline}}, \bibinfo {author} {\bibfnamefont {J.}~\bibnamefont
  {Borregaard}}, \bibinfo {author} {\bibfnamefont {K.}~\bibnamefont
  {De~Greve}}, \ and\ \bibinfo {author} {\bibfnamefont {M.~D.}\ \bibnamefont
  {Lukin}},\ }\href {\doibase 10.1103/PhysRevLett.123.070504} {\bibfield
  {journal} {\bibinfo  {journal} {Phys. Rev. Lett.}\ }\textbf {\bibinfo
  {volume} {123}},\ \bibinfo {pages} {070504} (\bibinfo {year}
  {2019})}\BibitemShut {NoStop}%
\bibitem [{\citenamefont {Shaklan}\ and\ \citenamefont
  {Roddier}(1987)}]{Shaklan1987}%
  \BibitemOpen
  \bibfield  {author} {\bibinfo {author} {\bibfnamefont {S.~B.}\ \bibnamefont
  {Shaklan}}\ and\ \bibinfo {author} {\bibfnamefont {F.}~\bibnamefont
  {Roddier}},\ }\href {\doibase 10.1364/AO.26.002159} {\bibfield  {journal}
  {\bibinfo  {journal} {Appl. Opt.}\ }\textbf {\bibinfo {volume} {26}},\
  \bibinfo {pages} {2159} (\bibinfo {year} {1987})}\BibitemShut {NoStop}%
\bibitem [{\citenamefont {Guyon}(2002)}]{Guyon2022}%
  \BibitemOpen
  \bibfield  {author} {\bibinfo {author} {\bibfnamefont {O.}~\bibnamefont
  {Guyon}},\ }\href@noop {} {\bibfield  {journal} {\bibinfo  {journal}
  {Astronomy and Astrophysics}\ }\textbf {\bibinfo {volume} {387}},\ \bibinfo
  {pages} {366} (\bibinfo {year} {2002})}\BibitemShut {NoStop}%
\bibitem [{\citenamefont {Jovanovic}\ \emph {et~al.}(2017)\citenamefont
  {Jovanovic}, \citenamefont {Schwab}, \citenamefont {Guyon}, \citenamefont
  {Lozi}, \citenamefont {Cvetojevic}, \citenamefont {Martinache}, \citenamefont
  {Leon-Saval}, \citenamefont {Norris}, \citenamefont {Gross}, \citenamefont
  {Doughty} \emph {et~al.}}]{Jovanovic}%
  \BibitemOpen
  \bibfield  {author} {\bibinfo {author} {\bibfnamefont {N.}~\bibnamefont
  {Jovanovic}}, \bibinfo {author} {\bibfnamefont {C.}~\bibnamefont {Schwab}},
  \bibinfo {author} {\bibfnamefont {O.}~\bibnamefont {Guyon}}, \bibinfo
  {author} {\bibfnamefont {J.}~\bibnamefont {Lozi}}, \bibinfo {author}
  {\bibfnamefont {N.}~\bibnamefont {Cvetojevic}}, \bibinfo {author}
  {\bibfnamefont {F.}~\bibnamefont {Martinache}}, \bibinfo {author}
  {\bibfnamefont {S.}~\bibnamefont {Leon-Saval}}, \bibinfo {author}
  {\bibfnamefont {B.}~\bibnamefont {Norris}}, \bibinfo {author} {\bibfnamefont
  {S.}~\bibnamefont {Gross}}, \bibinfo {author} {\bibfnamefont
  {D.}~\bibnamefont {Doughty}},  \emph {et~al.},\ }\href@noop {} {\bibfield
  {journal} {\bibinfo  {journal} {Astronomy and Astrophysics}\ }\textbf
  {\bibinfo {volume} {604}},\ \bibinfo {pages} {A122} (\bibinfo {year}
  {2017})}\BibitemShut {NoStop}%
\bibitem [{\citenamefont {Glindemann}(2011)}]{Glindemann2011}%
  \BibitemOpen
  \bibfield  {author} {\bibinfo {author} {\bibfnamefont {A.}~\bibnamefont
  {Glindemann}},\ }\href {\doibase 10.1007/978-3-642-15028-9} {\emph {\bibinfo
  {title} {Principles of Stellar Interferometry}}}\ (\bibinfo  {publisher}
  {Springer Berlin Heidelberg},\ \bibinfo {year} {2011})\BibitemShut {NoStop}%
\bibitem [{\citenamefont {Zheng}\ \emph {et~al.}(2022)\citenamefont {Zheng},
  \citenamefont {Sharma},\ and\ \citenamefont {Borregaard}}]{Zheng2022}%
  \BibitemOpen
  \bibfield  {author} {\bibinfo {author} {\bibfnamefont {Y.}~\bibnamefont
  {Zheng}}, \bibinfo {author} {\bibfnamefont {H.}~\bibnamefont {Sharma}}, \
  and\ \bibinfo {author} {\bibfnamefont {J.}~\bibnamefont {Borregaard}},\
  }\href@noop {} {\bibfield  {journal} {\bibinfo  {journal} {PRX Quantum}\
  }\textbf {\bibinfo {volume} {3}},\ \bibinfo {pages} {040319} (\bibinfo {year}
  {2022})}\BibitemShut {NoStop}%
\bibitem [{\citenamefont {Clements}\ \emph {et~al.}(2016)\citenamefont
  {Clements}, \citenamefont {Humphreys}, \citenamefont {Metcalf}, \citenamefont
  {Steven~Kolthammer},\ and\ \citenamefont {Walmsley}}]{Clements2016}%
  \BibitemOpen
  \bibfield  {author} {\bibinfo {author} {\bibfnamefont {W.~R.}\ \bibnamefont
  {Clements}}, \bibinfo {author} {\bibfnamefont {P.~C.}\ \bibnamefont
  {Humphreys}}, \bibinfo {author} {\bibfnamefont {B.~J.}\ \bibnamefont
  {Metcalf}}, \bibinfo {author} {\bibfnamefont {W.}~\bibnamefont
  {Steven~Kolthammer}}, \ and\ \bibinfo {author} {\bibfnamefont {I.~A.}\
  \bibnamefont {Walmsley}},\ }\href@noop {} {\bibfield  {journal} {\bibinfo
  {journal} {Optica, OPTICA}\ }\textbf {\bibinfo {volume} {3}},\ \bibinfo
  {pages} {1460} (\bibinfo {year} {2016})}\BibitemShut {NoStop}%
\bibitem [{\citenamefont {Goodman}(1985)}]{Goo85Statistical}%
  \BibitemOpen
  \bibfield  {author} {\bibinfo {author} {\bibfnamefont {J.~W.}\ \bibnamefont
  {Goodman}},\ }\href@noop {} {\emph {\bibinfo {title} {Statistical Optics}}}\
  (\bibinfo  {publisher} {John Wiley \& Sons},\ \bibinfo {year}
  {1985})\BibitemShut {NoStop}%
\bibitem [{\citenamefont {Mandel}\ and\ \citenamefont {Wolf}(1995)}]{MW95}%
  \BibitemOpen
  \bibfield  {author} {\bibinfo {author} {\bibfnamefont {L.}~\bibnamefont
  {Mandel}}\ and\ \bibinfo {author} {\bibfnamefont {E.}~\bibnamefont {Wolf}},\
  }\href@noop {} {\emph {\bibinfo {title} {Optical Coherence and Quantum
  Optics}}}\ (\bibinfo  {publisher} {Cambridge University Press, Cambridge},\
  \bibinfo {year} {1995})\BibitemShut {NoStop}%
\bibitem [{\citenamefont {Labeyrie}\ \emph {et~al.}(2006)\citenamefont
  {Labeyrie}, \citenamefont {Lipson},\ and\ \citenamefont
  {Nisenson}}]{Labeyrie2006}%
  \BibitemOpen
  \bibfield  {author} {\bibinfo {author} {\bibfnamefont {A.}~\bibnamefont
  {Labeyrie}}, \bibinfo {author} {\bibfnamefont {S.~G.}\ \bibnamefont
  {Lipson}}, \ and\ \bibinfo {author} {\bibfnamefont {P.}~\bibnamefont
  {Nisenson}},\ }\href@noop {} {\emph {\bibinfo {title} {An introduction to
  optical stellar interferometry}}}\ (\bibinfo  {publisher} {Cambridge
  University Press},\ \bibinfo {year} {2006})\BibitemShut {NoStop}%
\bibitem [{\citenamefont {Tsang}(2011)}]{Tsang2011}%
  \BibitemOpen
  \bibfield  {author} {\bibinfo {author} {\bibfnamefont {M.}~\bibnamefont
  {Tsang}},\ }\href {\doibase 10.1103/PhysRevLett.107.270402} {\bibfield
  {journal} {\bibinfo  {journal} {Phys. Rev. Lett.}\ }\textbf {\bibinfo
  {volume} {107}},\ \bibinfo {pages} {270402} (\bibinfo {year}
  {2011})}\BibitemShut {NoStop}%
\bibitem [{\citenamefont {Ram}\ \emph {et~al.}(2006)\citenamefont {Ram},
  \citenamefont {Ward},\ and\ \citenamefont {Ober}}]{Ram2006}%
  \BibitemOpen
  \bibfield  {author} {\bibinfo {author} {\bibfnamefont {S.}~\bibnamefont
  {Ram}}, \bibinfo {author} {\bibfnamefont {E.~S.}\ \bibnamefont {Ward}}, \
  and\ \bibinfo {author} {\bibfnamefont {R.~J.}\ \bibnamefont {Ober}},\ }\href
  {\doibase 10.1073/pnas.0508047103} {\bibfield  {journal} {\bibinfo  {journal}
  {Proc. Natl. Acad. Sci. U.S.A.}\ }\textbf {\bibinfo {volume} {103}},\
  \bibinfo {pages} {4457} (\bibinfo {year} {2006})}\BibitemShut {NoStop}%
\bibitem [{\citenamefont {Pawley}(2006)}]{pawley2006}%
  \BibitemOpen
  \bibfield  {author} {\bibinfo {author} {\bibfnamefont {J.}~\bibnamefont
  {Pawley}},\ }\href@noop {} {\emph {\bibinfo {title} {Handbook of biological
  confocal microscopy}}},\ Vol.\ \bibinfo {volume} {236}\ (\bibinfo
  {publisher} {Springer Science \& Business Media},\ \bibinfo {year}
  {2006})\BibitemShut {NoStop}%
\bibitem [{\citenamefont {Zmuidzinas}(2003)}]{Zmuidzinas2003CramrRaoSL}%
  \BibitemOpen
  \bibfield  {author} {\bibinfo {author} {\bibfnamefont {J.}~\bibnamefont
  {Zmuidzinas}},\ }\href@noop {} {\bibfield  {journal} {\bibinfo  {journal}
  {Journal of the Optical Society of America. A, Optics, image science, and
  vision}\ }\textbf {\bibinfo {volume} {20 2}},\ \bibinfo {pages} {218}
  (\bibinfo {year} {2003})}\BibitemShut {NoStop}%
\bibitem [{\citenamefont {Braunstein}\ and\ \citenamefont
  {Caves}(1994)}]{PhysRevLett.72.3439}%
  \BibitemOpen
  \bibfield  {author} {\bibinfo {author} {\bibfnamefont {S.~L.}\ \bibnamefont
  {Braunstein}}\ and\ \bibinfo {author} {\bibfnamefont {C.~M.}\ \bibnamefont
  {Caves}},\ }\href {\doibase 10.1103/PhysRevLett.72.3439} {\bibfield
  {journal} {\bibinfo  {journal} {Phys. Rev. Lett.}\ }\textbf {\bibinfo
  {volume} {72}},\ \bibinfo {pages} {3439} (\bibinfo {year}
  {1994})}\BibitemShut {NoStop}%
\bibitem [{\citenamefont {Barndorff-Nielsen}\ and\ \citenamefont
  {Gill}(2000)}]{Barndorff_Nielsen_2000}%
  \BibitemOpen
  \bibfield  {author} {\bibinfo {author} {\bibfnamefont {O.~E.}\ \bibnamefont
  {Barndorff-Nielsen}}\ and\ \bibinfo {author} {\bibfnamefont {R.~D.}\
  \bibnamefont {Gill}},\ }\href {\doibase 10.1088/0305-4470/33/24/306}
  {\bibfield  {journal} {\bibinfo  {journal} {Journal of Physics A:
  Mathematical and General}\ }\textbf {\bibinfo {volume} {33}},\ \bibinfo
  {pages} {4481} (\bibinfo {year} {2000})}\BibitemShut {NoStop}%
\bibitem [{\citenamefont {G.~A.~Paris}(2008)}]{Paris2008}%
  \BibitemOpen
  \bibfield  {author} {\bibinfo {author} {\bibfnamefont {M.}~\bibnamefont
  {G.~A.~Paris}},\ }\href {\doibase 10.1142/S0219749909004839} {\bibfield
  {journal} {\bibinfo  {journal} {International Journal of Quantum
  Information}\ }\textbf {\bibinfo {volume} {7}} (\bibinfo {year} {2008}),\
  10.1142/S0219749909004839}\BibitemShut {NoStop}%
\bibitem [{\citenamefont {{Hayashi}}\ and\ \citenamefont
  {{Matsumoto}}(2005)}]{2005atqs.book162H}%
  \BibitemOpen
  \bibfield  {author} {\bibinfo {author} {\bibfnamefont {M.}~\bibnamefont
  {{Hayashi}}}\ and\ \bibinfo {author} {\bibfnamefont {K.}~\bibnamefont
  {{Matsumoto}}},\ }\enquote {\bibinfo {title} {{Statistical Model with
  Measurement Degree of Freedom and Quantum Physics}},}\ in\ \href {\doibase
  10.1142/9789812563071_0014} {\emph {\bibinfo {booktitle} {Asymptotic Theory
  of Quantum Statistical Inference: Selected Papers.~Edited by HAYASHI
  MASAHITO.~Published by World Scientific Publishing Co.~Pte.~Ltd., 2005.~ISBN
  \#9789812563071, pp.~162-169}}},\ \bibinfo {editor} {edited by\ \bibinfo
  {editor} {\bibfnamefont {M.}~\bibnamefont {{Hayashi}}}}\ (\bibinfo
  {publisher} {World Scientific Publishing Co},\ \bibinfo {year} {2005})\ pp.\
  \bibinfo {pages} {162--169}\BibitemShut {NoStop}%
\bibitem [{\citenamefont {Gill}\ and\ \citenamefont
  {Massar}(2000)}]{PhysRevA.61.042312}%
  \BibitemOpen
  \bibfield  {author} {\bibinfo {author} {\bibfnamefont {R.~D.}\ \bibnamefont
  {Gill}}\ and\ \bibinfo {author} {\bibfnamefont {S.}~\bibnamefont {Massar}},\
  }\href {\doibase 10.1103/PhysRevA.61.042312} {\bibfield  {journal} {\bibinfo
  {journal} {Phys. Rev. A}\ }\textbf {\bibinfo {volume} {61}},\ \bibinfo
  {pages} {042312} (\bibinfo {year} {2000})}\BibitemShut {NoStop}%
\bibitem [{\citenamefont {Hayashi}(2006)}]{Hayashi2006}%
  \BibitemOpen
  \bibfield  {author} {\bibinfo {author} {\bibfnamefont {M.}~\bibnamefont
  {Hayashi}},\ }\href@noop {} {\emph {\bibinfo {title} {Quantum Information: An
  Introduction}}},\ \bibinfo {edition} {1st}\ ed.\ (\bibinfo  {publisher}
  {Springer-Verlag, Berlin Heidelberg},\ \bibinfo {year} {2006})\BibitemShut
  {NoStop}%
\bibitem [{\citenamefont {Sajjad}\ \emph {et~al.}(2021)\citenamefont {Sajjad},
  \citenamefont {Grace}, \citenamefont {Zhuang},\ and\ \citenamefont
  {Guha}}]{Sajjad2021}%
  \BibitemOpen
  \bibfield  {author} {\bibinfo {author} {\bibfnamefont {A.}~\bibnamefont
  {Sajjad}}, \bibinfo {author} {\bibfnamefont {M.~R.}\ \bibnamefont {Grace}},
  \bibinfo {author} {\bibfnamefont {Q.}~\bibnamefont {Zhuang}}, \ and\ \bibinfo
  {author} {\bibfnamefont {S.}~\bibnamefont {Guha}},\ }\href {\doibase
  10.1103/PhysRevA.104.022410} {\bibfield  {journal} {\bibinfo  {journal}
  {Phys. Rev. A}\ }\textbf {\bibinfo {volume} {104}},\ \bibinfo {pages}
  {022410} (\bibinfo {year} {2021})}\BibitemShut {NoStop}%
\bibitem [{\citenamefont {Fujiwara}(2006)}]{Fujiwara_2006}%
  \BibitemOpen
  \bibfield  {author} {\bibinfo {author} {\bibfnamefont {A.}~\bibnamefont
  {Fujiwara}},\ }\href {\doibase 10.1088/0305-4470/39/40/014} {\bibfield
  {journal} {\bibinfo  {journal} {Journal of Physics A: Mathematical and
  General}\ }\textbf {\bibinfo {volume} {39}},\ \bibinfo {pages} {12489}
  (\bibinfo {year} {2006})}\BibitemShut {NoStop}%
\bibitem [{Note1()}]{Note1}%
  \BibitemOpen
  \bibinfo {note} {In this paper, we will not consider the multi-parameter
  estimation problem, e.g., {\protect \em localization} of an unknown number of
  point emitters: ${\protect \boldsymbol \theta } = (\left \{x_s, b_s\right \},
  n_e)$, for which dynamically-adaptive modal measurement techniques is known
  to outperform direct imaging~\cite {Lee2022}}\BibitemShut {NoStop}%
\bibitem [{\citenamefont {Grace}\ \emph {et~al.}(2020)\citenamefont {Grace},
  \citenamefont {Dutton}, \citenamefont {Ashok},\ and\ \citenamefont
  {Guha}}]{Grace2020c}%
  \BibitemOpen
  \bibfield  {author} {\bibinfo {author} {\bibfnamefont {M.~R.}\ \bibnamefont
  {Grace}}, \bibinfo {author} {\bibfnamefont {Z.}~\bibnamefont {Dutton}},
  \bibinfo {author} {\bibfnamefont {A.}~\bibnamefont {Ashok}}, \ and\ \bibinfo
  {author} {\bibfnamefont {S.}~\bibnamefont {Guha}},\ }\href@noop {} {\bibfield
   {journal} {\bibinfo  {journal} {Journal of the Optical Society of America
  A}\ }\textbf {\bibinfo {volume} {37}},\ \bibinfo {pages} {1288} (\bibinfo
  {year} {2020})}\BibitemShut {NoStop}%
\bibitem [{\citenamefont {Boucher}\ \emph {et~al.}(2020)\citenamefont
  {Boucher}, \citenamefont {Fabre}, \citenamefont {Labroille},\ and\
  \citenamefont {Treps}}]{Boucher2020}%
  \BibitemOpen
  \bibfield  {author} {\bibinfo {author} {\bibfnamefont {P.}~\bibnamefont
  {Boucher}}, \bibinfo {author} {\bibfnamefont {C.}~\bibnamefont {Fabre}},
  \bibinfo {author} {\bibfnamefont {G.}~\bibnamefont {Labroille}}, \ and\
  \bibinfo {author} {\bibfnamefont {N.}~\bibnamefont {Treps}},\ }\href@noop {}
  {\bibfield  {journal} {\bibinfo  {journal} {Optica}\ }\textbf {\bibinfo
  {volume} {7}} (\bibinfo {year} {2020})}\BibitemShut {NoStop}%
\bibitem [{\citenamefont {Goodman}(2005)}]{Goodman2005}%
  \BibitemOpen
  \bibfield  {author} {\bibinfo {author} {\bibfnamefont {J.~W.}\ \bibnamefont
  {Goodman}},\ }\href@noop {} {\emph {\bibinfo {title} {Introduction to Fourier
  Optics}}},\ \bibinfo {edition} {3rd}\ ed.\ (\bibinfo  {publisher} {Roberts
  and Company},\ \bibinfo {year} {2005})\BibitemShut {NoStop}%
\bibitem [{\citenamefont {Nair}\ and\ \citenamefont {Tsang}(2016)}]{Nair2016}%
  \BibitemOpen
  \bibfield  {author} {\bibinfo {author} {\bibfnamefont {R.}~\bibnamefont
  {Nair}}\ and\ \bibinfo {author} {\bibfnamefont {M.}~\bibnamefont {Tsang}},\
  }\href {\doibase 10.1364/OE.24.003684} {\bibfield  {journal} {\bibinfo
  {journal} {Opt. Express}\ }\textbf {\bibinfo {volume} {24}},\ \bibinfo
  {pages} {3684} (\bibinfo {year} {2016})}\BibitemShut {NoStop}%
\bibitem [{LBT()}]{LBT}%
  \BibitemOpen
  \href {https://www.lbto.org/overview.html} {\enquote {\bibinfo {title} {Large
  binocular telescope observatory},}\ }\BibitemShut {NoStop}%
\end{thebibliography}%
\end{document}